\documentclass[a4paper,fleqn,usenatbib]{mnras}

\PassOptionsToPackage{hyphens}{url}
\usepackage{times}
\usepackage[T1]{fontenc}
\usepackage{ae,aecompl}

\usepackage{graphicx}	
\usepackage{amsmath}	
\usepackage{amssymb}	
\usepackage{pdflscape}

\newcommand{\kms}{km\,s$^{-1}$}
\newcommand{\rsun}{$R_\odot$}

\newcommand{\bell}{$B_\ell$}

\newcommand{\teff}{$T_{\rm eff}$}
\newcommand{\logg}{$\log(g)$}
\newcommand{\vsini}{$v\sin i$}
\newcommand{\vmac}{$v_{\rm mac}$}

\title[MiMeS: Magnetic analysis of O-type stars]{The MiMeS survey of magnetism in massive stars: Magnetic analysis of the O-type stars}

\author[J.H. Grunhut et al.]{
J.H. Grunhut$^{1,2}$\thanks{E-mail: jason.grunhut@gmail.com},
G.A. Wade$^{3}$,
C. Neiner$^{4}$,
M.E. Oksala$^{4,5}$,
V. Petit$^6$,
E. Alecian$^{4,7}$,
\newauthor
D.A. Bohlender$^8$,
J.-C. Bouret$^9$,
H.F. Henrichs$^{10}$,
G.A.J. Hussain$^1$,
O. Kochukhov$^{11}$,
\newauthor
and the MiMeS Collaboration
\\
$^{1}$European Southern Observatory, Karl-Schwarzschild-Str. 2, 85748 Garching bei M{\" u}nchen, Germany\\
$^{2}$Dunlap Institute for Astronomy and Astrophysics, University of Toronto, 50 St. George Street, Toronto, ON, M5S 3H4, Canada\\
$^{3}$Department of Physics, Royal Military College of Canada, PO Box 17000, Kingston, Ontario, K7K 7B4, Canada\\
$^{4}$LESIA, Observatoire de Paris, PSL Research University, CNRS, Sorbonne Universit{\' e}s, UPMC Univ. Paris 06, Univ. Paris Diderot,\\
 Sorbonne Paris Cit{\' e}, 5 place Jules Janssen, 92195 Meudon, France \\
$^{5}$Department of Physics, California Lutheran University, 60 West Olsen Road \#3700, Thousand Oaks, CA, 91360\\
$^{6}$Department of Physics \& Space Sciences, Florida Institute of Technology, Melbourne, FL 32901, USA\\
$^{7}$UJF-Grenoble 1/CNRS-INSU, Institut de Plan{\' e}tologie et d'Astrophysique de Grenoble, UMR 5274, 38041 Grenoble, France\\
$^{8}$Dominion Astrophysical Observatory, Herzberg Astronomy and Astrophysics Program, National Research Council of Canada, \\ 5071 West Saanich Road, Victoria, BC V9E 2E7, Canada\\
$^{9}$Aix Marseille Universit{\' e}, CNRS, LAM (Laboratoire d'Astrophysique de Marseille) UMR 7326, 13388, Marseille, France\\
$^{10}$Anton Pannekoek Institute for Astronomy, University of Amsterdam, Science Park 904, 1098 XH Amsterdam, The Netherlands\\
$^{11}$Department of Physics and Astronomy, Uppsala University, Box 516, SE-75120 Uppsala, Sweden\\
}

\date{Accepted XXX. Received YYY; in original form ZZZ}

\pubyear{2016}

\begin{document}
\label{firstpage}
\pagerange{\pageref{firstpage}--\pageref{lastpage}}
\maketitle

\begin{abstract}
We present the analysis performed on spectropolarimetric data of 97 O-type targets included in the framework of the MiMeS (Magnetism in Massive Stars) Survey. Mean Least-Squares Deconvolved Stokes $I$ and $V$ line profiles were extracted for each observation, from which we measured the radial velocity, rotational and non-rotational broadening velocities, and longitudinal magnetic field $B_\ell$. The investigation of the Stokes $I$ profiles led to the discovery of 2 new multi-line spectroscopic systems (HD\,46106, HD\,204827) and confirmed the presence of a suspected companion in HD\,37041. We present a modified strategy of the Least-Squares Deconvolution technique aimed at optimising the detection of magnetic signatures while minimising the detection of spurious signatures in Stokes $V$. Using this analysis, we confirm the detection of a magnetic field in 6 targets previously reported as magnetic by the MiMeS collaboration (HD\,108, HD\,47129A2, HD\,57682, HD\,148937, CPD-28\,2561, and NGC\,1624-2), as well as report the presence of signal in Stokes $V$ in 3 new magnetic candidates (HD\,36486, HD\,162978, HD\,199579). Overall, we find a magnetic incidence rate of $7\pm3$\%, for 108 individual O stars (including all O-type components part of multi-line systems), with a median uncertainty of the $B_\ell$ measurements of about 50\,G. An inspection of the data reveals no obvious biases affecting the incidence rate or the preference for detecting magnetic signatures in the magnetic stars. Similar to A- and B-type stars, we find no link between the stars' physical properties (e.g. $T_{\rm eff}$, mass, age) and the presence of a magnetic field. However, the Of?p stars represent a distinct class of magnetic O-type stars.
\end{abstract}




\section{Introduction}
\defcitealias{wade16}{Paper~I}
Stars of spectral type O are the most massive and luminous stars in the Universe. Due to their intense UV luminosities, dense and powerful stellar winds, and rapid evolution, they exert an  impact on the structure, chemical enrichment, and evolution of galaxies that is disproportionate to their small relative numbers.

O-type stars are the evolutionary progenitors of neutron stars and stellar-mass black holes. The rotation of the cores of red supergiants \citep{meynet14}, the characteristics of core collapse supernova explosions \citep{heger05}, and the relative numbers, rotational properties and magnetic characteristics of neutron stars (and their exotic component of magnetars) may be sensitive to the magnetic properties of their O-type progenitors. Low-metallicity Oe-type stars have also been associated with the origin of long-soft gamma-ray bursts \citep[e.g.][]{martayan10}.

Considering the importance of O stars as drivers of galactic structure and evolution, and the significance of magnetic fields in determining their wind structure \citep[e.g.][]{shore90, babel97a, uddoula02, townsend05}, rotation \citep[e.g.][]{uddoula09, mikulasek08,townsend10}, and evolution \citep[e.g.][]{meynet11, maeder14}, understanding the magnetic characteristics of O stars is of major current interest.

The sample of known magnetic O stars is currently very small - less than a dozen are confidently-identified \citep{wade15b}. The first magnetic O-type star - the young O dwarf $\theta ^1$\,Ori\, C - was discovered to be magnetic by \citet{donati02}. Measurements of $\theta^1$\,Ori\,C by \citet{wade06} showed that the field is well-described by a dipole configuration with longitudinal magnetic field strength ($B_\ell$) ranging from about $-100$ to 600\,G. Modelling of those measurements revealed that the dipolar magnetic field strength is between 1-2\,kG and that the magnetic field is oblique to the rotation axis by an angle of $\sim$30-70\degr. Only one other O star was confidently detected to be magnetic prior to the start of the Magnetism in Massive Stars (MiMeS) survey (the Of?p star HD\,191612; \citealt{donati06a}). 

Within the context of the MiMeS project, HD\,191612 was re-observed and found to show $B_\ell$ variations from about $-600$ to 100\,G. Similarly to $\theta^1$\,Ori\,C, the field is well-described by a dipole, with a polar field strength of about 2.5\,kG, with a magnetic axis oblique to the rotation axis by about 70\degr. The O supergiant $\zeta$\,Ori\,A is another O-type star with a highly suspected magnetic field \citep{bouret08}. \citet{bouret08} observed this star and found marginal evidence for the detection of a Zeeman signature in their observations; however, based on the temporal variability of these signatures, they were able to establish with more confidence that this star hosted the weakest magnetic field of O stars known at this time, with a surface dipolar field strength of about 60\,G. The field was also found to be oblique to the rotation axis by about 80\degr. This result has recently been confirmed within the context of the MiMeS project by \citet{blazere15}, who identified the $\zeta$\,Ori\,Aa component as the magnetic star with a field strength of $\sim$140\,G. Measurements of another Of?p star, HD\,148937, were reported to find a detected $B_\ell$ ($B_\ell/\sigma=3.1$) by \citet{hubrig08}, but a reanalysis of this observation by \citet{bagnulo12} found a slightly reduced $B_\ell$ value with a correspondingly reduced detection significance of about 2.9, resulting in only a marginal detection of a magnetic field.

The number of confidently detected magnetic O stars has significantly increased since the start of the MiMeS Project. The MiMeS survey alone was responsible for discovering (or confirming the suspicion of) magnetic fields in 6 O stars: HD\,108 \citep{martins10}, HD\,57682 \citep{grunhut09, grunhut12}, HD\,148937 \citep{wade12a}, NGC\,1624-2 \citep{wade12b}, HD\,47129A2 \citep{grunhut13}, CPD-28\,2561 \citep{wade15}. Sufficient data exists, and has been reported, for three of these stars (HD\,57682, HD\,148937, CPD-28\,2561) to characterise their magnetic field properties (further details of these observations are discussed in Sect.~\ref{det_samp_sect}). Similar to the previously known magnetic O stars, the magnetic fields in these stars are well described by a mainly centred dipole field, with a polar surface field strength ranging from about 1-3\,kG, and a magnetic axis inclined to the rotation axis by about 35-80\degr. 

Other authors \citep{hubrig07, hubrig08, hubrig11b, hubrig12b, hubrig13, hubrig14} have also claimed the detection of a magnetic field in 20 other O-type stars, primarily based on low-resolution FORS data. The validity of several of these and other magnetic claims for different classes of stars based on FORS1 observation were investigated by \citet{bagnulo12}. In particular, \citeauthor{bagnulo12}, using the same FORS data but a different analysis, could not confirm the detection of a significant number of the reported FORS1 detections. In light of this result, there are serious doubts about the robustness of the reported magnetic claims based on low-resolution data. Despite these many refuted claims, magnetic field detections have been obtained with low-resolution FORS data. \citet{naze12,naze14} discovered and confirmed the presence of a magnetic field in the cluster star Tr16-22 from a survey consisting of 21 massive stars (including 8 O-type stars). Furthermore, in their study of 50 massive stars (including 28 O-type stars), the BOB collaboration announced the detection of a magnetic field in the O star HD\,54879 \citep{castro15, fossati15}, using a combination of low-resolution FORS2 and high-resolution HARPSpol observations. 

The occurrence of magnetic fields amongst O stars is still debated. Based on the complete sample of known magnetic stars in their study (including non O-type stars), \citet{fossati15} found a magnetic incidence rate of $6\pm4$\%, but they only identified one magnetic detection out of 28 O stars, leading to a slightly smaller magnetic incidence fraction of $\sim$4\%. Although based on a much smaller sample, the study by \citet{naze12} found one magnetic star out of 8 O stars, leading to a much higher incidence rate of $\sim$13\%. These studies, however, deal with small number statistics. Inclusion of any of the previously mentioned studies with refuted claims would also drastically change these statistics. 

The MiMeS survey (\citealt{wade16}; hereinafter `Paper~I') collected over 4800 high resolution circular polarization spectra of roughly 560 bright stars of spectral types B and O. The aim of the survey is to provide critical missing information about field incidence and statistical field properties for a large sample of hot stars, and to provide a broader physical context for interpretation of the characteristics of known magnetic B and O stars. 

In this paper (Paper~II), we report the results obtained for all 97 O-type stars (or multiple star systems) obtained within the survey. In Sect. 2, we summarise the target sample, and review the characteristics of the observations. Sect. 3 discusses the Least-Squares Deconvolution analysis of the spectropolarimetric data, including line mask selection and tuning, line profile fitting to derive line broadening and binary parameters, and ultimately the magnetic field diagnosis. In Sect. 4 we report our results, summarising the magnetic detections obtained for the previous MiMeS discoveries, the possible magnetic detections, and the probable spurious detections. In Sect. 5, we discuss tests performed to investigate the reliability of our results, examine the characteristics of the observations and details pertaining to possible trends or subsamples of stars, and compare our results with previous reports of magnetic stars in the literature. Finally, Sect. 6 provides a summary of this study.


\section{Sample and observations}\label{obs_sect}

As described by \citetalias{wade16}, high-resolution circular polarization (Stokes $V$) spectra of 110 Wolf-Rayet (WR) and O-type targets were collected in the context of the MiMeS Project. Of these targets, 3 magnetic stars ($\theta^1$\,Ori\,C, \citealt{donati02}; $\zeta$\,Ori\,A, \citealt{bouret08}; and HD\,191612, \citealt{donati06a}) were previously known or highly suspected to host a magnetic field and were observed as part of the Targeted Component (TC). The 11 WR stars were previously discussed by \citet{delachevrotiere13} and \citet{delachevrotiere14} and are not further discussed here with the exception of HD\,190918, which also contains a spectroscopic O star companion that is included in this study. In this paper, we focus on the 97 Survey Component (SC) systems that host an O-type star. 

A total of 879 Stokes $V$ observations of these 97 targets were obtained with the ESPaDOnS, Narval and HARPSpol echelle spectropolarimeters. As described by \citetalias{wade16}, these instruments acquire high-resolution ($R=65,000$ for ESPaDOnS and Narval, $R=115,000$ for HARPSpol) spectra spanning the optical spectrum (from 370 nm to 1~$\mu$m for ESPaDOnS and Narval, and from 380-690 nm for HARPSpol). A majority (57\%) of these spectra were obtained in the context of the MiMeS Large Programs (LPs). The remainder (43\%) were collected from the Canada-France-Hawaii Telecsope (CFHT), T\'elescope Bernard Lyot (TBL) and European Southern Observatory (ESO) archives. While a large number of polarimetric sequences were obtained from the archives, some data for all but 5 targets were acquired from the LPs. 

The observed sample of O-type stars is best described as an incomplete, magnitude-limited sample. Approximately 50 bright O stars for which high-resolution IUE spectra exist were identified to be observed during the ESPaDOnS LP, and form the core of the sample. The sample contains a number of stellar subgroups of particular interest for magnetic field investigations, including the peculiar Of?p stars, Oe stars, and weak-wind stars. The Of?p stars were systematically included in the survey (all known Galactic Of?p stars were observed), but other classes of stars (e.g. Oe, weak-wind, etc.) were not systematically targeted, unless specific stars were claimed to be magnetic in the literature.  In most cases, stars for which better magnetic sensitivity was likely to be obtained were prioritised. Hence we preferentially observed brighter stars with lower projected rotational velocities.

Fig.~3 of \citetalias{wade16} illustrates the distribution of apparent $V$-band magnitudes of the entire SC sample. Among the O stars, the brightest star of the sample is $V=1.8$, while the faintest has $V=11.8$. The median magnitude of the O star SC sample is about 6.7, which is about 0.5 mag fainter than the combined sample. 

\citetalias{wade16} discusses the completeness of the SC sample (illustrated in their Fig.~5), and reported that approximately 7\% of all stars with B or O spectral types and brighter than $V=8$ were observed in the survey. However, due to the smaller absolute numbers of bright ($V<8$) O-type stars, the magnitude-limited completeness of the O-type SC sample is much higher: we observed about 43\% of all O stars brighter than $V=8$. This is a natural result of the rapid increase of the total number of bright stars toward late B spectral types, combined with our survey focus on the hottest (hence most massive) objects. So even though we observed only one-quarter the number of O stars as B stars, our sampling of the complete population of bright O stars is actually much better

Often, to increase the signal-to-noise ratio (S/N) sufficiently to reach the desired magnetic sensitivity, we acquired multiple successive Stokes $V$ spectra of a target during an observing night. We ultimately co-added the un-normalized spectra obtained on a given night for each star, which led to 432 individual polarized spectra of the 97 targets. For some stars, only one nightly-averaged observation exists, while for others we have several nightly-averaged observations obtained over the course of the project. The analysis for each star was carried out on the co-added nightly averages. Individual polarimetric sequences were also investigated for those stars with high $v \sin i$ or that were previously known to show variations on time-scales shorter than the timespan of the co-added sequence of observations. In each case we found the results were consistent with the nightly averaged spectra.

The S/N of the co-added spectra ranged from about 50 to 6200, with a median of 1005, as computed from the peak S/N per 1.8\,\kms\ pixel of each spectrum, in the 500-650\,nm range. The large range in obtained S/N is largely a consequence of varying weather conditions, varying brightness of the targets, and differences due to the adopted exposure times (further discussed below). The 210 ESPaDOnS spectra of 87 individual targets were generally of the highest S/N (1059), but they span a large range in precision (the standard deviation of the sample S/N is 722). The 214 Narval spectra of 23 individual targets were of the next highest precision (median S/N of 980, with a standard deviation of 272). Only a small number (7) of O stars were observed with HARPSpol, yielding a median S/N of 470 for 8 co-added spectra (per $\sim$1.8\,\kms\ velocity bin). 

Exposure times for spectra acquired in the context of the LPs were computed using the MiMeS exposure time calculation, which predicts the S/N (and hence exposure time) required to reach a desired ``magnetic sensitivity" (see \citetalias{wade16}, Sect.~3.5). Archival observations, on the other hand, adopted their own strategy for determining exposure times based on the requirements of their individual programmes. Despite the different strategies that may have been adopted, the S/N of the archival data (median S/N$\sim$1100, with a standard deviation of 323) is slightly higher than the data obtained within the MiMeS LPs (median S/N$\sim$950, with a standard deviation of 482).

The sample of SC O-type stars and their basic properties are summarised in Table~5 of \citetalias{wade16}.

\section{Analysis}
\subsection{Least-Squares Deconvolution}\label{lsd_sect}

The Least-Squares Deconvolution technique \citep[LSD;][]{donati97} was applied to all polarimetric spectra to increase the effective S/N in order to detect weak magnetic Zeeman signatures. This multi-line procedure combines information from many metallic and He lines in the spectrum to extract a mean unpolarized intensity profile (Stokes $I$), a mean circularly polarized profile (Stokes $V$), and a mean diagnostic null profile (that characterises spurious signal; e.g. \citealt{bagnulo09}). As input, the procedure requires a ``line mask", which contains the predicted central wavelength, the line depth, and the predicted or measured Land{\' e} factor. The mean Stokes $I$ profile was constructed from the central line depth-weighted average of all lines included in the line mask, while the mean Stokes $V$ profile was constructed from weighting of the product of the central depth, the central wavelength and the Land{\' e} factor of each line in the line mask. Because of this weighting, the LSD procedure is somewhat sensitive to the input line mask \citep[e.g.][]{donati97}. In particular, the presence of emission lines and lines that fail the self-similarity assumption of the LSD procedure (i.e. that are not well represented by the average shape of the majority of the other lines), can add destructively to the final line profile. Thus, care must be taken in the construction of the line mask to reduce the effects of these lines, since a relatively small number of lines are available for LSD in the spectra of hot stars (in contrast to the thousands of lines potentially available in the spectra of cool stars, for example). 

The primary tool used in this study was the {\sc iLSD} code of \citet{kochukhov10} and an {\sc idl} front-end developed by one of us (JHG) to extract all profiles on to a velocity grid with a resolution of 1.8\,\kms. We adopted LSD scaling weights corresponding to a Land{\' e} factor of 1.2 and wavelength of 500\,nm. To further increase the S/N, we also took advantage of the regularisation capabilities of {\sc iLSD} by setting the regularisation parameter ($\lambda$) to a value of 0.2 (see Sect.~\ref{mask_comp_sect} for further details). 

In order to construct optimal line masks, we first utilised the Vienna Atomic Line Database \citep[VALD2;][]{piskunov95, kupka99} to create the initial ``full" line list. The input used appropriate values for their effective temperature (\teff) and surface gravity (\logg), which were based on the spectral type of each star using the corresponding calibration of \citet{martins05a}, and assumed solar abundances. The full mask included all lines retrieved via an {\it extract stellar} request to VALD2 in the range of 370 to 980\,nm, with a line-depth cut-off of 1\% the continuum. This yielded between 1000 to 2500 lines for each mask, decreasing in number with increasing temperature. 

Using an interactive {\sc idl} code that compares the LSD model (the convolution of the LSD profile with the line mask) with the observed spectrum, we proceeded to develop a ``clean" line mask for each observation that excluded all H lines, strong emission lines, lines blended with these lines, and lines blended with strong telluric absorption bands. Finally, we continued to remove all lines that poorly represented the average line profile (e.g. broad He lines) and thus did not satisfy the self-similarity assumption of the LSD procedure.

We next created a ``tweaked" mask, whereby we automatically adjusted the depths of the remaining lines to provide the best fit between the LSD model and the observed Stokes $I$ spectrum. This was carried out using the Levenberg-Marquardt, non-linear least-squares algorithm from the {\sc mpfit} library \citep{more78, markwardt09}. The line depths were constrained to have positive values (i.e. absorption lines). The line mask resulting from the successive procedures of cleaning and tweaking was considered the optimal line mask. We found that, with typically only a few hundred lines in the final optimal line mask for each star, the tweaking procedure can greatly improve the quality of fit between the observed spectrum and the LSD model and also improve our ability to detect Zeeman signatures (see Sect.~\ref{mask_comp_sect} for further details). This last step, which essentially assigns empirical depths to each of the remaining lines, also reduces our sensitivity to the choice of input line mask, which may have a slightly different model \teff, \logg\ or abundances from the observed star. One of the main results of the tweaking process is to increase the strength of the He lines relative to the metallic lines. The process of cleaning and tweaking greatly reduced the total number of lines in the optimal line mask, to about 200 to 1200 lines. In general, stars with lower effective temperature and narrower line widths had the most lines remaining in their masks; there was no correlation between \logg\ (or luminosity class) and the remaining number of lines from the optimisation procedure. Typically, all elements lighter than Cerium remained in the list, with the majority of the lines comprised of He, C, N, O, Ne, and Fe. All data, LSD profiles, and masks for each star in this study are hosted at a dedicated MiMeS page at the Canadian Astronomy Data Centre (CADC\footnote{\tiny{\url{http://www.cadc.hia.nrc.gc.ca/data/pub/VOSPACE/MiMeS/MiMeS_O_stars.html}.}}).

While all stars of the same spectral type and luminosity class (independent of other factors such as line width) used the same initial line mask, we optimised the line mask for each star and each observation separately (the same initial cleaned mask was used for each observation, but each observation was tweaked separately). This strategy essentially treats all observations independently (even for the same star), which, in principle, should maximise our ability to detect weak Zeeman signatures from individual observations for stars with multiple observations and a varying spectrum; however, the LSD profiles extracted from a single mask for stars with multiple observations were very similar to the LSD profiles from the individually tailored masks (the usable lines for the LSD procedure did not vary too substantially). Therefore, a single mask per star could have been used and the results presented here would not differ by much.

From each optimal line mask we also used the multi-profile capability of {\sc iLSD} to simultaneously extract representative mean, unblended profiles of both He and metallic lines. This was accomplished by providing {\sc iLSD} with two input line masks, one entirely composed of He lines and the other consisting of all other remaining lines in the mask. 

In addition to the optimal line mask and its derivatives, we also extracted LSD profiles using the line mask employed by \citet{donati06a} for Of?p star HD\,191612. This line mask contains only 12 lines between 400 and 600\,nm, most of which are He lines, in addition to some CNO lines. Despite the relatively few lines employed in this mask, it has proven to yield the most significant Zeeman detections in the discovery of many recent magnetic O-type stars \citep[e.g.][]{wade11, wade12a, grunhut13}. From hereon out, this line mask is referred to as the Of?p mask. 

The extraction of the final LSD profiles utilised a $\sigma$-clipping procedure (applied to pixel-by-pixel differences between the observed Stokes $I$ spectrum and LSD model). All pixels that differed by more than 50$\sigma$ from the model were rejected and not used in the calculation of the LSD profile. We found this was necessary to reduce the impact of blended telluric features, cosmic rays, echelle ripples, and other general cosmetic issues or spectral contributions that were not of stellar origin.

In a few situations, we encountered extracted LSD Stokes $V$ and diagnostic null $N$ profiles with continuum levels that were systematically offset from zero. This was only observed for observations that were extracted from several co-added high-S/N spectra and the offset appeared to be the same in both Stokes $V$ and $N$. This offset may be due to remnant pseudo-continuum polarization that was not fully subtracted during the {\sc libre-esprit} reduction process. In order to correct for this effect, we fit a linear function of the form $y=mx+b$ to the LSD diagnostic null profile and subtracted this fit from both the LSD Stokes $V$ and $N$ profiles.

The last step in the calculation of the final LSD profiles was to renormalize each profile to its apparent intensity continuum. A line of the form $y=mx+b$ was fit to the continuum regions (determined interactively) about the Stokes $I$ profiles. We then divided all Stokes profiles ($I$, $V$ and $N$) by this fit.

In addition to using the {\sc iLSD} code of \citet{kochukhov10}, we also extracted LSD profiles using the LSD code of \citet{donati97} as a consistency check, as it remains the most commonly used code. Unlike the LSD code of \citeauthor{donati97}, {\sc iLSD} only performs the deconvolution procedure, leaving the user to implement additional operations (some of which have been implemented via our wrapper code, as previously discussed). While the results of the two codes are generally in excellent agreement, the noise characteristics of the LSD profiles can differ in some cases. Furthermore, the use of regularisation, as discussed by \citealt{kochukhov10}, can improve the S/N, which is important in this work as we are searching for weak signals. However, the potentially higher S/N and the difference in the noise characteristics may lead to an increase in spurious signal and the apparent detection of a Zeeman signature (see Sects.~\ref{mag_diagnosis_sect} and \ref{mask_comp_sect} for further details). As discussed by \citealt{donati97}, several factors can lead to a spurious signal, especially for high S/N observations (e.g. rapid variability of the target, spectrograph drifts, inhomogeneities in CCD pixel sensitivities). We suspect that spurious signals in our sample are most likely caused by small variations in the shape of the line profiles (likely due to stellar variability) from one sub-exposure to the next and small differences in the line profile shape between the two polarization spectra (possibly due to differential optical aberrations or non-uniform fibre illumination). Due to the different treatment of the data by each code, in important specific cases, we also mention the results obtained using the \citealt{donati97} code in this paper.

\subsection{Profile fitting}

\subsubsection{Single Stars}
Each of the final LSD Stokes $I$ profiles were fit following the same procedure as discussed by \citet{neiner15}. From the fitting procedure, we derived for each profile the radial velocity $v_r$, the projected equatorial rotational broadening \vsini, contributions remaining from non-rotational broadening, which we consider as macroturbulent broadening \vmac, and the line depth. We emphasise that our goal here is to determine reliable \vsini\ and total line width measurements, and, as discussed by \citet{simon-diaz14}, inclusion of \vmac\ is important to avoid over-estimating \vsini. We warn the reader against over-interpreting the \vmac\ results, as the inclusion of He\,{\sc i} lines and the LSD technique itself, can introduce additional broadening to the final mean profile \citep{kochukhov10}.

Following the strategy adopted by \citeauthor{simon-diaz14}, each observed profile was compared to a synthetic profile that was computed from the convolution of a rotationally-broadened profile with that of a radial-tangential (RT) macrotubulence broadened profile following the parametrisation of \citet{gray05}, assuming equal contributions from the radial and tangential component. A linear limb-darkening law was also used to compute the synthetic profiles, with a limb-darkening coefficient of 0.3, which is appropriate for O-type stars \citep[e.g.][]{claret00}. We adopted the RT macroturbulent formalism in our modelling as \citeauthor{simon-diaz14} have shown a good agreement between their similar profile fitting technique and the more time-consuming (and believed to be more accurate) Fourier technique \citep{gray81}. Furthermore, as the RT broadening does not contribute significantly in the region of the line core, this method should maximise the contribution of rotational broadening to the line profile compared to the more commonly used Gaussian profile for hot OB stars \citep[e.g.][]{martins15}. The total line broadening $v_{\rm tot}$ is obtained by adding the $v\sin i$ and \vmac\ in quadrature.

The fitting procedure uses the {\sc mpfit} library \citep{more78, markwardt09} to find the best fit solution. To further maximise the contribution of rotational broadening, we set the initial guess of $v \sin i$ to the full width half maximum of the profile (identified interactively), and the macroturbulent contribution to one-half of this value. It is certainly possible that the contribution from rotational broadening may be overestimated with this approach and hence our profiles correspond more to ``maximal" rotation profiles. Typical uncertainties for the measurements are on the order of 10-20\%. Results are presented in Appendix~\ref{online_tables_sec}.  

To assess the reliability of our measurements, we compared our results obtained for single stars to those presented by \citeauthor{simon-diaz14}. In total, 44 stars were found to be in common between both studies, with our \vsini\ measurements being about 6\% higher on average, with a standard deviation of 20\%. We noticed a slight trend between the two different subsets of these measurements. Generally, we achieved a poorer agreement with the results of \citeauthor{simon-diaz14} for stars with $v\sin i<100$ (on average our results are 16\% larger compared \citeauthor{simon-diaz14}, with a 20\% standard deviation), compared to stars with $v\sin i > 100$ (our measurements are on average 10\% lower, with a 5\% standard deviation). While there are some differences between the results of the two studies, the agreement appears consistent within our estimated uncertainties (of 10-20\%). 

\begin{figure}
\centering
\includegraphics[width=3.2in]{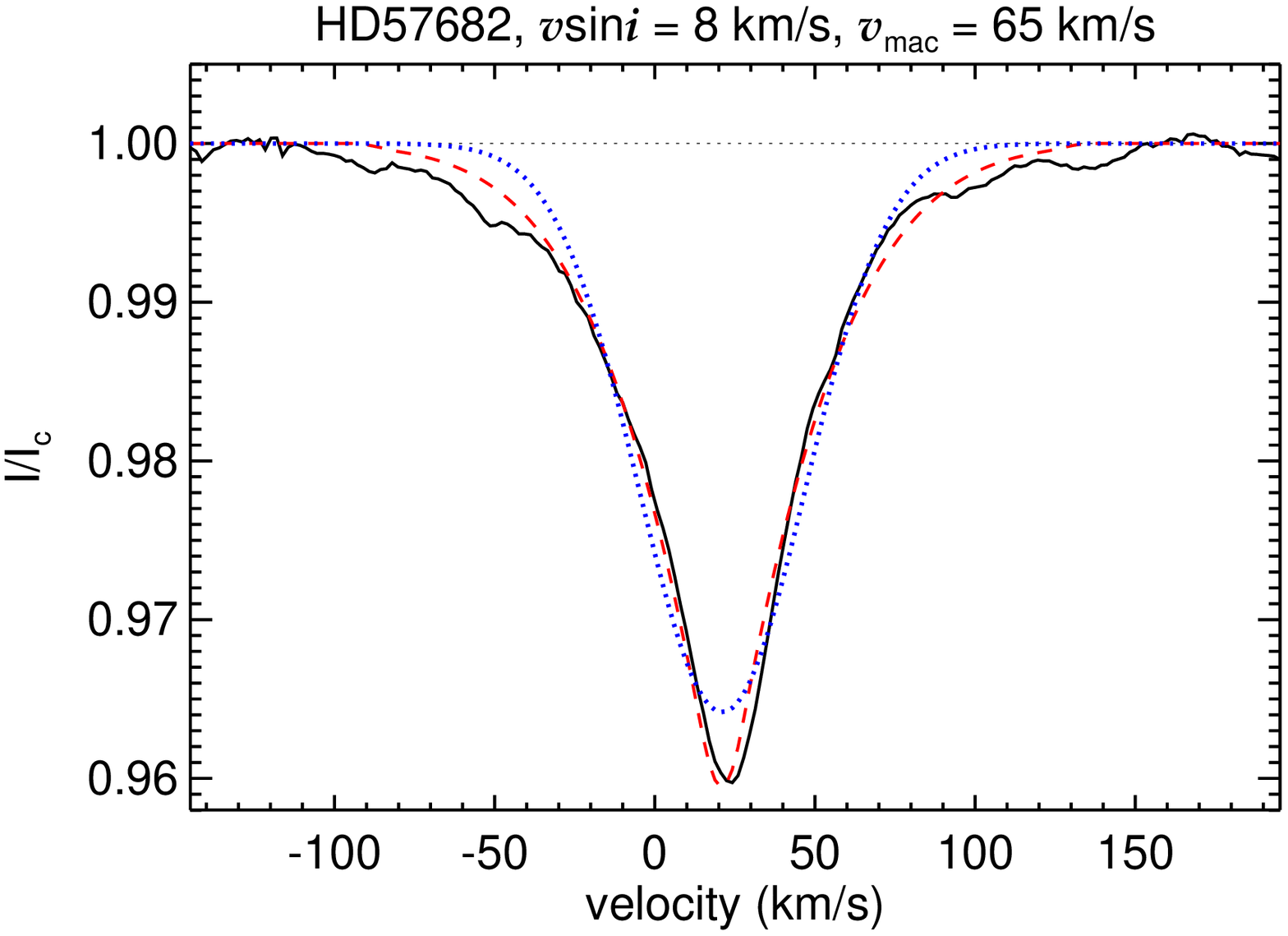}
\par
\vspace{2.5mm}
\includegraphics[width=3.2in]{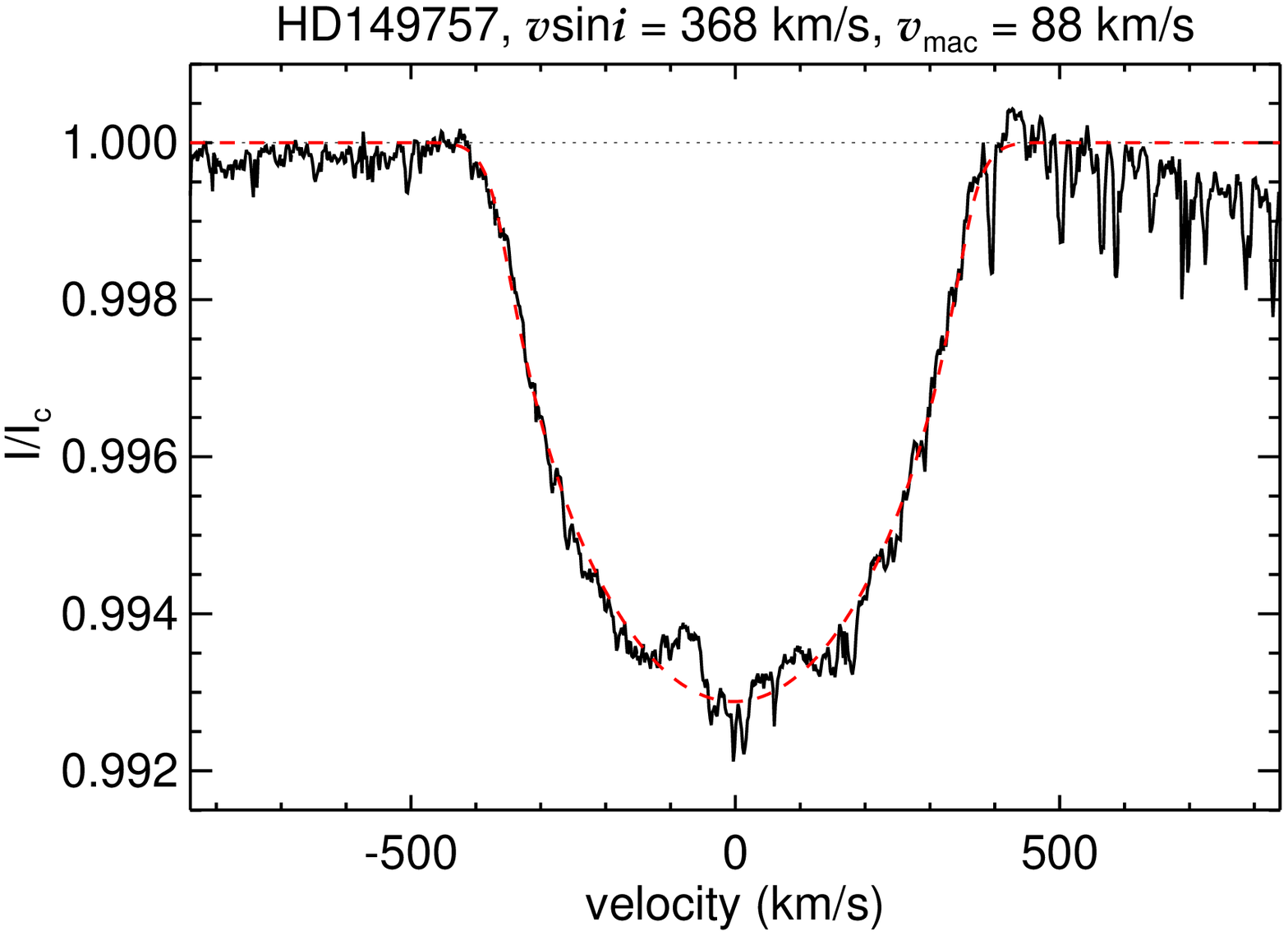}
\caption{Example LSD profiles illustrating the quality of fit of the profile fitting procedure. The observed LSD profile (solid black) is compared with the best-fitting model profile (dashed red) for one profile dominated by macroturbulence (top panel) and another profile that is dominated by rotational broadening (bottom panel). In the macroturbulence dominated case (top panel), we also illustrate the poorer quality of the fit achieved when using a Gaussian (dotted blue) instead of the radial-tangential formulation for macroturbulence, as adopted in this study.} 
\label{single_prof_fit_fig}
\end{figure}

In Fig.~\ref{single_prof_fit_fig}, we illustrate the achieved quality of fit for two examples: one profile that is dominated by macroturbulent broadening (the magnetic star HD\,57682), and one profile with a very high relative contribution of rotational broadening (HD\,149757). In the case of HD\,57682, the fitted parameters (and the quality of the fit) are in better agreement with values derived using the Fourier technique, and additional constraints derived from the measured rotation period as reported by \citet{grunhut12c}, than would be the case using a Gaussian profile to represent macroturbulence (\vsini$\sim$13\,\kms\ when using a Gaussian profile vs 8\,\kms\ using the RT formulation; according to \citealt{grunhut12c}, the \vsini\ should be on the order of 5\,\kms).

\subsubsection{Spectroscopic multiple systems}
For LSD profiles that show signs of multiple spectroscopic components, and for stars that are known spectroscopic binaries, we attempted to simultaneously fit multiple single-star absorption profiles to the observed LSD profile. The individual synthetic fits follow the same description as for the single star case previously discussed, and an overall best fit was determined using {\sc mpfit}. The simultaneous fitting of multiple profiles for a single observation is a difficult task and the solution is often degenerate. We therefore attempted to constrain each fit based on previously published parameters (e.g. \vsini, radial velocity), whenever possible. The details of the fitting attempts are further discussed in Appendix~\ref{binary_appendix}. The best fitting parameters are available in Appendix~\ref{online_tables_sec}. These results are simply used to derive the profile fitting parameters (such as radial velocity, line broadening, and line depth), and are not meant to infer any other (physical) parameter of the systems (such as radius/luminosity ratios).

\begin{figure}
\centering
\includegraphics[width=3.2in]{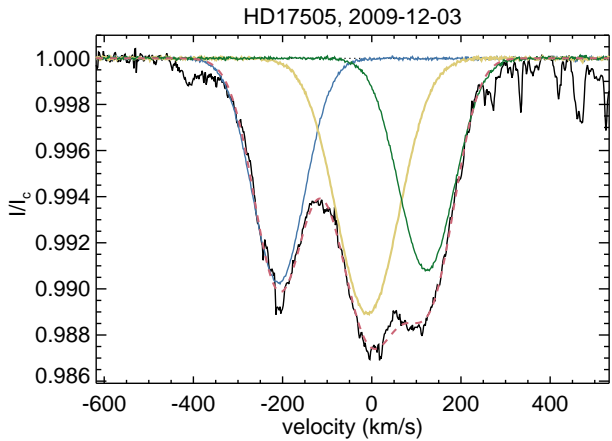}
\caption{Example LSD profile showing the quality of fit of the multi-profile fitting procedure. The observed LSD profile (solid black) is compared with the best-fitting profiles for each component (indicated by different colours). The thick dashed line shows the co-added profile of the individual components, while the thin horizontal dashed line indicates the continuum level. The additional features in the observed LSD profile reflect telluric blends from some line regions that persist into the final LSD profile.}
\label{binary_prof_fit_fig}
\end{figure}

We next constructed semi-empirical `disentangled' profiles for each component in the observed LSD profile by combining the best-fitting Stokes $I$ profile model for each component (random Gaussian noise is also added, in accordance with the S/N of the observation, to preserve the relative noise contribution from each profile for future calculations; however, residual telluric features are the dominant source of ``noise" in most profiles, which is not accounted for) with the observed Stokes $V$ and diagnostic $N$ profiles. In the case of non-detections, the use of the fitted profiles better enabled us to determine spectroscopic and magnetic measurements (such as the longitudinal field) for each component separately. This is due to the fact that the velocity limits and a more representative equivalent width measurement could be determined from the separated profiles. In Fig.~\ref{binary_prof_fit_fig}, we provide an example of the achieved quality of fit for an LSD profile showing multiple components. A comparison for all stars is provided in Fig.~\ref{bin_profs_fig}, in Appendix~\ref{binary_appendix}.

\subsection{Magnetic diagnosis}\label{mag_diagnosis_sect}
As discussed in Sect.~3.4 of \citetalias{wade16}, our primary method for establishing the presence of a magnetic field relies on the detection of excess signal in the LSD Stokes $V$ profile, resulting from the longitudinal Zeeman effect, based on the calculation of the False Alarm Probability (FAP), as described by \citet{donati92}. We quantify the likelihood that a Zeeman signature was detected by measuring the FAP computed from each LSD Stokes $V$ profile, within the confines of the Stokes $I$ line profile \citep[as determined visually][]{donati92}. Following \citet{donati97}, we consider a Zeeman signature to be definitely detected (DD) if the excess signal within the line profile results in a FAP $<10^{-5}$. If the FAP is greater than $10^{-5}$ but less than $10^{-3}$, a signature is considered marginally detected (MD). A FAP greater than $10^{-3}$ is considered a non-detection (ND). In addition to establishing the presence of excess signal within the line profile, we further require that no excess signal is measured outside of the line profile. In the case of some strongly magnetic stars, residual incoherent polarization signal may remain outside of the line profile. However, in such cases the magnetic signal is sufficiently strong that there is no ambiguity concerning its detection; in such cases signatures are usually detectable in individual spectral lines as well. An additional criterion for the evaluation of the reality of the signal is that no excess signal is detected in the null profile. However, radial velocity motions or other line profile variations that occur on timescales of a single polarimetric sequence can result in residual uncancelled signal in the diagnostic null for stars with Zeeman signatures in Stokes $V$. The magnetic signal is typically only slightly affected and sufficiently strong that there is no ambiguity concerning its detection. This problem is common for pulsating stars \citep[e.g.][]{neiner12e}. In the event that a FAP leads to a detection within the line profile (FAP $<10^{-5}$), but fails one or more of the other criteria, we consider this to be a marginal detection. Visual inspection of the detected profiles is further carried out to confirm the detection status.

We note that this adopted approach is sensitive to any deviations of the Stokes $V$ profile within the confines of the line profile. In principle, many systematics could result in spurious detections (e.g. rapid variability of the target, spectrograph drifts, inhomogeneities in CCD pixel sensitivities), in addition to random noise. Quality control checks carried out using the large number of null detections (see Sect.~\ref{quality_sect}) or analyses performed on the TC (see, for example, \citetalias{wade16}) lead us to understand that the incidence of such artefacts is quite low. Furthermore, examination of the shape, and the coherence of the temporal variation of the Stokes $V$ profile is the best method for verification. Using this guideline as a basis, we consider a detection spurious when the Stokes $V$ profile does not reveal any obvious Zeeman signature and/or the coherence of the temporal variation of this signature is inconsistent with expectations (e.g. the signal is statistically detected in only a few of many observations of similar S/N). {\it A priori}, we do not know which stars are magnetic, but, in general, the stars for which we confidently detect magnetic fields show clear evidence of a Zeeman signature in several observations, and, furthermore, the temporal variations of these signatures behave within expectations. The sample of confidently detected magnetic stars have been previously reported by the MiMeS collaboration and consists of: HD\,108 \citep{martins10}, HD\,57682 \citep{grunhut09, grunhut12}, HD\,148937 \citep{wade12a}, NGC\,1624-2 \citep{wade12b}, HD\,47129A2 \citep{grunhut13}, CPD-28\,2561 \citep{wade15}. For some stars, there is clear evidence for a Zeeman signature in at least one observation, but we lack a sufficient number of observations to confirm this detection. We consider these stars to be potential magnetic candidates.

Since the total velocity width of the line profiles varies substantially from one star to another, we devised a procedure to determine the optimal width of the velocity bin (yielding the most precise magnetic diagnosis) for each individual extracted LSD profile. This was accomplished by maximising the likelihood of detecting a magnetic field by searching for the bin width that provided the lowest FAP. This optimisation requires a delicate balance between increasing the bin width (thereby increasing the S/N per bin) and at the same time decreasing the amplitude of any potential Zeeman signature. To avoid the latter, the maximum allowed bin width was chosen such that the line profile must span a minimum of 20 bins (where possible, limited by the adopted minimum velocity width of 1.8\,\kms\ for the LSD profiles extracted from all instruments - which corresponds to the spectral pixel width for ESPaDOnS and Narval - and the intrinsic width of the line profile). This value was chosen based on our experience of modelling Stokes $V$ profiles resulting from large-scale magnetic fields.

In addition to quantifying the detection of a Zeeman signature using the FAP, we also computed the mean longitudinal magnetic field $B_\ell$ using the unbinned profiles from each observation. The longitudinal field was determined using the first-order moment of the Stokes $V$ profile \citep{rees79, mathys89, donati97, wade00}:
\begin{equation}
\centering
B_\ell  = -21.4\times10^11 \frac{\int(v-v_0)V(v)dv}{\lambda z c \int[1-I(v)]dv}.
\end{equation}

\noindent In this equation, $V(v)$ and $I(v$) represent the continuum normalized Stokes $I$ and $V$ profiles. The mean Land{\' e} factor ($z$) and mean wavelength ($\lambda$) correspond to the LSD weights adopted in our analysis (1.2 and 500\,nm, respectively), while $c$ is the speed of light. The integration limits are the same as the ones used for the FAP analysis. The uncertainties were computed by propagating the individual uncertainties of each pixel following standard error propagation rules (see \citealt{landstreet15} Eq.~3 for further details). We also computed similar measurements from the diagnostic null profile $N_\ell$ using the same integration limits. Results are available in Appendix~\ref{online_tables_sec}. The $B_\ell$ measurements were not used to establish the presence of a magnetic field, as it is possible that a particular magnetic geometry could lead to a net null $B_\ell$ measurement, but the velocity-resolved Stokes $V$ profile still shows a clear Zeeman signature due to the combination of the Zeeman and Doppler effects for large-scale fields.

The same analysis described above for single stars was also performed on the disentangled profiles extracted from observations of systems with multiple components. This allowed us to establish magnetic measurements and detection criteria for each component individually; however, this procedure naturally does not account for any possible magnetic contamination in the Stokes $V$ signal from overlapping profiles, i.e. it assumes that the other components are not magnetic. Results are available in the Appendix~\ref{online_tables_sec}.

\subsection{Mask comparison}\label{mask_comp_sect}

For each observation, we extracted at least 6 LSD profiles using each of the different line masks discussed in Sect.~\ref{lsd_sect}:

\begin{itemize}
\item the original line mask derived from the VALD request
\item the `cleaned' version of the VALD line mask
\item the optimal `cleaned and tweaked' VALD line mask
\item the He-line only `cleaned and tweaked' VALD line mask
\item the metal-line only `cleaned and tweaked' VALD line mask
\item the Of?p line mask.
\end{itemize}

For each mask, we examined the binned and unbinned versions of the resulting LSD profiles. When comparing the results for the optimal line mask, we found a noticeable difference in the number of detections among the known magnetic sample. In this case, the optimally binned profiles resulted in about 3 times more detected Zeeman signatures (61) compared to the unbinned profiles (22). This result emphasises the importance of this procedure for such a large sample of stars with different line widths. From this point forward, all discussion of the detection criteria corresponds to the optimally binned profiles, unless otherwise specified.

We also extracted additional LSD profiles with varying values of the regularisation parameter. Regularisation is important, since, as discussed by \citet{kochukhov10}, it can improve the achievable S/N, which is important in this study as we are searching for weak Zeeman signatures. To assess the performance of the different masks and the procedures, we investigated the number of detections, both real and presumably spurious (i.e. formal detections obtained from observations from the unconfirmed magnetic star sample). The regularisation parameter was modified between 0 and 0.5 (where a higher value increases the amount of regularisation) and the results are presented in Table~\ref{reg_res_tab}. As we increased the amount of regularisation, we found an increase in the number of detections among the confirmed magnetic sample, but this also led to an even larger fraction of apparently spurious detections. Ultimately, we adopted a value of 0.2 as it provided a reasonable balance between the number of detections belonging to the confirmed magnetic stars and the number of potentially spurious detections. Finally, we note that several of the previously reported magnetic stars would not have been detected in this analysis without regularisation (HD\,148937, CPD-28\,2561).

\begin{table}
\centering
\caption{Summary of regularisation tests conducted with {\sc iLSD}. Listed are the regularisation value used for the given test, the number of observations of confirmed magnetic stars that resulted in detections, and the number of observations from non-magnetic stars that resulted in potentially spurious detections.}
\label{reg_res_tab}
\begin{tabular}{ccc}
\hline
Regularisation value & Confirmed & Spurious \\
\hline	
0.00 & 29 & 0 \\				
0.05 & 39 & 3 \\
0.10 & 58 & 4 \\
0.20 & 61 & 9 \\
0.30 & 71 & 15 \\
0.40 & 72 & 38 \\
0.50 & 78 & 56\\ 
\hline
\end{tabular}
\end{table}

We next attempted to assess the performance of the different masks. In particular, we compared the FAP and the detection status from the sample of confirmed magnetic stars. The results are listed in Table~\ref{mask_res_tab}. The main conclusion from this comparison is that the optimal line mask provided the largest number of detected Stokes $V$ signatures.  Compared to the original VALD line mask, the optimal line mask provided about a 50\% increase in the number of detected profiles (61 vs 39). We conclude that `tweaking' is an important step to improve the ability to detect weak Zeeman signatures in O stars, since results from the `cleaned' line mask did not increase the total number of detections compared to the original line mask. To further emphasise the importance of this procedure, we note that the increased number of detections when using the additional step of  `tweaking' are not limited to just a larger number of MDs. In fact, we found a much larger improvement in the number of DDs (41 vs 20) compared to a small increase in MDs (20 vs 19), when comparing these two categories of line masks.

\begin{table}
\centering
\caption{Performance comparison of different masks. Included for each mask is its identifier, the number of MDs and DDs, and the total number of detected observations for the known magnetic star sample. Lastly, we list the total number of potentially spurious detected observations among the presumably non-magnetic stars.}
\label{mask_res_tab}
\begin{tabular}{ccccc}
\hline
Mask & \multicolumn{3}{c}{Confirmed magnetic} & Spurious \\
\ & MD & DD & total & total \\
\hline	
Original & 14 & 25 & 39 & 7\\
Cleaned & 19 & 20 & 39 & 4\\
Optimal & 20 & 41 & 61 & 9\\
He only & 20 & 39 & 59 & 16\\
Metal only & 11 & 27 & 38 & 20 \\
Of?p & 14 & 29 & 43 & 13\\
\hline
\end{tabular}
\end{table}

The LSD profiles extracted from the He-only line mask provided the next largest number of detections (59), which likely reflects the fact that strong He lines dominate the Zeeman signal; however, we still found a large number of detected profiles with the metal line only line mask (38). While the Of?p line mask has proven to yield the most significant Zeeman detections in the recent discovery of a number of magnetic O stars, our study finds that this mask resulted in considerably fewer detections compared to some of the other line masks (43); however, in some situations, this mask provided a marked improvement compared to the optimal line mask (e.g. HD\,47129A2, CPD-28\,2561).  

Using the observations of the magnetic star sample is one way to evaluate the line masks, but it is also important to consider the non-magnetic sample. In this respect we are interested in the number of apparently spurious detections resulting from the use of a given line mask. From this comparison we found that the metal line only line mask resulted in the largest number of spurious detections, while a similar number of spurious detections were also found from the  He-only line mask. The Of?p line mask and optimal line mask had a similar number of potentially spurious detections. We do note that some of the apparently spurious detections are potential magnetic candidates (as further discussed below).

As previously mentioned, different strategies adopted by different LSD codes can also affect these results. If instead we used the LSD code of \citet{donati97}, we found fewer spurious detections using the optimal line mask (5 vs 9), but {\sc iLSD} also resulted in a larger number of detected profiles among the confirmed magnetic star sample (61 vs 54; see Table~\ref{lsd_code_res_tab} for a summary). Some of the spurious detections were the same between both codes (HD\,34078 - both MD; HD\,162978 - both MD; HD\,199579 - DD with {\sc iLSD}, MD with \citeauthor{donati97} code), some had spurious detections for the same star, but with different observations (HD\,24912 - 2006-12-14 resulted in a MD with {\sc iLSD}, 2007-09-10 resulted in a MD with \citeauthor{donati97} code; HD\,47129A1 - 2012-09-28 resulted in a MD with {\sc iLSD}, 2012-02-09 resulted in a MD with \citeauthor{donati97} code), while others were only detections with {\sc iLSD} (HD\,36486, HD\,66811, HD\,167264, HD\,209975). Some of the discrepancies may be attributed to the differences in the achieved S/N between the two codes, likely a result of the improvement afforded by the use of regularisation with {\sc iLSD}. In the cases where {\sc iLSD} resulted in a lower FAP (and a different detection threshold), the S/N achieved with {\sc iLSD} was anywhere between 2\% lower to 55\% higher than what was found with the \citeauthor{donati97} code, with a median improvement of about 10\%. Furthermore, some of the apparently spurious detections are in fact considered possible magnetic stars, and the achieved detection status sometimes differed between both codes (HD\,162978, HD\,199579 - detected with both codes; HD\,36486 - only detected with {\sc iLSD}; see Sect.~\ref{det_samp_sect} for further details). Further discussion of all stars and observations with formal detections is provided in the following sections.

\begin{table}
\centering
\caption{Comparison of results obtained with {\sc iLSD} and the LSD code of \citet{donati97}. Listed are the total number of MDs and DDs obtained for the previously known magnetic stars, and for the sample of presumably non-magnetic stars.}
\label{lsd_code_res_tab}
\begin{tabular}{lcccc}
\hline
Code & \multicolumn{2}{c}{Magnetic} & \multicolumn{2}{c}{Non-magnetic} \\
\ & MD & DD & MD & DD \\
\hline	
{\sc iLSD} & 20 & 41 & 5 & 4 \\
\citet{donati97} & 17 & 37 & 5 & 0\\
\hline
\end{tabular}
\end{table}

Based on this comparison of all LSD profiles extracted from the various line masks, we conclude that the optimal line mask is the most suitable choice for the aims of this survey. This line mask generally provides the highest S/N for the resulting LSD line profiles and also results in the highest success of detecting Zeeman signatures. In the following, all results, unless otherwise stated, are based on the optimal line mask.

Comparing the results obtained here with previous studies of O stars performed using the same data \citep[e.g.][]{grunhut09, martins10, grunhut12c, wade12a, wade15,wade12b}, we note that there are differences in the details of the measurements due to the use of different masks. However, all stars that were previously detected remain detected (in fact, for most stars the quality of the profiles and the statistical significance of the detections is improved thanks to the optimal binning, regularisation, or sometimes a better line mask). Any basic parameters determined from the published magnetic measurements (rotational periods, magnetic dipole field strengths/geometries), are in good agreement with similar measurements determined from the homogeneous analysis presented here.

\section{Results}

Of the 97 O star targets, we identified 28 targets belonging to spectroscopic multiple star systems. Two of these are newly suspected multi-line spectroscopic systems (HD\,46106, HD\,204827). \citet{simon-diaz06} presented evidence for the possible presence of a spectroscopic companion in HD\,37041, which we confirm with greater certainty in this work. The rest of systems were previously known to exhibit multi-line spectra. Twelve of these systems contain at least one O-type star companion. A spectral classification was not carried out for the newly suspected multi-line systems, so we cannot establish if the companions in these systems are also O stars. 

From the 28 systems with evidence for multi-line profiles, we could not reliably disentangle 6 systems, although we note (as discussed in Appendix~\ref{binary_appendix}) that mean profiles of a few of these systems are likely dominated by just a single component. Therefore, we evaluated the magnetic properties of 69 presumably single O-type stars, 18 systems with only a single O-type star, 9 systems with two O-type stars, and 1 system with 3 O-type stars, leading to a total of 108 O stars analysed in this study.

Table~\ref{det_samp_tab} summarises the basic characteristics of the sample of stars for which we obtained a detection of a Zeeman signature in at least one observation. From this analysis, we confirm that 6 out of our 97 survey targets are confidently detected to be magnetic. Furthermore, our data suggest the possibility that an additional 3 targets could also host detectable magnetic fields, although we lack sufficient evidence for confirmation. Lastly, we find marginal evidence for the formal detection of a signal in the Stokes $V$ profiles of 6 additional stars in our sample. A careful inspection of these profiles does not reveal any obvious Zeeman signature, and we ultimately conclude that the excess signal is of spurious origin. 

None of the other 82 O stars (or systems) evaluated here result in a formal detection of signal in the mean Stokes $V$ profile based on the FAP analysis with the optimal mask. The incidence of detected magnetic fields in our survey sample is 6 over 97 star systems, i.e. $6.2\pm 2.6$\% of the O star systems we observed are confirmed to be magnetic, where the uncertainties are derived from counting statistics\footnote{The uncertainty derived from counting statistics assumes that the uncertainty on $N$ counts is $\sqrt N$. The uncertainties are then propagated according to standard rules.}. Including all individual O stars that are part of multiple star systems, we find an incidence rate of 6 out of 108 stars, or $5.6\pm2.3$\%. Finally, including the potential magnetic candidates, we obtain an incidence fraction between 5.6 and 8.3\%. From this range, we arrive at a final magnetic incidence fraction of $7\pm3$\%, where the uncertainty takes into account the additional uncertainty stemming from counting statistics.

\subsection{The detected sample}\label{det_samp_sect}

\begin{table*}
\centering
\caption{Summary of the observations of stars for which we obtain a formal detection of signal in the Stokes $V$ profile based on the FAP analysis discussed in the text. The first set of stars show overwhelming evidence for the presence of a magnetic field and are thus considered confirmed magnetic stars. The next group of stars exhibit clear significant structure that is qualitatively consistent with the Zeeman effect in their Stokes $V$ profiles, but we lack sufficient evidence to confirm the presence of a magnetic field in these stars. The last group of stars have a formal detection of signal in Stokes $V$ in some of the analysed LSD profiles, but show no strong evidence that this signal is a result of a magnetic field. These are therefore considered to be spurious detections. Listed for each star is its HD, CPD, or NGC designation, common name, spectral type, $B$-band magnitude, the total number of nightly-combined observations, the number of marginal detections (MD), the number of definite detections (DD), the range of variation of the measured longitudinal field ($B_\ell$), the median uncertainty of the $B_\ell$ measurements ($\sigma$), the median $v\sin i$, the median $v_{\rm mac}$ of all observations, and the median total line broadening (computed from adding $v\sin i$ and $v_{\rm mac}$ in quadrature) of all the observations, each inferred from the LSD profiles.}
\label{det_samp_tab}
\begin{tabular}{lccc|rrrrr|rrr}
\hline
Name	& Common &	Spec	&	$B$	&	\#	&	\#	&	\#	&	\multicolumn{1}{c}{$B_\ell$ range} & $\sigma$ & $v\sin i$	&	$v_{\rm mac}$	&	$v_{\rm tot}$\\
\	& name	&	type	&	mag	&	Obs	&	MD	&	DD	& \multicolumn{1}{c}{(G)} & \multicolumn{1}{c}{(G)} &(\kms)	&	(\kms)	&	(\kms)\\
\hline																							
\multicolumn{12}{c}{{\bf Confirmed magnetic stars}}\\
HD\,108$^{1,2}$	&	\	&	 O8f?p	&	7.58	&	37	&	5	&	16	&	-152,+27	&	20	&	122	&	93	&	153	\\
HD\,47129A2$^{3,4}$	&	Plaskett's star	&	O7.5\,III	&	6.11	&	21	&	4$^*$	&	6$^*$	&	-1235,+807	&	296	&	370	&	183	&	413	\\
HD\,57682$^5$	&	\	&	O9.5\,IV	&	6.24	&	20	&	3	&	17	&	-121,+246	&	12	&	10	&	62	&	63	\\
HD\,148937$^6$	&	\	&	O6f?p	&	7.12	&	17	&	1	&	6	&	-736,+361$^{**}$	&	268	&	71	&	130	&	148	\\
CPD-28\,2561$^7$	&	\	&	O6.5f?p	&	10.13	&	21	&	6$^{***}$	&	1$^{***}$	&	-1752,+981	&	411	&	20	&	197	&	198	\\
NGC\,1624-2$^{8,9}$	&	\	&	O7f?p	&	12.4	&	12	&	2	&	7	&	8,+4378	&	448	&	14	&	80	&	81	\\
\hline																									
\multicolumn{12}{c}{{\bf Possible magnetic stars}}\\																									
HD\,36486	&	$\delta$ Ori A	&	O9.5\,IINwk	&	2.02	&	2	&	0	&	1	&	-151,+48	&	64	&	121	&	105	&	160	\\
HD\,162978	&	63 Oph 	&	O8\,II(f)	&	6.24	&	2	&	1	&	0	&	-12,+113	&	17	&	78	&	111	&	136	\\
HD\,199579	&	HR\,8023	&	O6.5\,V((f))z	&	6.01	&	1	&	0	&	1	&		-159,+183		&	17	&	60	&	126	&	140	\\
\hline																									
\multicolumn{12}{c}{{\bf Spurious detections}}\\																									
HD\,24912	&	$\xi$ Per	&	O7.5\,III(n)((f))	&	4.08&	13	&	1	&	0	&	-62,+105	&	20	&	203	&	90	&	222	\\
HD\,34078	&	AE\,Aur	&	O9.5\,V	&	6.18	&	4	&	1	&	0	&	-252,+19	&	10	&	17	&	55	&	58	\\
HD\,47129A1	&	Plaskett's star	&	O8\,III/I	&	6.11	&	21	&	1	&	0	&	-88,+522	&	34	&	77	&	75	&	107	\\
HD\,66811	&	$\zeta$ Pup	&	O4\,If	&	1.98	&	2	&	1	&	0	&	-4,+30	&	18	&	187	&	178	&	258	\\
HD\,167264	&	15 Sgr	&	O9.7\,Iab	&	5.42	&	9	&	0	&	1	&	-17,+105	&	24	&	59	&	108	&	123	\\
HD\,209975	&	19 Cep	&	O9\,Ib &	5.19	&	10	&	0	&	1	&	-95,+32	&	12	&	71	&	105	&	127	\\
\hline
\multicolumn{12}{l}{Notes: $^*$Using the Of?p star line mask; $^{**}$Ignoring the two poorest quality observations; $^{***}$Using the Of?p star line mask and the}\\
\multicolumn{12}{l}{\citet{donati97} LSD code. Additional references: $^1$\citet{martins10}; $^2$Shultz et al. (in prep); $^3$\citet{grunhut13}; $^4$Grunhut et al.}\\
\multicolumn{12}{l}{(in prep); $^5$\citet{grunhut12c}; $^6$\citet{wade12a}; $^7$\citet{wade15}; $^8$\citet{wade12b}; $^9$Macinnis et al. (in prep).}
\end{tabular}
\end{table*}

\begin{figure*}
\centering
\includegraphics[width=2.3in]{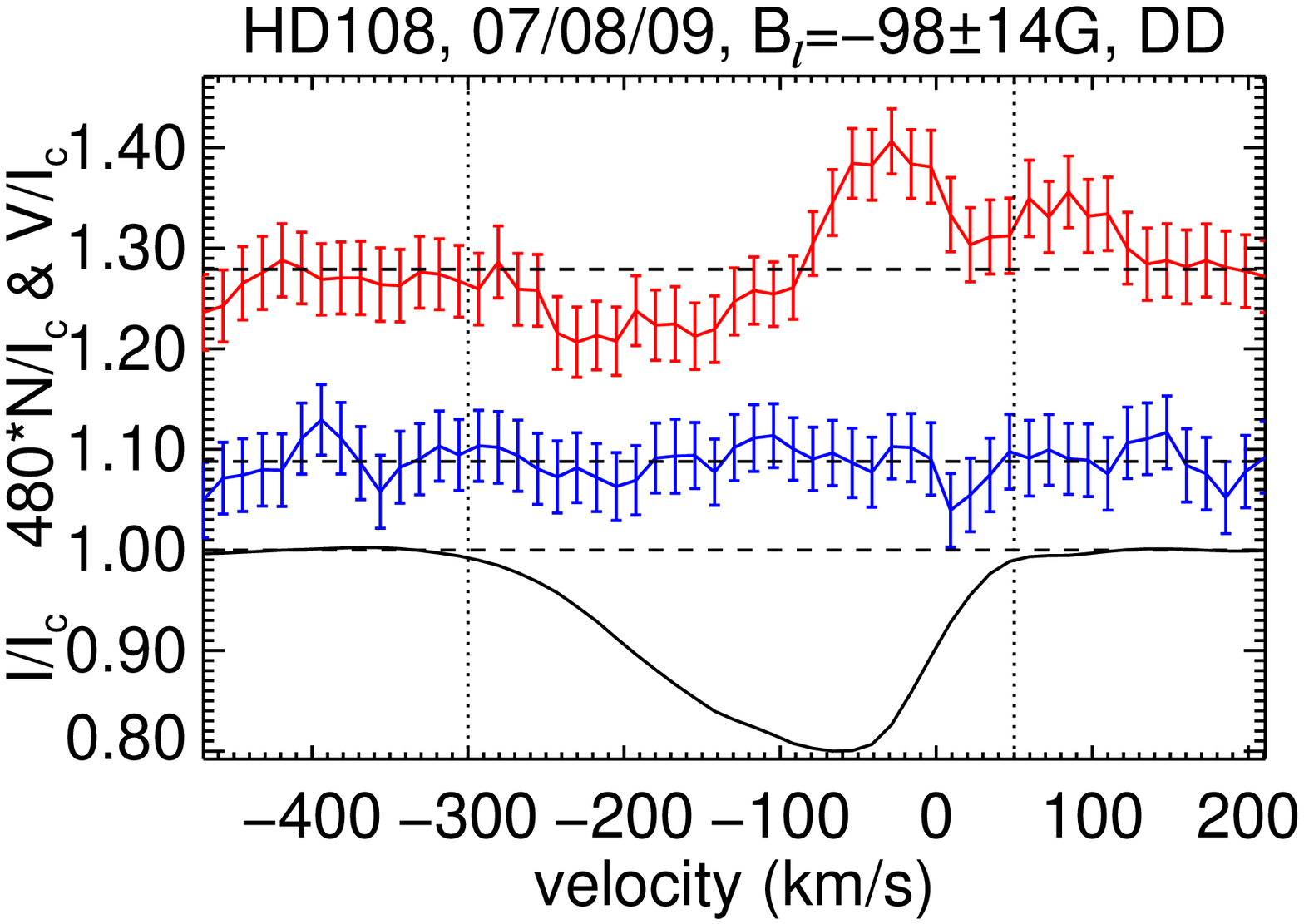}
\includegraphics[width=2.3in]{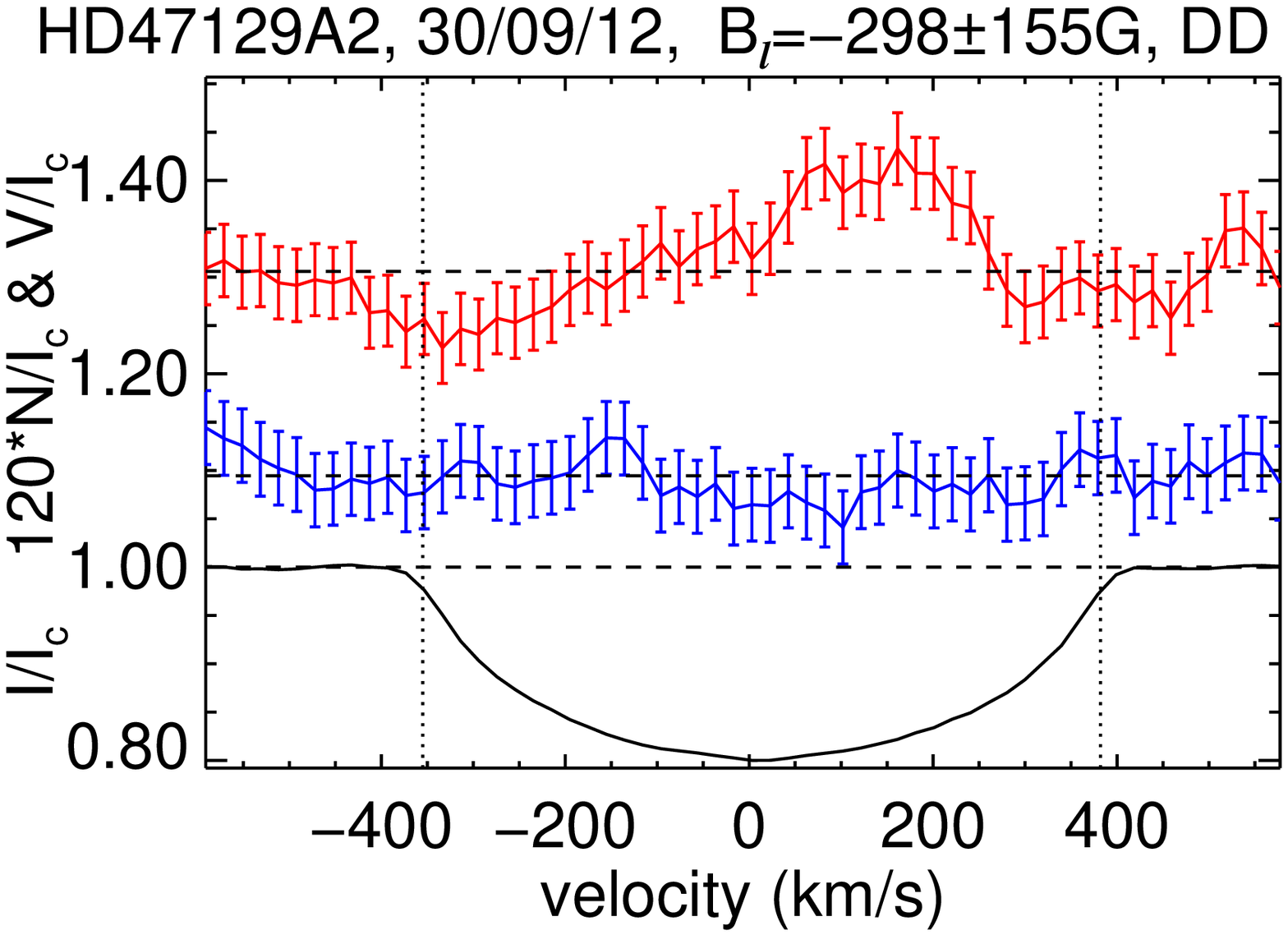}
\includegraphics[width=2.3in]{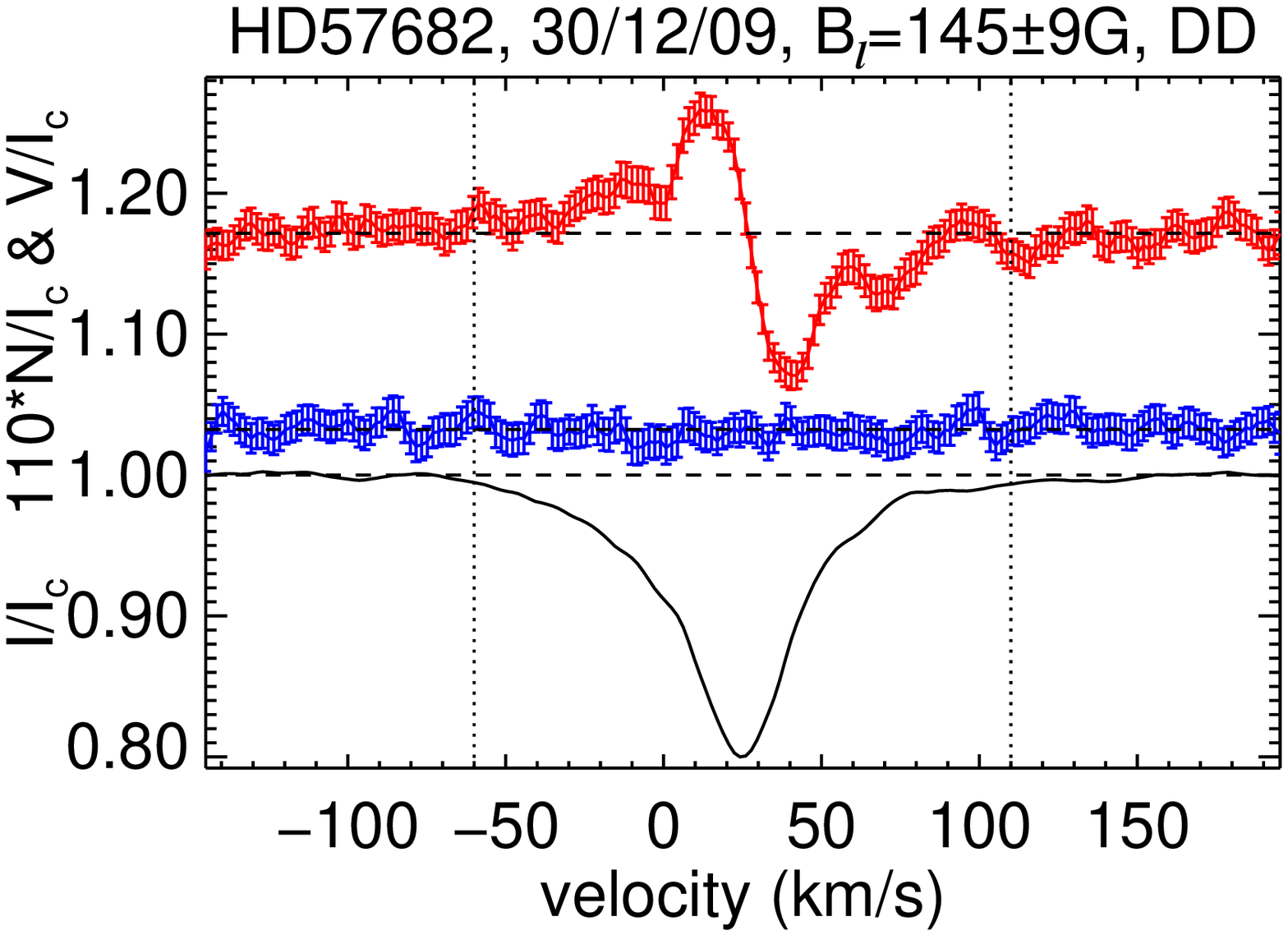}\\
\includegraphics[width=2.3in]{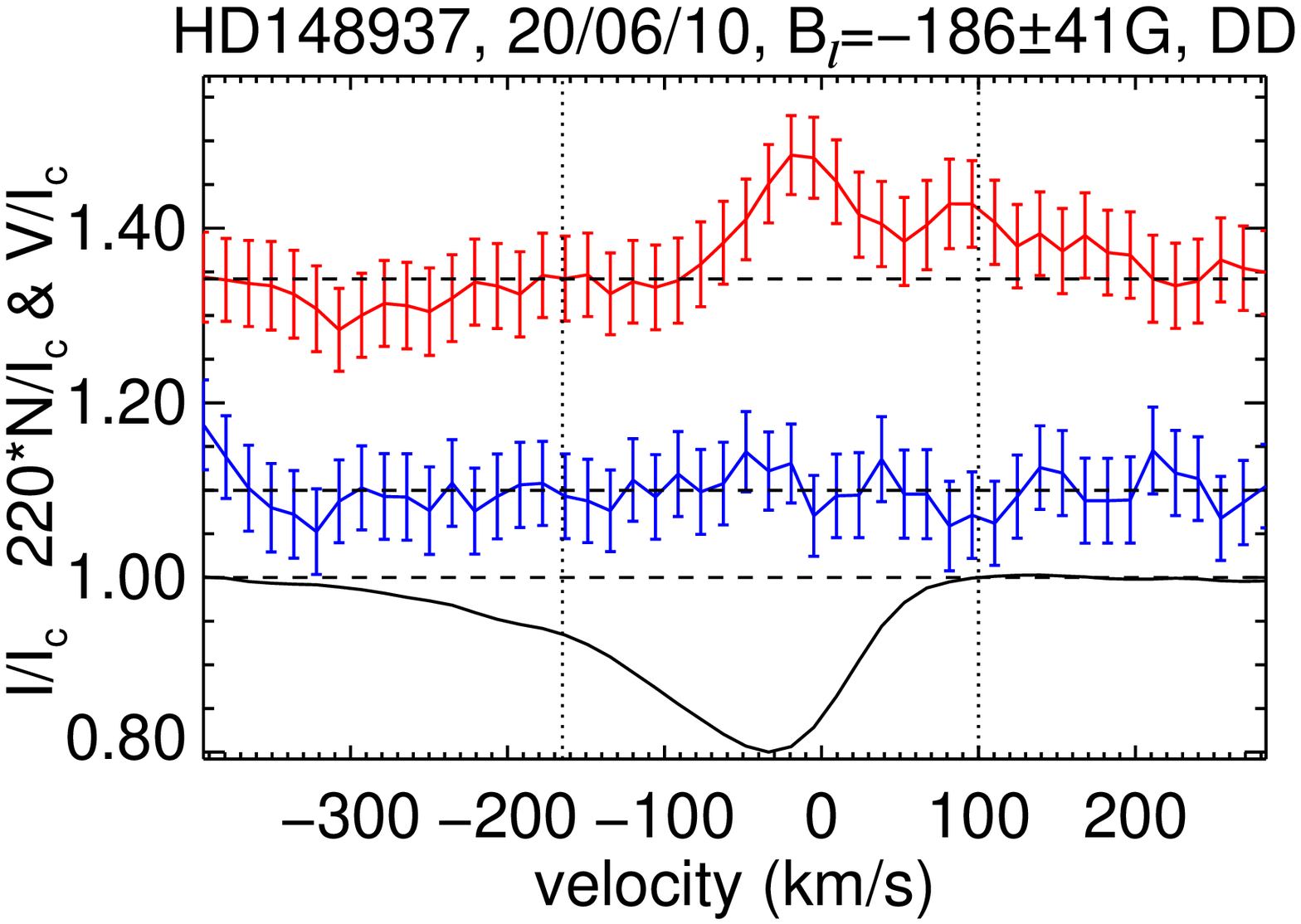}
\includegraphics[width=2.3in]{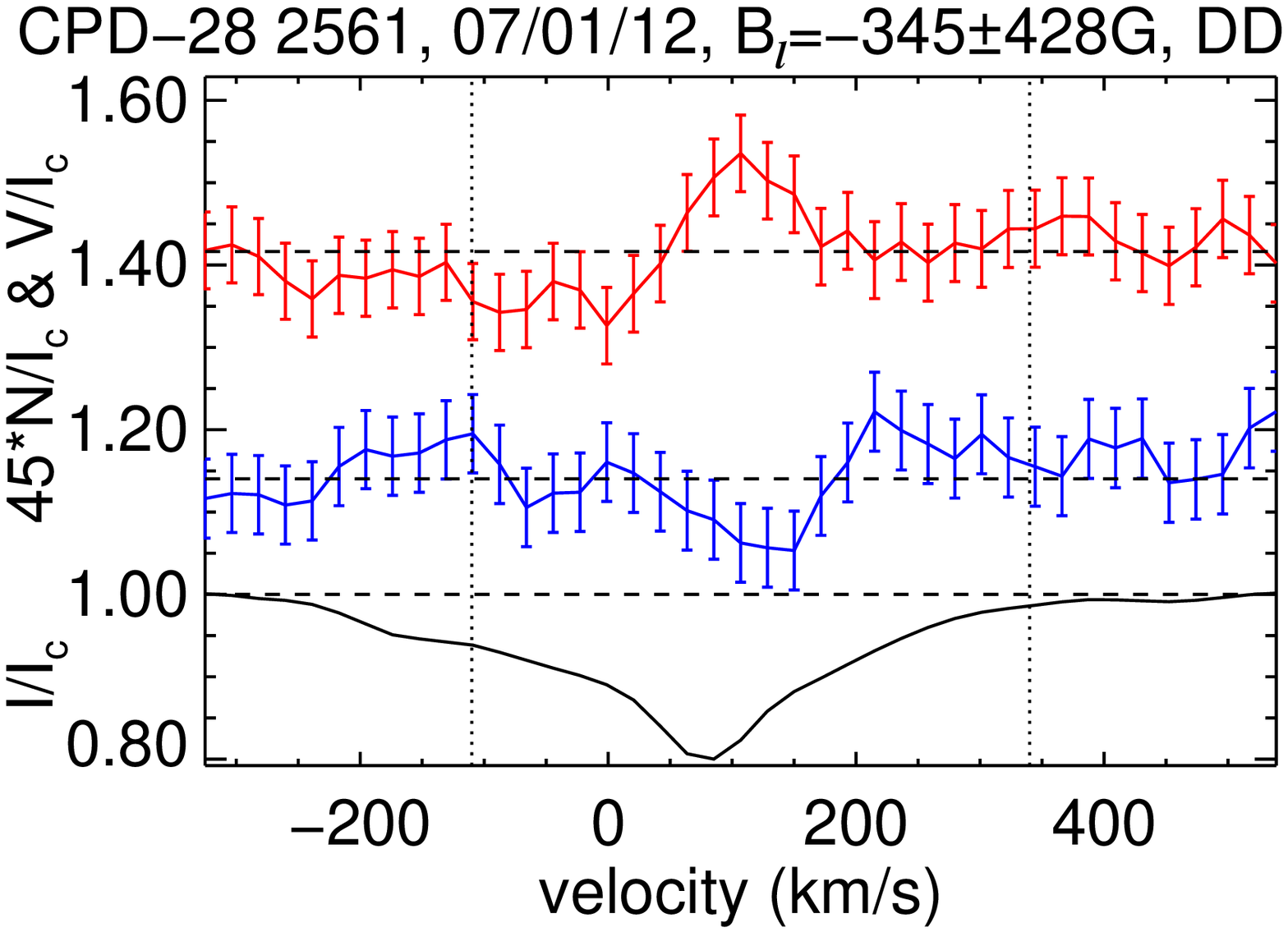}
\includegraphics[width=2.3in]{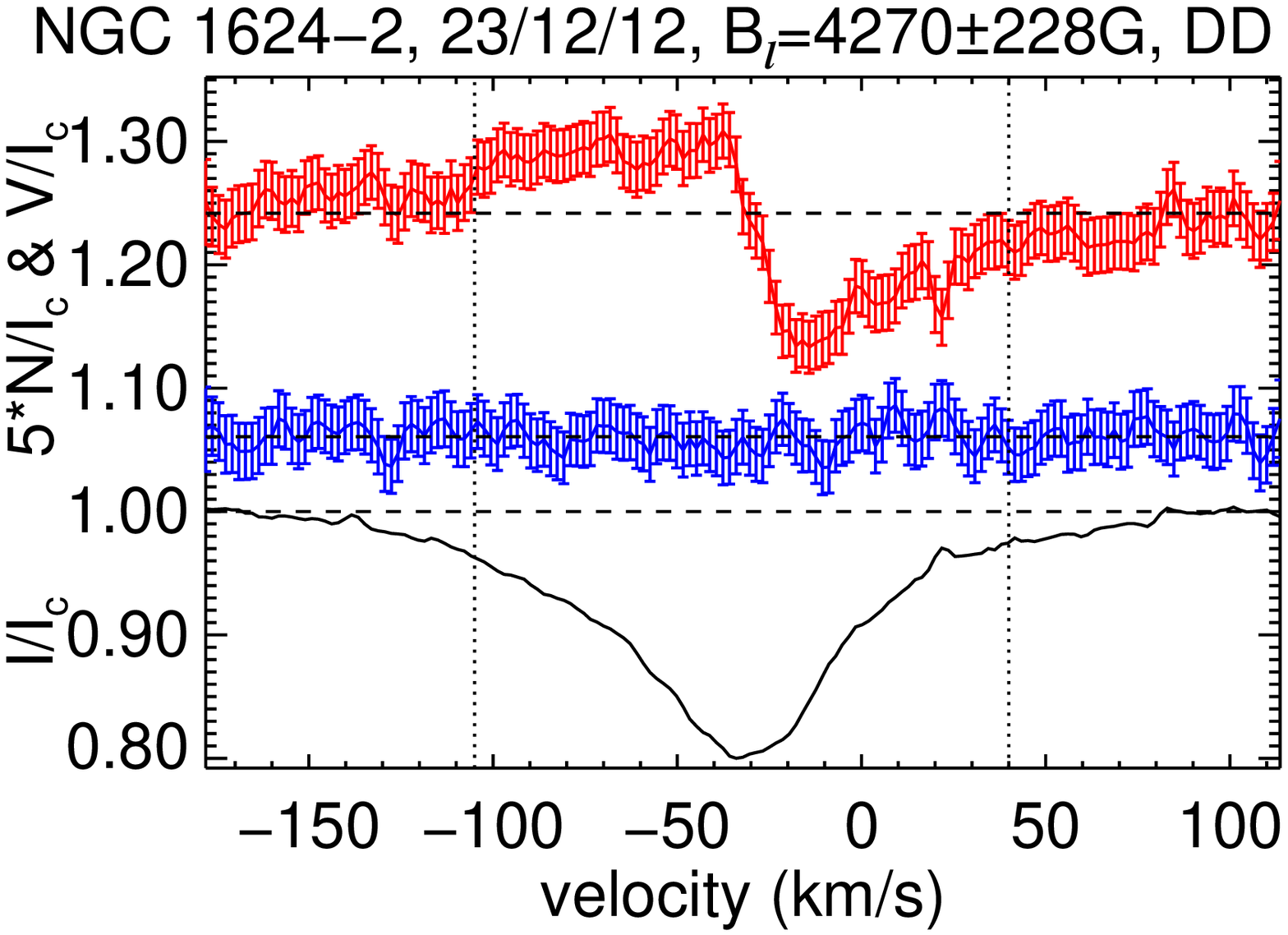}
\caption{Example unpolarized Stokes $I$ (bottom), diagnostic null (middle) and circularly polarized Stokes $V$ (top) LSD profiles for each star considered magnetic. The detection of a Zeeman signature within the line profile of each Stokes $V$ profile with the accompanying lack of excess signal in the diagnostic null profile or outside of the line profile is used to qualify an observation as a detection.  The profiles have all been rescaled such that the Stokes $I$ profile reaches a depth of 20\% of the continuum and the Stokes $V$ and diagnostic null profiles have been amplified by the indicated factor. The LSD profiles for HD\,47129A2 and CPD-28\,2561 were constructed from the Of?p line mask discussed in the text. All other profiles were constructed from the optimal line mask. The Stokes $I$ profile for HD\,47129A2 is the best-fitting profile. The name, observation date, longitudinal magnetic field with corresponding uncertainty, and detection diagnosis is indicated for each profile. Vertical dotted lines are included to show the adopted integration range for each profile. The profiles have been smoothed over 3 pixels and expanded by the indicated amount for display purposes.}
\label{det_profs_fig}
\end{figure*}

Fig.~\ref{det_profs_fig} shows example LSD profiles for each of the confirmed magnetic stars: HD\,108, HD\,47129A2, HD\,57682, HD\,148937, CPD-28\,2561, NGC\,1624-2. The detection of a Zeeman signature is significant in each case. In addition to these confirmed magnetic detections, three O stars known to be magnetic prior to the MiMeS survey were also observed as part of the Project: $\theta^1$\,Ori\,C \citep{donati02, wade06, chuntonov07}, $\zeta$\,Ori\,Aa \citep{bouret08, blazere15}, and HD\,191612 \citep{donati06a, hubrig10, wade11}. We confirm a magnetic detection for each of these stars.

\begin{figure*}
\centering
\includegraphics[width=2.3in]{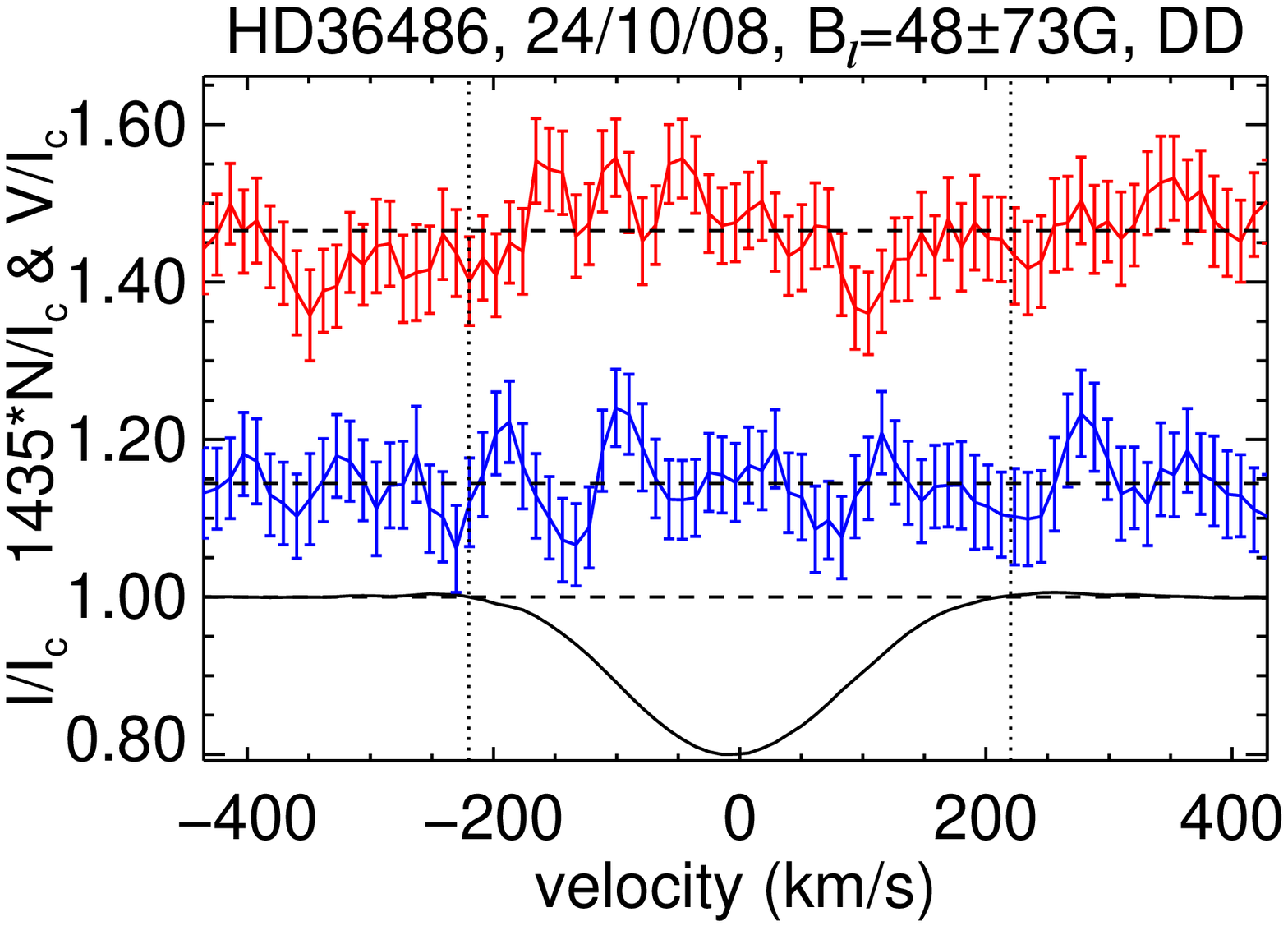}
\includegraphics[width=2.3in]{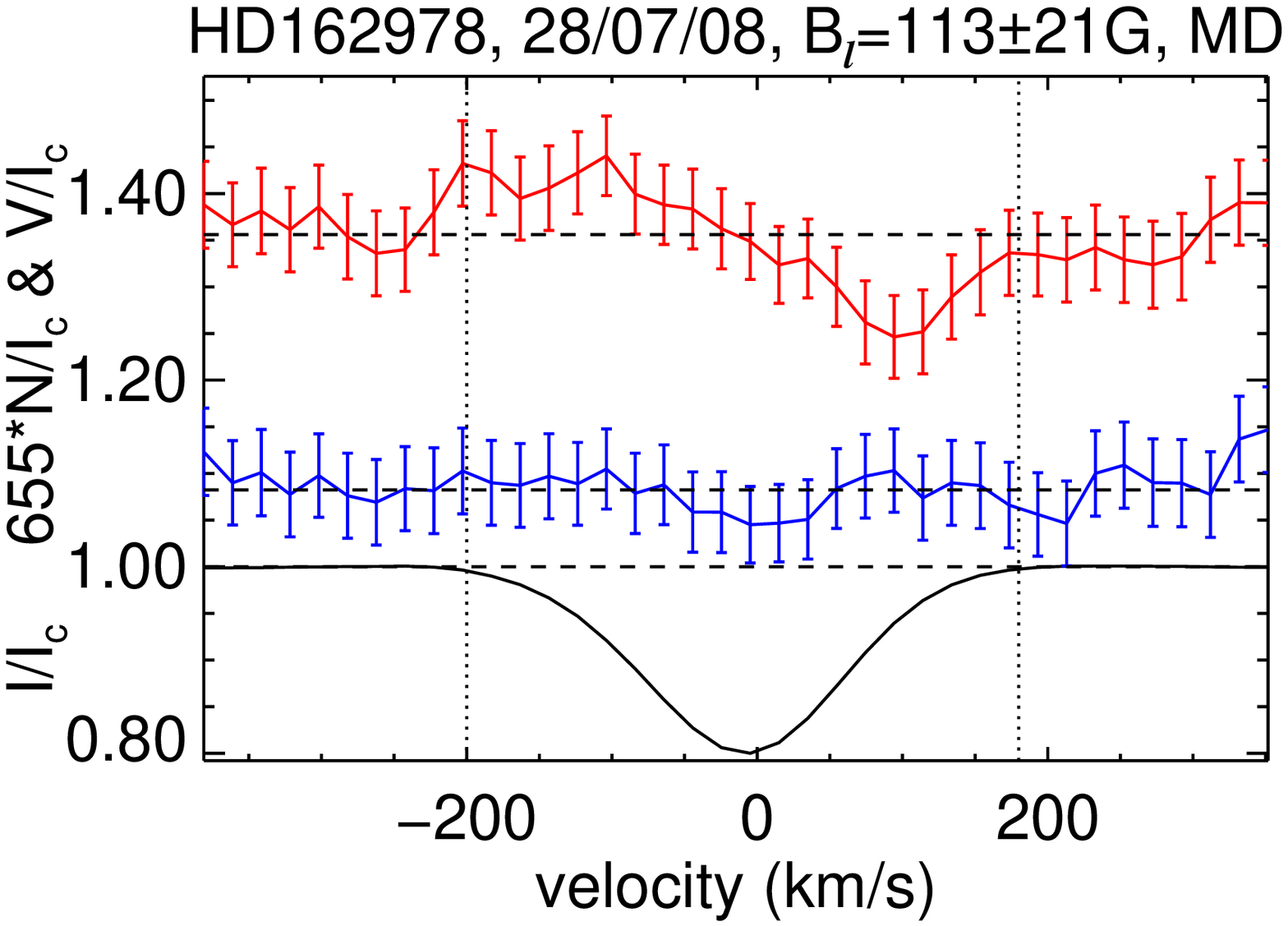}
\includegraphics[width=2.3in]{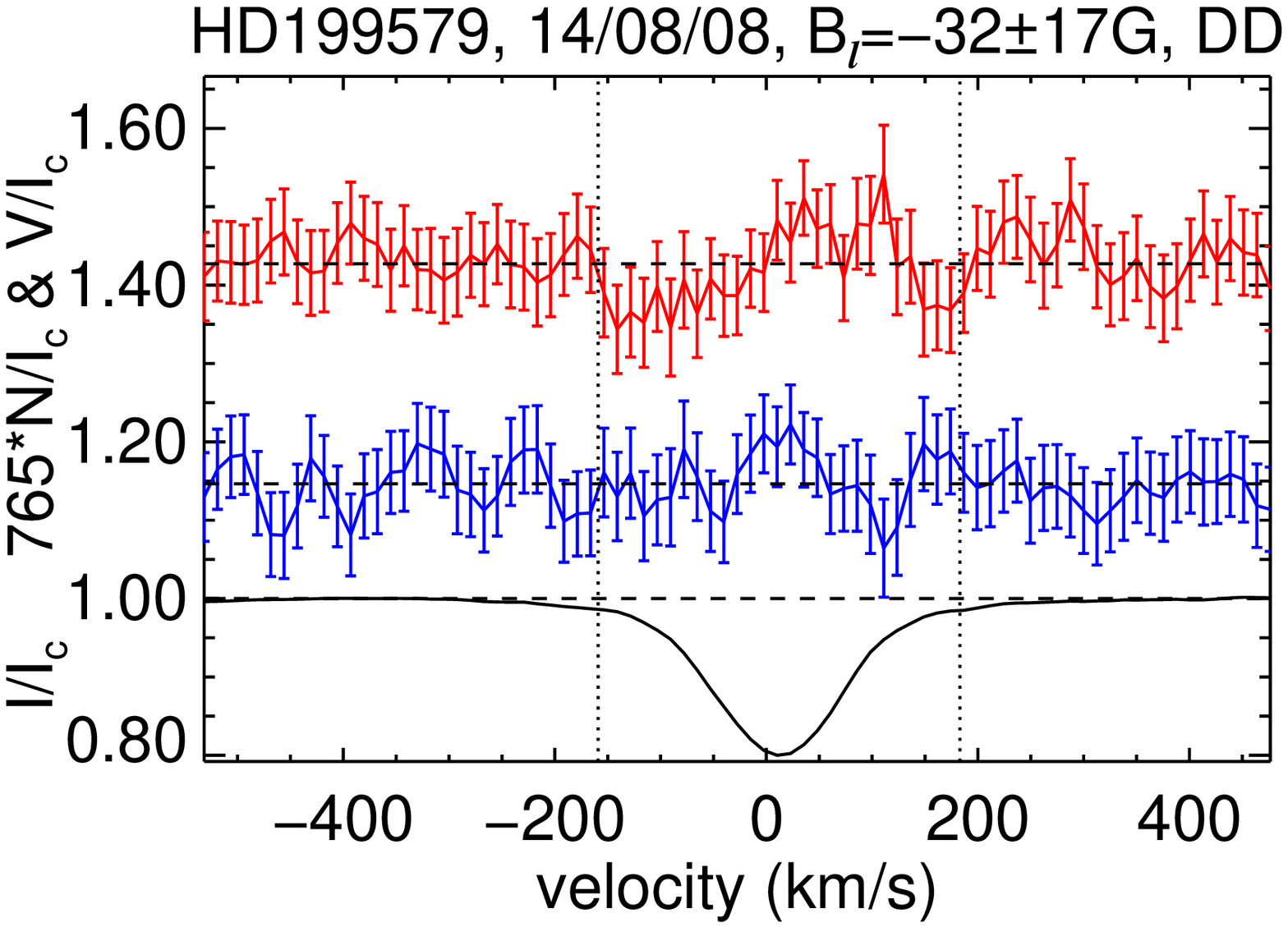}
\caption{Same as Fig.~\ref{det_profs_fig} for each star that may possibly be magnetic. See Fig.~\ref{det_profs_fig} for further details.}
\label{pos_det_profs_fig}
\end{figure*}

Fig.~\ref{pos_det_profs_fig} provides example LSD profiles for each of the possible magnetic stars: HD\,36486, HD\,162978, HD\,199579. For each of these stars we obtain a formal detection of signal in at least one observation, and the Stokes $V$ profile presents a coherent variation across the line that is apparent; however, insufficient data exists to confirm the presence of a magnetic field in these stars.

\begin{figure*}
\centering
\includegraphics[width=2.3in]{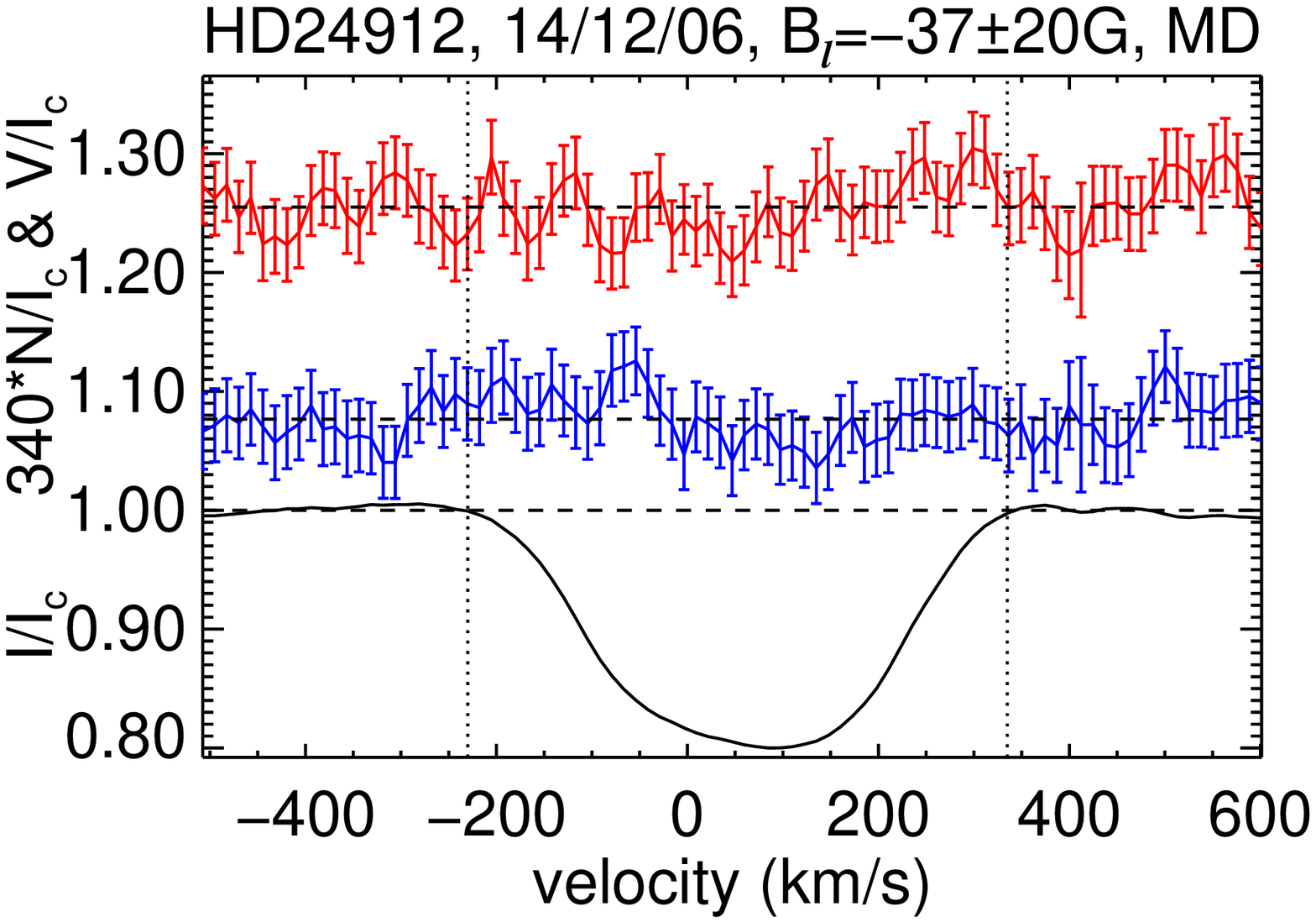}
\includegraphics[width=2.3in]{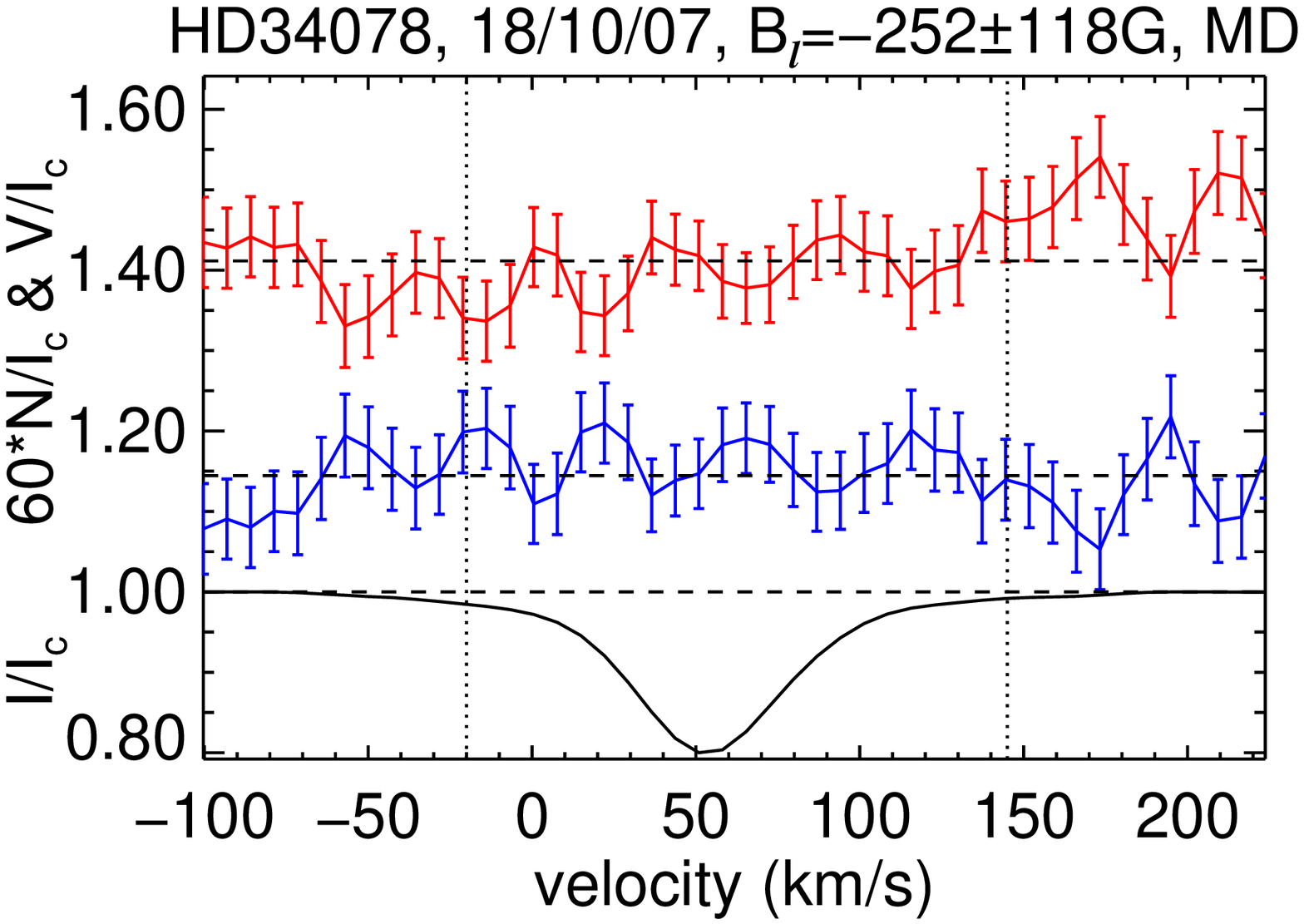}
\includegraphics[width=2.3in]{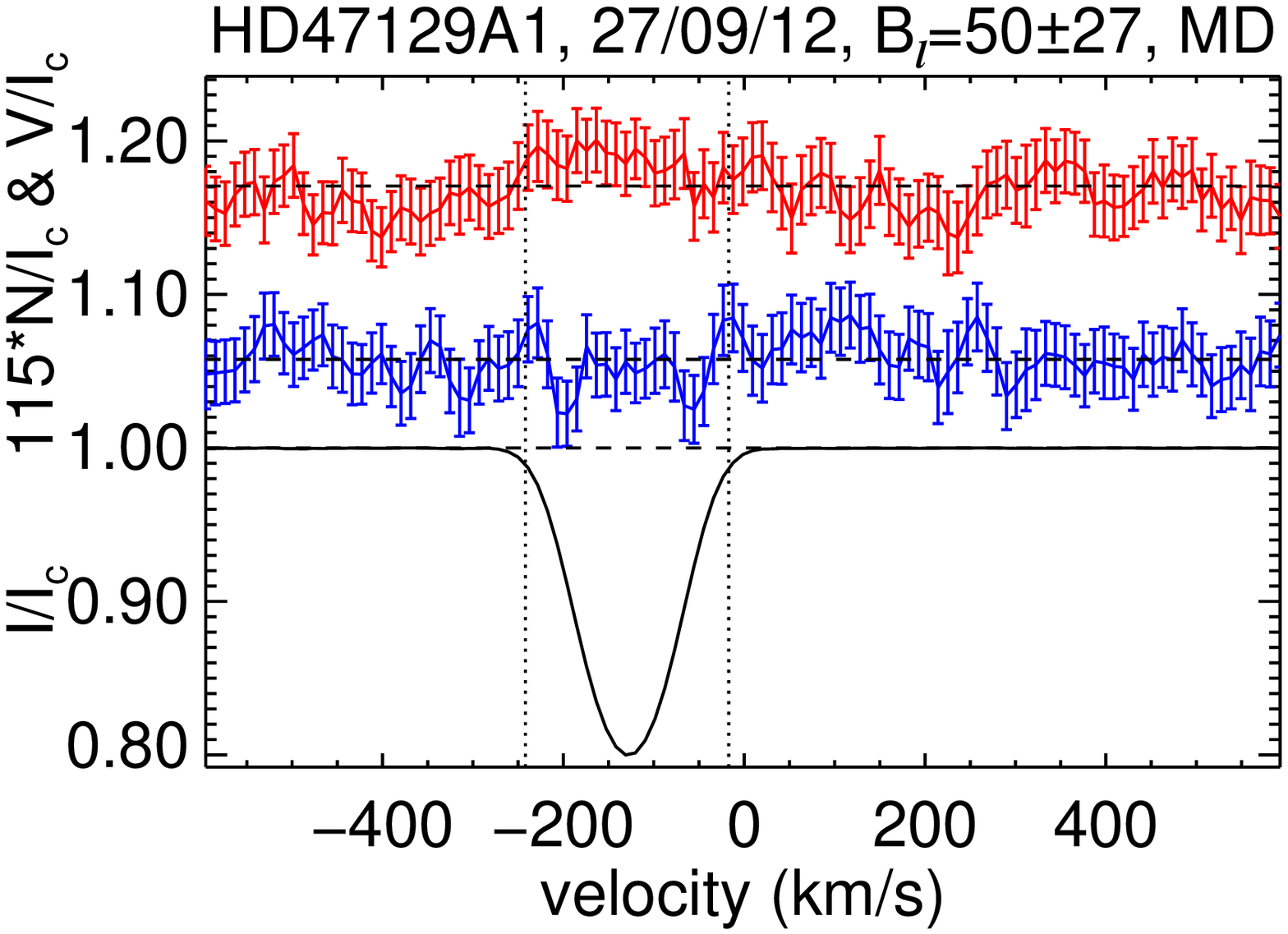}\\
\includegraphics[width=2.3in]{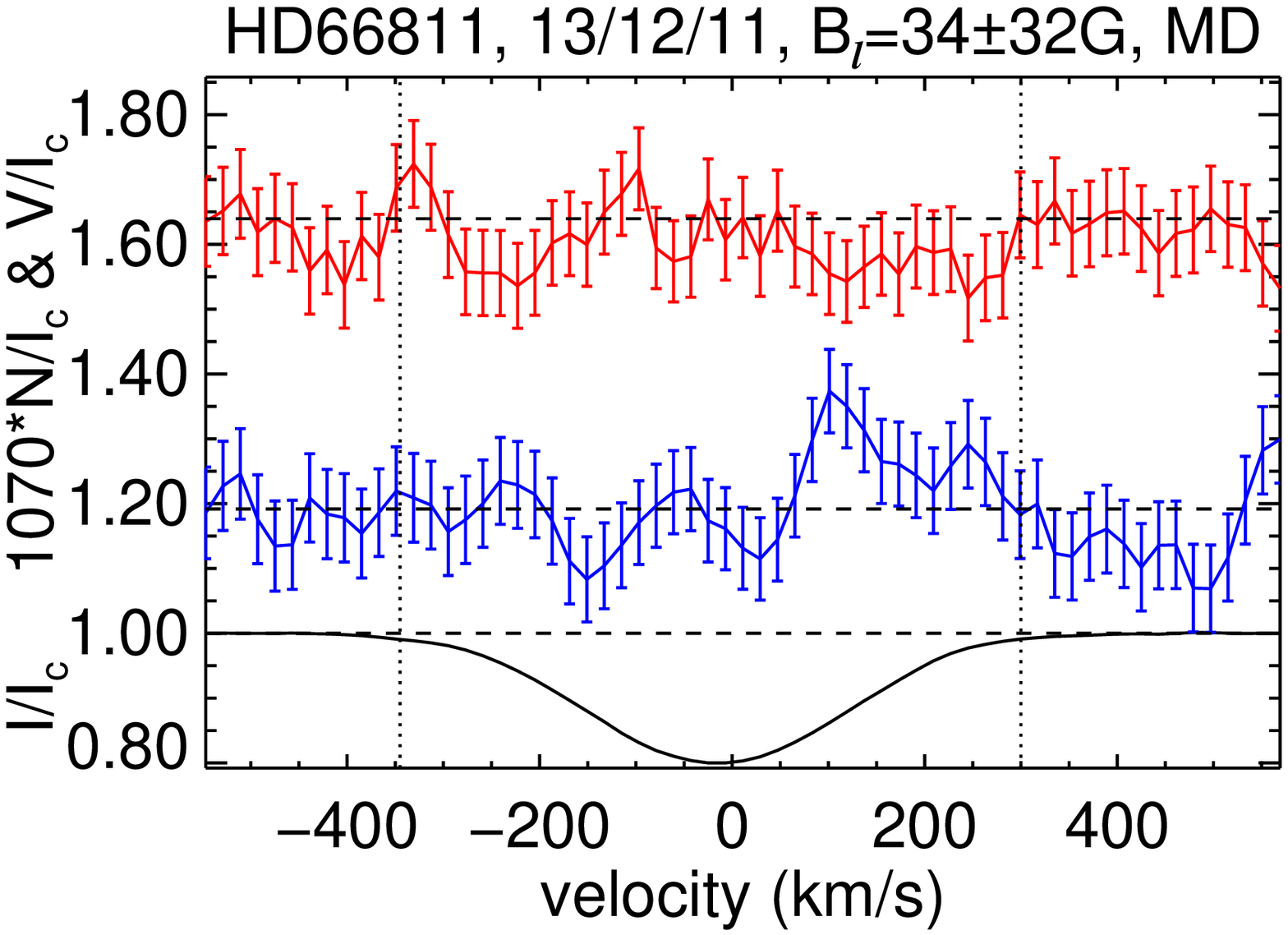}
\includegraphics[width=2.3in]{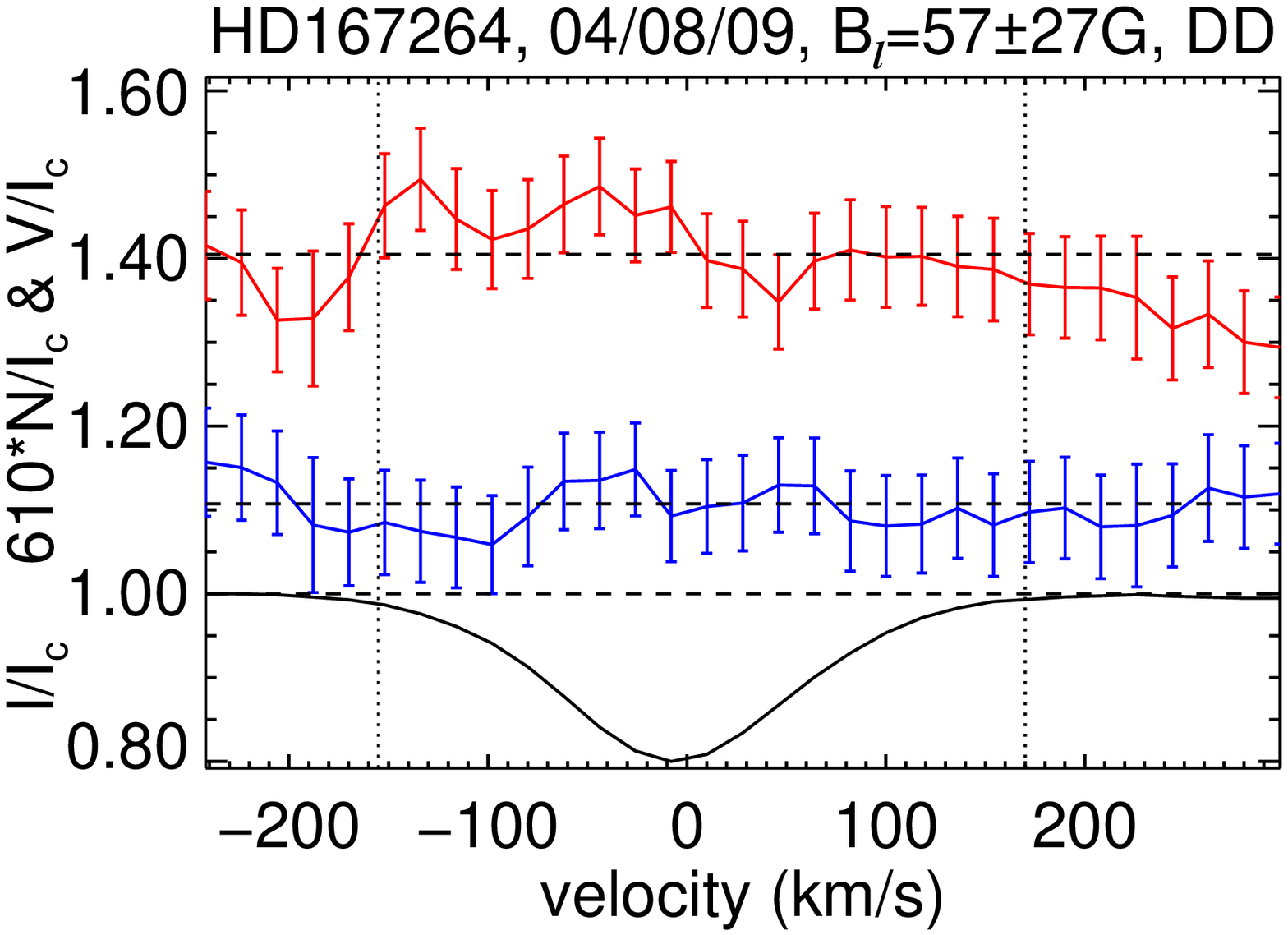}
\includegraphics[width=2.3in]{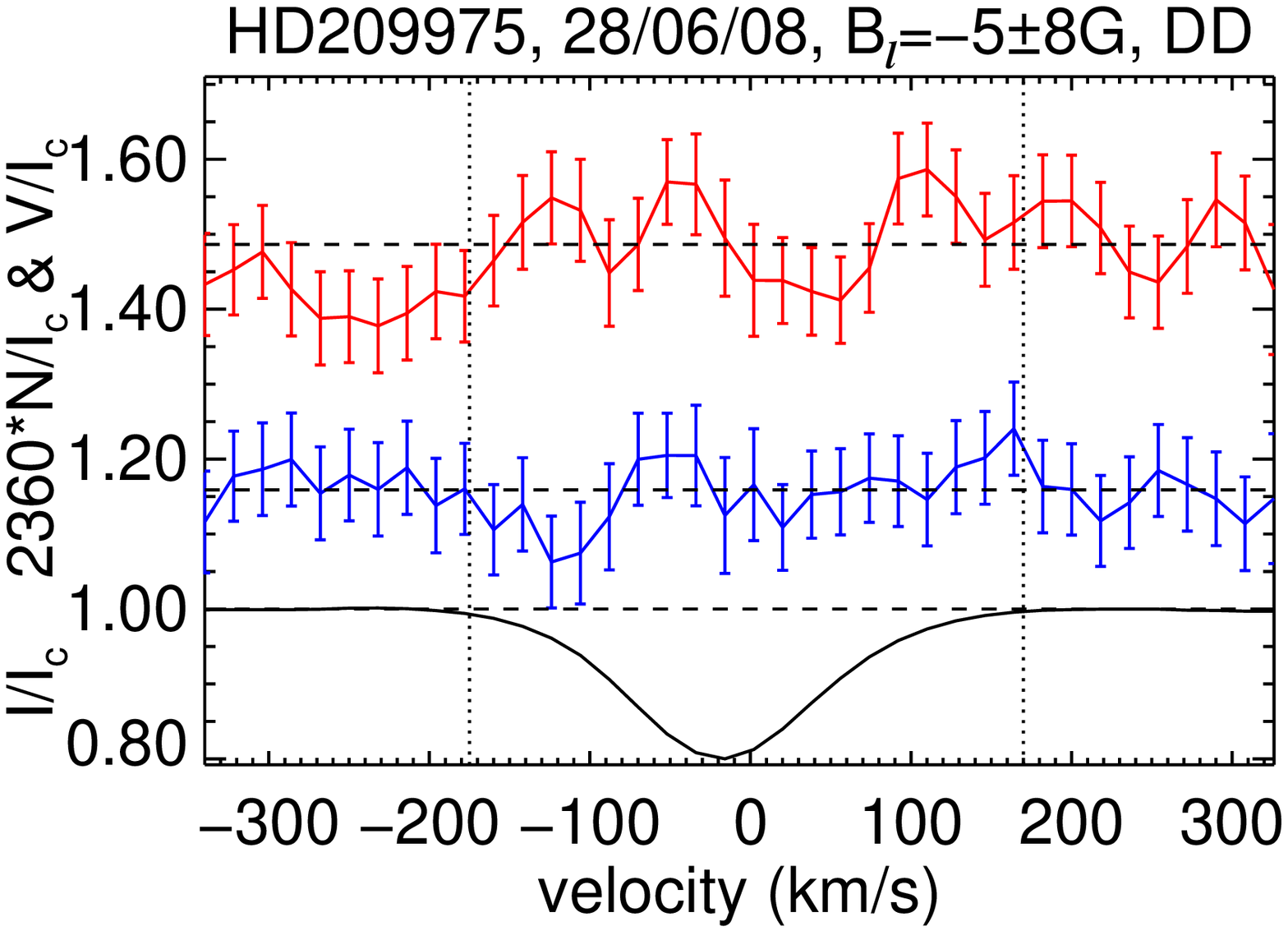}
\caption{Same as Fig.~\ref{det_profs_fig} for each star that is considered to be a spurious detection. The Stokes $I$ profile for HD\,47129A1 is the best-fitting profile. See Fig.~\ref{det_profs_fig} for further details.}
\label{spurious_det_profs_fig}
\end{figure*}

Fig.~\ref{spurious_det_profs_fig} presents example LSD profiles of each of the probable spuriously detected stars. In this case, a formal detection of signal is obtained in at least one observation, but upon closer examination of the data, or when considering the entirety of the data, we conclude that the star is not magnetic and that the excess signal detected in the observations is of spurious origin (i.e. the signal is not a consequence of an organised field on the surface of the star).

As previously discussed, in a few of the known/confirmed magnetic stars, the Of?p line mask results in a systematic improvement in the detection of signal. We therefore list the supposedly non-magnetic stars for which we obtained detections using the Of?p line mask: HD\,37041, HD\,47839, HD\,48099, HD\,153426. A visual inspection of all of the detected profiles did not reveal any clear evidence of a Zeeman signature.

It should be noted that the thresholds set by \citet{donati97} for the designation of a polarization signal to be a DD or a MD are somewhat arbitrary. If the threshold, primarily for a MD, were set to a lower value, many of the spurious detections would cease to qualify. In some of these cases, random noise within the Stokes $V$ line profile results in signal, which could be why the higher S/N achieved for these observations with {\sc iLSD} is more likely to result in a MD than with the LSD code of \citeauthor{donati97}. In most situations, the majority of the spurious detections result in a MD. Furthermore, these MDs, in general, only represent a small fraction of the total number of observations obtained for an individual target, which stands in strong contrast to the much larger proportion of DDs achieved for the confirmed magnetic stars (except for CPD-28\,2561) and, to a lesser extent, the possible magnetic stars.

Notes for particular stars (following the order presented in Table~\ref{det_samp_tab}) are provided below.

\begin{itemize}
\item {\bf HD\,108}: Excess signal is detected outside of the line profile of one of the MDs (2008-10-26). In this case we suspect that the excess signal is residual incoherent polarization left over from the imperfect profile extraction.

\item {\bf HD\,47129A2} (Plaskett's star): is a well-known multiple star system consisting of two similar O-type stars, one that exhibits relatively narrow lines (the component A1) and one component that hosts relatively broad lines (the component A2; see Appendix~\ref{binary_appendix} for additional details). From the FAP analysis, we obtain 3 MD with the optimal line mask. Alternatively, using the simpler Of?p line mask (as employed by \citealt{grunhut13} and Grunhut et al. in prep), we obtain 6 DD and 4 MD. This likely reflects the fact that the optimal line mask is better tailored for the component with stronger lines - in this case the narrow-line component. The Of?p line mask generally includes the strongest lines of the broad line component, and so the polarization signal is less diluted by contamination from the narrow line component. No magnetic field is confidently detected for the narrow line star (see discussion below).

\item {\bf CPD-28\,2561} (CD-28\,5104):  With the FAP analysis, we obtain 1 MD using the optimal line mask. Using the Of?p line mask we obtain somewhat better results with 3 MD, but using the \citealt{donati97} code results in 1 DD and 6 MD. Other than HD\,47129A2, this is the only star for which adoption of the Of?p line mask results in a systematic improvement in our ability to detect a Zeeman signature.

\item {\bf HD\,36486} ($\delta$\,Ori\,A): is an eclipsing binary system with a close visual companion \citep[e.g.][]{harvin02}. Two observations were obtained of this system on consecutive nights in 2008-10, from which we could not disentangle the individual components. We therefore discuss the results from the blended profile. A DD is obtained for the second observation (2008-10-24), which has a significantly higher S/N relative to the first observation (2008-10-23). We note that the profiles extracted with the \citet{donati97} LSD code both result in a ND. With only two observations, we cannot reliably confirm nor deny the presence of a magnetic field in this star. However, the lack of obvious emission in the typically strong magnetospheric emission lines of magnetic O-type stars (e.g. Balmer lines such as H$\alpha$, He\,{\sc ii} $\lambda$4686, see e.g. \citealt{grunhut12c}) could argue against this star hosting a large-scale magnetic field. If the global magnetic field is sufficiently weak though, we may not expect strong emission. While the well-known magnetic star $\delta$\,Ori\,C is nearby, it is sufficiently separated such that we do not expect any contamination in our observations of this star.

\item {\bf HD\,162978} (63\,Oph): was observed 2 times: the first observation (2008-07-28) results in a MD and a $\sim$5$\sigma$ detection significance of the longitudinal field ($B_\ell=111\pm23$\,G). The star was reobserved several years later (2012-06-21) with approximately twice better S/N ($B_\ell=-13\pm14$\,G), but we do not confirm the presence of signal in the Stokes $V$ profile from that observation (the FAP analysis results in a ND). This star also exhibits uncharacteristically weak emission in key lines in which magnetospheric emission would be expected. However, we do note that the first observation in which signal was detected shows a higher level of emission relative to the second observation with a non-detected field. This behaviour could be naturally explained by a magnetic star if the magnetic pole was oriented closer to our line of sight in the first observation compared to the second, and the magnetically confined wind, which is more confined to the magnetic equator, were viewed more face-on \citep[e.g.][]{sundqvist12}. We therefore consider this star a highly probable magnetic candidate.

\item {\bf HD\,199579} (HR\,8023): Is a binary system that was observed a single time (2008-08-15). The profiles of this observation are sufficiently entangled that we could not extract individual profiles for each component and instead discuss the results of the blended profile. The FAP analysis results in a DD (a MD is obtained with the \citeauthor{donati97} LSD code). This star exhibits an asymmetric absorption profile in H$\alpha$, which could be indicative of weak magnetospheric emission.

\item {\bf HD\,24912}  ($\xi$\,Per): The FAP measured from one of the profiles provides a MD (2006-12-14). Several other observations were obtained with similar S/N as the MD observation, yet no formal detections are found in any of those profiles. The profile with the MD does not present any obvious Zeeman signature in Stokes $V$. Considering this fact and the larger number of NDs, we conclude that the detection is spurious. HD\,24912 is also a known non-radial pulsator, with a 3.5\,h pulsation period and weak spectroscopic variability associated with this phenomenon \citep{dejong99}. The pulsation is likely responsible for the asymmetric Stokes $I$ line profile (see Fig.~\ref{spurious_det_profs_fig}), and may be responsible for the spurious signal, if it were accompanied by other issues, as previously discussed. Similar results and conclusions for this star were obtained by \citet{david-uraz14}. 

\item {\bf HD\,34078} (AE\,Aur): The FAP of one of the binned profiles (2007-10-18) results in a MD; however, a MD with a similar FAP is obtained  outside of the line profile in Stokes $V$ as well. No obvious Zeeman signature can be seen in the Stokes $V$ profile.  

\item{\bf HD\,47129A1} (Plaskett's star): One of the binned profiles of the narrow line component (A1; 2012-09-27) results in a MD based on the FAP analysis. The broad line profile of that same night also results in a MD. Given that the narrow line profile is blended with the broad line profile at all phases, its Stokes $V$ profile is contaminated by the broad line component. In fact, the detected signal appears to be part of a broader Zeeman signature that extends well outside the limits of the line profile, which is attributed to the broad line component (as shown in Fig.~\ref{spurious_det_profs_fig}). A more thorough analysis by Grunhut et al. (in prep) shows that there is no apparent signal associated with the narrow line component.

\item {\bf HD\,66811} ($\zeta$\,Pup): A MD is obtained from the FAP analysis for the first observation (2011-12-13) (a ND is found using the \citeauthor{donati97} LSD code). A DD is also obtained within the null profile for the first observation. The second observation taken a few months later (2012-02-13) had a similar S/N but results in a ND. Since a detection is also found in the null profile, this leads us to consider the Stokes $V$ detection to be spurious. Similar results were also obtained by \citet{david-uraz14}.

\item {\bf HD\,167264} (15\,Sgr): Despite the similar precision of all the observations, only the FAP analysis of the profile of a single night (2009-08-04) results in a DD (a ND is obtained using the LSD code of \citeauthor{donati97}). Visual inspection of the Stokes $V$ LSD profile reveals some coherent structure that could be evidence of a Zeeman profile, although the potential signature extends well outside the line profile, suggesting it is likely spurious.

\item {\bf HD\,209975} (19\,Cep): One observation (2008-06-27) results in a $\sim$$3.1\sigma$ significance ($B_\ell=24\pm8$\,G); however, the FAP analysis of this observation results in a ND. The observation obtained on 2008-06-28 results in a DD (a ND is obtained using the LSD code of \citeauthor{donati97}). The Stokes $V$ profile shows structure that could be indicative of a Zeeman signature, but this structure also extends outside of the line profile and therefore is likely spurious. Similar results were also reported by \citet{david-uraz14}.

\end{itemize}

\section{Discussion}
\subsection{Quality assessment}\label{quality_sect}
In this section, we discuss the quality and reliability of our data, and how that may relate to spurious detections, and possible biases. We do not address the potential for missed fields, as this will be discussed in the forthcoming paper in this series.

\begin{figure}
\centering
\includegraphics[width=3.2in]{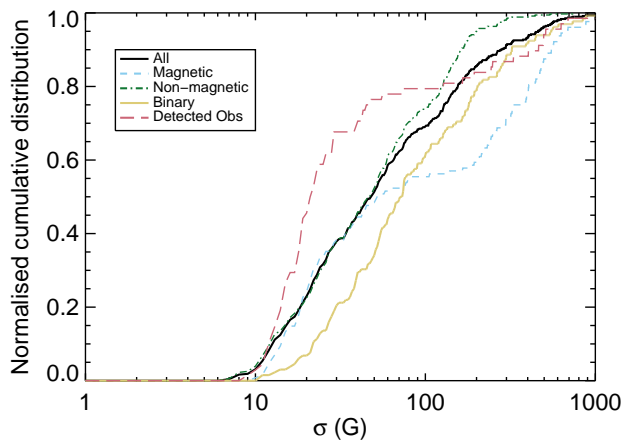}
\caption{Cumulative distribution of longitudinal field uncertainty $\sigma$ for different samples of observations (as indicated by the included legend). The magnetic sample includes all observations corresponding to the stars deemed confidently detected as magnetic. The non-magnetic sample includes all observations of stars that are not considered confidently detected. The binary sample includes all binary stars for which the different line profiles could be disentangled. The detected sample includes only those observations for which a signal was detected in Stokes $V$.}
\label{cum_sigma_fig}
\end{figure}

In Fig.~\ref{cum_sigma_fig}, we show the cumulative distribution of the longitudinal field uncertainty ($\sigma$) as a representation of the precision achieved in this study. We find that about 25\% of our observations achieved a $\sigma$ of 20\,G or better. As some stars were observed more than once, we note that this precision was achieved for 34 different targets. Approximately 50\% of our sample of observations (76 different targets) was acquired with a $\sigma$ of 50\,G or less, or about 70\% (101 targets) with 100\,G or better. The quality achieved in this study represents the most magnetically sensitive probe of the largest sample of O-type stars to date.
 
\begin{figure}
\centering
\includegraphics[width=3.2in]{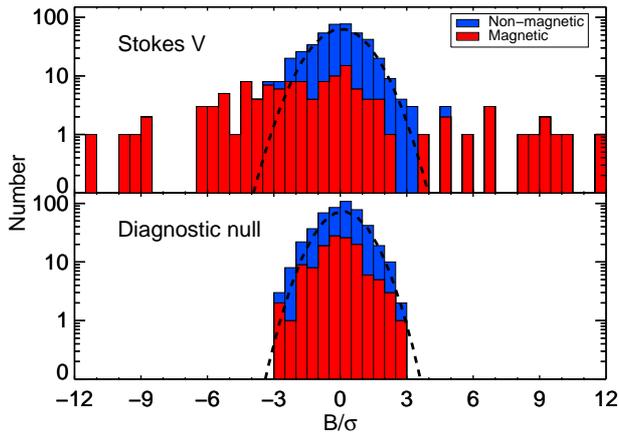}
\caption{Top: histogram of the detection significance of the longitudinal field measurement ($B_\ell/\sigma$) obtained from all observations of individual stars (i.e. all non-entangled stars). Bottom: same as the top but measured from the diagnostic null profile *$N_\ell/\sigma$). The dashed curves represent a pure Gaussian distribution for comparison. The magnetic sample only includes all observations from stars that were confidently detected as magnetic. The non-magnetic sample includes all observation from stars that were not considered confidently detected as magnetic.}
\label{z_hist_fig}
\end{figure}

One quality check that we performed was to investigate the precision of the data (including instrumental, reduction, and measurement systematics), by analysing the distribution of significances of the longitudinal magnetic field measurements relative to the estimated uncertainties. In Fig.~\ref{z_hist_fig}, we show the distributions for both the confirmed magnetic sample and the unconfirmed/non-detected sample (which includes potential and spurious detections) for all stars. The measurements computed from the Stokes $V$ profile ($B_\ell/\sigma$) and the diagnostic null profiles ($N_\ell/\sigma$) are both included in this figure. 

The first important conclusion is that the measurements obtained from the diagnostic null profiles from both the magnetic and non-magnetic sample (a total of 483 profiles) are consistent with a Gaussian distribution centred around a significance of 0, and all values are within $\pm3\sigma$. The next important conclusion is that the $B_\ell$ measurements of the non-magnetic sample (including unconfirmed magnetic stars) are consistent with the null measurements. A two-sided Kolmogorov-Smirnov (KS) test supports the hypothesis that the non-magnetic sample of observations is drawn from the same underlying distribution - this hypothesis is not ruled out at about 2$\sigma$ confidence. This is not true for the $B_\ell$ measurements from the magnetic sample, which show clear differences from the null distribution. We warn the reader to avoid over-interpreting the preference for negative $B_\ell$ values from the magnetic star sample in this figure. This is a result of a large number of negative measurements for a few stars and should not be interpreted as a statistical preference for a given orientation of magnetic fields in O-type stars. We underscore that while the quality control checks show that the $B_\ell$ measurements are well behaved, these values are not used to establish whether an observation is considered a magnetic detection.

Another quality check that we performed was to assess the overall reliability of the LSD noise level of the polarimetric profiles. This is particularly important as it could provide insight into the small number of spurious detections that we encountered in our analysis, especially if we find that the noise is underestimated. We carried out a series of tests using the diagnostic null profiles, the details of which are provided in Appendix~\ref{lsd_noise_sect}. The overall conclusion is that, in general, the LSD uncertainties appear to be {\it over-estimated} by about 20\% for the unbinned profiles. On the one hand, reducing the uncertainties by about 20\% does increase the number of detections among the magnetic sample when using the unbinned profiles (37 vs 22), but this is still not as efficient as the binning strategy (61 detections). On the other hand, the uncertainty of the binned profiles are in good agreement with our expectations (they are over-estimated by $\sim$3\%), although the tests show that the binning does introduce a small increase ($\sim$1.5\%) in the number of FAPs from the null profiles that would be considered detections, and so some extra care must be taken when using this procedure and assessing the presence of a detectable Zeeman signature. In fact, 1.5\% of the non-magnetic sample (353 observations) corresponds to about 5 observations, which is in excellent agreement with the number of detections we highly suspect as being spurious.  

\subsection{Biases}

\begin{figure}
\centering
\includegraphics[width=3.2in]{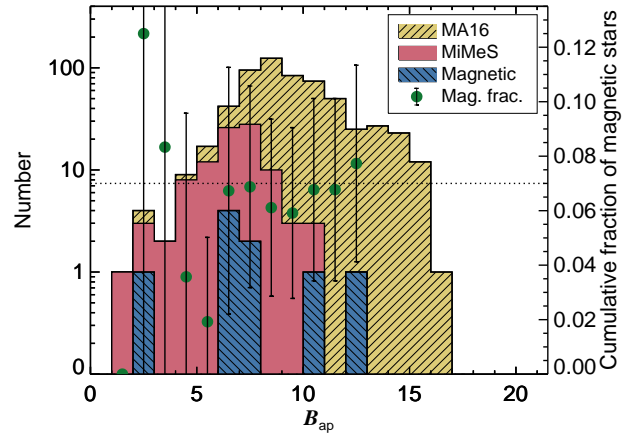}
\caption{Histograms comparing the number of targets observed at a given approximate $B_{\rm ap}$ magnitude. Included are histograms according to the GOSSS \citep[MA16]{maiz16}, the targets observed in this study, and the magnetic and potentially magnetic stars identified in this work. Also shown is the cumulative magnetic incidence fraction as a function of $B_{\rm ap}$. Each bin provides the total magnetic incidence for all targets (97 in total) with a $B_{\rm ap}$ less than or equal to the indicated $B_{\rm ap}$ of the bin. The adopted magnetic incidence fraction (7\%) is indicated by the horizontal dotted line, but note that the magnetic incidence fraction derived from the 97 targets, but including the potential candidates, is $8\pm4$\%.}
\label{magnitude_dist_fig}
\end{figure}

In this section we discuss the reliability of our findings when taking into account observational biases. The first bias we consider is the reliability of our incidence rate with respect to the brightness of the selected stars in our sample. This is particularly important as two of the six  confirmed magnetic stars are faint targets, and much fainter than the brightness of the general population of stars included in this study. To assess this potential issue, we compare the approximate $B$ magnitude ($B_{\rm ap}$) of our sample \citep[e.g.][]{maiz16} to the distribution presented by the Galactic O-Star Spectroscopic Survey (GOSSS; \citealt{maiz16}. Fig.~\ref{magnitude_dist_fig} compares the histograms of the $B_{\rm ap}$ obtained from the GOSSS sample, our entire sample, and the sample of magnetic stars (including the potential candidates). We also present the cumulative magnetic incidence fraction achieved in our study, taking into account the potential magnetic candidates in this figure. In this comparison, we treat each system as an individual target to avoid potential discrepancies associated with determination of the relative brightness of each component in multiple star systems. Hence, the statistics correspond to only the 97 targets and not the total 108 individual O stars, as previously mentioned. The most obvious conclusion to be drawn from this analysis is that the incidence fraction converges to $\sim$7\% with increasing magnitude. Maximum completeness of our sample of stars ($\sim$86\%) is reached for $B_{\rm ap}<4$, but this only includes 6 stars. A more statistically relevant sample is reached for $B_{\rm ap}<7$, for which we achieve a completeness of 66\% that includes 52 total targets in our sample, and between 2 and 4 magnetic stars (including the potential candidates). From this sample, we derive an incidence fraction of $7\pm5$\%, which is consistent with our adopted value. Including the fainter bins with a much higher magnetic incidence fraction relative does not appear to significantly affect the final incidence fraction, which leads us to conclude that our results are not sensitive to this particular observational bias.

Given the previous findings, the next outstanding question that we address here is whether there were any characteristics of the detected stars that favoured the detection of signal in the magnetic star sample versus the non-magnetic sample. This is of particular interest because it would introduce a detection bias, and it would also affect the inferred properties of the magnetic stars. One such possibility could be a result of systematically obtaining higher magnetic precision with the detected sample. We return to Fig.~\ref{cum_sigma_fig}, which also shows cumulative distributions for a number of different sub-populations, to address this issue. Our first conclusion is that the distribution of uncertainties measured from the individual components of the binary sample shows a definite trend compared to the general sample of all observation - it systematically achieves a lower precision for the same relative population. This is to be expected since the S/N that we aimed for, according to the exposure time calculator described in \citetalias{wade16}, did not take into account binarity, so the typical magnetic precision that we achieve for each component of a binary system would naturally be lower. 

Comparing the sample of observations of the magnetically confirmed stars to the non-magnetic sample, we also find some obvious differences. The samples are in excellent agreement up to about $\sigma=30$\,G, before the magnetic sample's cumulative distributions start to diverge. Surprisingly, the magnetic sample achieves substantially poorer precision than the non-magnetic sample for more than 50\% of its population. Taking a closer look at the magnetic results, we find that this distribution is bimodal. The low-$\sigma$ population is dominated by many observations of HD\,108 and HD\,57682. HD\,57682 is unique among the magnetic O-star sample as it has a rich spectrum of sharp lines, which results in higher magnetic precision at the same S/N as the other magnetic stars. The high-$\sigma$ end is populated by observations of the hot, faint stars CPD-28\,2561 and NGC\,1624-2, with the other stars filling in the values in between. While this comparison shows that the general population of observations of the magnetic stars achieved poorer precision than the non-magnetic sample, there are also several observations of the magnetic stars for which we did not detect excess signal. If instead we only look at the subsample of the observations with detected signatures, we do find a trend towards higher precision for about 80\% of that subsample; however, these observations are again dominated by many observations of just two stars: HD\,108 and HD\,57682, and it is therefore difficult to draw any conclusions.

\begin{figure}
\centering
\includegraphics[width=3.2in]{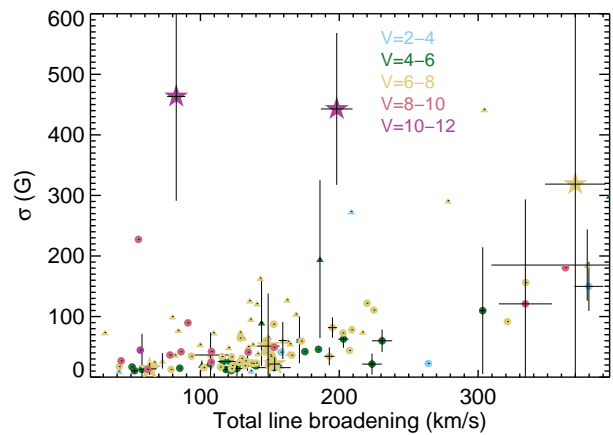}
\caption{Measured longitudinal field error bar $\sigma$ vs total line width. The plotted data points correspond to the median value obtained for each star, and the error bars correspond to the standard deviation of the values. Different colours correspond to different magnitude bins (as indicated), while different symbols correspond to different sub-populations (circles: single stars, triangles: binaries, stars: confirmed magnetic stars). The two poorest observations for CPD-28\,2561 were ignored.}
\label{sig_vtot_fig}
\end{figure}

Another way to investigate any potential biases related to magnetic precision is to search for any systematic trends in the two most important factors for estimating the original exposure times: the apparent brightness and line broadening. In Fig.~\ref{sig_vtot_fig}, we compare $\sigma$ to the measured total line broadening and apparent brightness for each star. In general, we find no obvious relationship between the line width and the achieved magnetic precision for most stars, except for the few stars with the very broadest lines ($v_{\rm tot}\ga300$\,\kms), which have larger uncertainties relative to the narrower line stars, and some of the individual components of binary systems. Furthermore, we also find no obvious relationship between the obtained precision and the brightness of the star. This result is expected as the Survey aimed to detected Zeeman signatures for field strengths between about 100 to 1000\,G, which roughly translates into expected $\sigma$$\sim$$10-100$\,G, for all stars in the sample.

The magnetic stars essentially fall into two groups: some stars with average precision that is in good agreement with other stars of similar brightness and line width; and the other group of stars with a magnetic precision that is generally worse than that of stars with similar line width. This former group includes most of the magnetic stars (HD\,108, the broad line component of HD\,47129 (A2), HD\,57682, and HD\,148937). The latter group includes the two fainter magnetic stars (CPD-28\,2561 and NGC\,1624-2), which are among the faintest stars observed in the Survey. Thus the poorer precision is a reflection of their apparent brightness and the lower achieved S/N (recall that integration times were generally kept to less than two hours). One other aspect of the magnetic stars is that several stars have a large range in their obtained precision. The relatively large range is primarily a reflection of changes in observing conditions and the corresponding S/N that was achieved for these observations, and not, for example, due to variable emission that would reduce the strength of Stokes $I$.

Therefore, based on the above discussion, we conclude that there are no obvious biases that would account for the detection of excess signal in the magnetic sample versus the non-magnetic sample. 

\subsection{Comparison with magnetic results obtained in other works}\label{comp_other_works_sect}

In this paper we list the confirmation of a magnetic field in six O stars previously reported as magnetic by the MiMeS collaboration (HD\,108, the broad line component of HD\,47129 (A2), HD\,57682, HD\,148937, CPD-28\,2561, and NGC\,1624-2), in addition to the three already known magnetic O stars discovered before this Survey ($\theta^1$\,Ori\ C, $\zeta$\,Ori\,A, and HD\,191612). Three of the six new magnetic O stars have been observed by other authors and confirmed to be magnetic \citep{hubrig08,hubrig10, hubrig11a, hubrig12, hubrig13, hubrig15}.

As previously discussed, a magnetic field detection has also been obtained with low-resolution FORS data for the O star Tr16-22 by \citet{naze12} and confirmed by \citet{naze14}, and for the O star HD\,54879, which was also confirmed with high-resolution spectropolarimetry \citep{castro15}. These two targets have not been observed within the MiMeS Survey, but their magnetic detections are convincing and therefore there is little doubt that they are indeed magnetic. 

While a number of magnetic field detections have been claimed in 20 other O-type stars \citep{hubrig07, hubrig08, hubrig11b, hubrig12b, hubrig13, hubrig14}, doubts had already been cast by other authors on the validity of these claims. In particular, \citet{bagnulo12} showed, using the same FORS data as \citet{hubrig08}, and focusing in particular on the evaluation of realistic uncertainties, that 4 O stars previously claimed to be magnetic by \citeauthor{hubrig08} were not magnetic: HD\,36879, HD\,152408, HD\,155806, and HD\,164794. In the same way, again using the same FORS data, \citet{bagnulo12, bagnulo15} showed that the claims of a field in $\zeta$\,Oph \citep{hubrig11b} and 15\,Mon \citep{hubrig13} are also spurious. 

Of those 20 O stars claimed to be magnetic, 12 were analysed in this work (the results for each star can be found in Appendix~\ref{online_tables_sec}). In most cases, the magnetic precision in this study exceeded that of previous measurements that led to claimed detections. No evidence of a large-scale magnetic field was found for any of these stars. For 6 of these stars, the FORS data were also re-analysed by other teams, and all were refuted as magnetic stars \citep{bagnulo15}.

Only 5 O stars ($\theta^1$\,Ori\,C, HD\,148937, CPD-28\,2561, Tr16-22, and HD\,54879) have been measured to be magnetic with FORS data and are confirmed to be magnetic O stars. This means that about 75\% of the claimed magnetic detections among O stars using FORS are considered spurious. This percentage is consistent with the 80\% spurious detections among O stars observed with FORS, as found by \citet{bagnulo12}. Eight stars claimed to be magnetic from FORS observations have not been observed or analysed yet by independent teams, but considering the statistics exposed above, the claimed magnetic fields in these 8  stars should be considered with great caution. 

Notes for particular stars previously claimed as magnetic and included in this Survey are provided below.

\begin{itemize}

\item {\bf HD\,47839} (15\,Mon) is a multi-line spectroscopic binary that was observed 8 times between 2006-12 and 2012-03, with the majority of these observations being acquired between 2007-09 to 2007-11. A ND is obtained for each component, and for each observation, but we note that the primary component dominates the line flux. 

\item {\bf HD\,66811} ($\zeta$ Pup) was previously discussed in Sect.~\ref{det_samp_sect}.

\item {\bf HD\,153426} is a spectroscopic binary that was observed twice between 2011-07 and 2012-07. A ND is obtained for each component. The primary component dominates the line profile.

\item {\bf HD\,155806} was observed 4 times between 2008-06 and 2008-08. No formal detection of excess signal was found for any of the observations (see also \citealt{fullerton11}). The $B_\ell$ measurement from one of our observations results in a 3.2$\sigma$ detection ($B_\ell=-44\pm14$\,G; 2008-06-25), but no evidence of nonzero $B_\ell$ is obtained from the other observations with similar precision.

\item {\bf HD\,164794} (9 Sgr) is a spectroscopic binary that was observed 5 times between 2005-06 and 2011-07. We were not able to disentangle the two components and therefore we report the results for the blended profile. \citet{hubrig13} presented results based on publicly available HARPS data collected within the MiMeS HARPSpol LP. Their analysis resulted in a $B_\ell=210\pm42$\,G ($B_\ell/\sigma\sim5$) from this data, while our analysis of the same data results in a substantially lower field measurement of $B_\ell=-1\pm48$\,G ($B_\ell/\sigma\sim0$), despite having similar uncertainties. 

\end{itemize}

The total number of known and confirmed magnetic O stars at the time of writing is thus 11: HD\,108, the broad line component of HD\,47129 (A2), HD\,54879, HD\,57682, HD\,148937, HD\,191612, CPD-28\,2561, NGC\,1624-2, Tr16-22, $\theta^1$\,Ori\ C, and $\zeta$\,Ori\,Aa. 

\subsection{Searching for trends}
\begin{figure}
\centering
\includegraphics[width=3.2in]{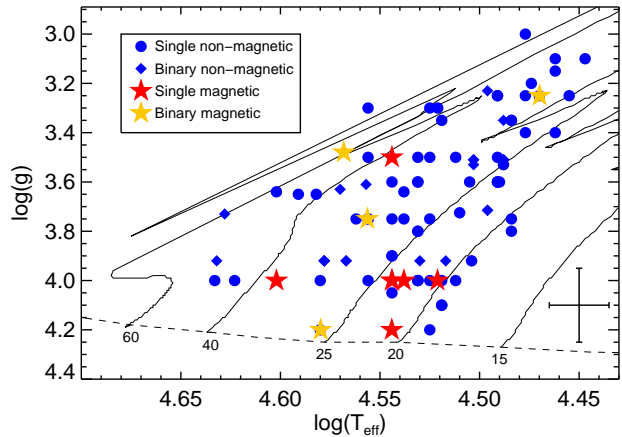}
\caption{$\log(g)-T_{\rm eff}$ diagram showing the position of all stars studied here in addition to other confirmed magnetic O stars ($\theta^1$\,Ori\,C, $\zeta$\,Ori\,Aa, Tr16-22, HD\,54879, HD\,191612). Evolutionary tracks of \citet{ekstrom12} are also included. The blue loops of some of the evolutionary tracks are neglected for display purposes. Note that there are several overlapping symbols due to different stars with similar properties. The size of the typical error bars is indicated in the figure.}
\label{hr_diag}
\end{figure}

One of the ultimate goals for a study of this nature is to identify a link between a physical or phenomenological property and massive O-type stars hosting large-scale, strong magnetic fields. This idea is well established for late-type stars as there is a direct correlation between the observed magnetic field strength and certain activity indicators (e.g. Ca\,{\sc ii} H\&K core emission, X-ray emission, etc.), which probe the coronal or chromospheric emission that is modified by the presence of a magnetic field \citep[e.g.][]{noyes84}. Similarly, among the intermediate-mass A and B stars that host strong magnetic fields, distinct chemical peculiarities are observed that make these stars easily identifiable - the so called Ap/Bp stars \citep[e.g.][]{donati09}. \citet{martins12, martins15} already investigated the correlation between CNO abundance peculiarities among the magnetic O-type stars compared to the non-magnetic O-type stars and found no obvious distinctions between the two different populations. We therefore look into other physical or phenomenological properties.

The first aspect we investigate is whether there are any correlations between physical properties and magnetism in the attempt to identify regions of the Hertzsprung-Russel (HR) diagram that would be more likely to host magnetic stars, similar, for example, to the pulsational instability strips. In Fig.~\ref{hr_diag}, we present a $\log(g) - T_{\rm eff}$ diagram that contains all the stars studied here in addition to other confirmed magnetic O stars ($\theta^1$\,Ori\,C, $\zeta$\,Ori\,Aa, Tr16-22, HD\,54879, HD\,191612). The $T_{\rm eff}$ and $\log(g)$ values for each star are taken from \citet{martins15} (when available) or were estimated based on their spectral type according to the study of \citet{martins05a}. The parameters adopted for HD\,54879 were taken from \citet{castro15}, while the parameters for NGC\,1624-2 adopted the results of \citet{wade12b}, and the parameters for $\zeta$\,Ori\,Aa were taken from \citet{blazere15}. This figure is an update to that presented by \citet{martins15}, now containing all known magnetic O stars (11) and all the O stars studied in this work. Even with the inclusion of several additional magnetic O stars, our conclusion mirrors that of \citeauthor{martins15} - there appears to be no correlation between physical properties (e.g. $M$, $T_{\rm eff}$, $\log(g)$ and therefore $L$, $R$ and age) and the presence of magnetic fields in O-type stars. While there are no obvious correlations, we do point out that majority of known magnetic O-type stars are concentrated on the first half of the main sequence. The only star that is more evolved is the supergiant $\zeta$\,Ori\,Aa, which is close to the terminal age main sequence. This could be related to the decrease of the surface field strength with increasing radius as the star evolves, due to magnetic flux conservation.

Motivated by the results of \citet{landstreet07} and \citet{fossati16} that show evidence for magnetic field decay, we attempted to search for any correlation between age and surface magnetic field strength. To do so, we followed a similar strategy as adopted by \citet{landstreet07} and looked at the correlation between the $B_{rms}$ parameter (derived from \bell\ measurements found in this work, and from \bell\ measurements reported in the dedicated studies of these individual stars) and $\log(g)$, which we used as a proxy for age. We also looked at the correlation of $\log(g)$ with surface dipole field strength, based on results of the dedicated studies of the individual magnetic O stars. Unfortunately, the majority of the magnetic stars have $log(g)\sim4$ and host a large range of field strengths (several hundreds to thousands of G). Other than $\zeta$\,Ori\,Aa, the magnetic stars with $log(g)\la4$ have similar field strengths to those measured from stars with $log(g)\sim4$, so no obvious conclusions can be made. The only star that shows evidence for a decrease of magnetic field strength with age is $\zeta$\,Ori\,Aa, which is the most evolved magnetic star in this sample and also hosts the weakest field. 

\begin{figure}
\centering
\includegraphics[width=3.2in]{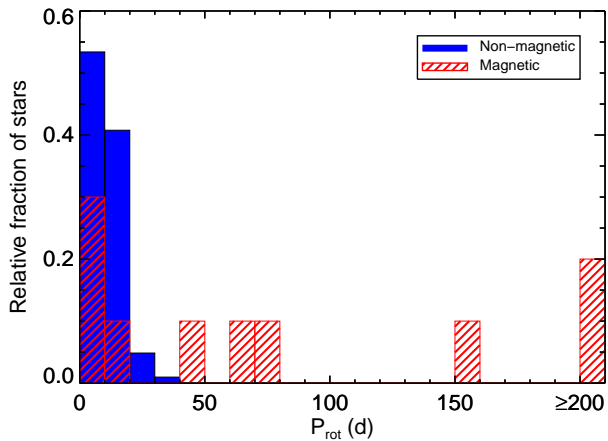}
\caption{Histogram of the O star rotation periods. The rotation periods of the magnetic stars were taken from the literature, while the periods of the non-magnetic stars were estimated from their \vsini\ measurements and radii according to their spectral types.}
\label{prot_hist_fig}
\end{figure}

Another important aspect that we investigate is the correlation between rotation rate and the presence of a magnetic field. It is well established that main sequence and pre-main sequence magnetic A and B stars generally rotate at a fraction of the speed of non-magnetic A and B stars \citep[e.g.][]{donati09, alecian13b}, likely as a result of the shedding of angular momentum due to magnetic coupling to the stellar wind and/or circumstellar disk during star formation \citep[e.g.][]{stepien00}. To address this aspect, we present a histogram of the rotation periods for each O-type star in our sample, in addition to all confirmed magnetic O stars (see Fig.~\ref{prot_hist_fig}). For the magnetic stars, we adopted the rotation periods inferred from their photometric, spectroscopic, or $B_\ell$ variations from previous studies. Since we do not have independent measurements of the rotation periods for the non-magnetic stars, we therefore estimated the rotation periods of these stars assuming rigid rotation using our \vsini\ measurements and typical radii according to their spectral types, based on the study by \citet{martins05a}. For stars with multiple measurements, we used the median \vsini. The rotation period was also estimated in this way for HD\,54879, using the radius ($6.7^{1.0}_{-0.9}$\,\rsun) and \vsini\ ($7\pm2$\,\kms) of \citet{castro15}. We assumed $\sin i=1$ when obtaining the rotation periods to obtain maximal values (this also reduces the potential bias from our profile fitting that could over-estimate the contribution of \vsini\ to the line profile). While the individual rotation periods of the non-magnetic stars are highly uncertain, as a collective they are more robust, and the distribution is sufficient for the purposes of this discussion.

The full interpretation of the results presented in Fig.~\ref{prot_hist_fig} is outside the scope of this paper, but we can make some general conclusions. The first conclusion is that the distribution of rotation periods for the magnetic stars is very different from that of the non-magnetic stars (a two-sided KS test supports the hypothesis that the magnetic sample is not drawn from the same underlying distribution as the non-magnetic sample - the null hypothesis that the two distributions are drawn from the same sample is rejected at 99.9\% confidence). The most significant conclusion is that the majority (60\%) of the magnetic stars have rotation periods longer than the longest periods found in the population of non-magnetic stars (stars with rotation periods $\ga$50\,d all appear to be magnetic). We can therefore conclude that magnetic fields play a very important role in explaining the rotation among the slowest rotating O-type stars. However, there still exists a population of magnetic stars with rotation periods comparable to the periods found from the non-magnetic stars. As discussed by \citet{uddoula09}, angular momentum loss depends on several key physical parameters of the star such as the magnetic field strength, the mass-loss rate, the rotation rate, and the radius. Therefore, to interpret the current rotation period of the star one needs to take all these factors into account, in addition to the age of the star (the older the star, the longer the star may have been affected by angular momentum loss).

The magnetic stars with the shortest rotation periods are: the broad line component of Plaskett's star (1.21551\,d - Grunhut et al. in prep); $\zeta$\,Ori\,Aa (6.83\,d - \citealt{blazere15}); HD\,148937 (7.03\,d - \citealt{naze08, wade12a}); $\theta^1$\,Ori\,C (15.442\,d - see \citealt{wade06}, and references therein). The more rapid rotation of these stars relative to the other magnetic O stars is likely a reflection of one (or several) of the following reasons: the age (an insufficient amount of time has passed to carry away enough angular momentum); the magnetic field strength is much weaker than other magnetic O-type stars and therefore couples more weakly to the stellar wind, which reduces the angular momentum loss; some form of angular momentum transport has occurred due to binary interaction, which has rejuvenated the star. This topic will be addressed in a future paper in this series.

Given the tendency to overestimate the rotational contribution to line profiles of magnetic stars \citep[e.g.][]{sundqvist13b}, it may very well be that the rotation rates of the non-magnetic stars are also over-estimated and therefore the rotation periods are under-estimated; however, if we compare to the \vsini\ distribution of all stars instead of their rotation periods, we arrive at a very similar result, except that the \vsini\ of HD\,108  is considerably over-estimated in this study \citep[see e.g.][Shultz et al. in prep]{martins10}.

We next attempt to link an observable phenomenon to the presence of magnetic fields, by looking at the incidence rate of magnetism among a number of different subsamples.

\subsubsection{Oe stars}
Oe stars are a subset of O-type stars that currently exhibit, or at some point in their history have exhibited, emission in their spectra, typically in the Balmer lines. As recent studies have suggested a much higher incidence of magnetic fields among early-type emission line stars, this group is of particular interest \citep[e.g.][]{hubrig07b, hubrig09b}. 

A careful examination of all our spectra revealed evidence of emission in the H$\alpha$ Balmer line for 47 of our targets, including all stars with a confirmed magnetic field. This results in a somewhat higher magnetic incidence fraction among emission line O stars of $13\pm6$\%, but by no means indicates a direct link between emission and magnetic fields. This result is not surprising as a number of different physical mechanisms could be responsible for producing emission in these sources (e.g. wind emission, wind-wind collision, magnetosphere, decretion disks, etc). 

Of particular interest among the Oe stars are the classical Oe stars \citep{conti74b}, which have been suggested to be a magnetic phenomenon (for example, the magnetically torqued disk model; \citealt{cassinelli02}). The characteristics of their line emission is generally morphologically distinct from other emission line stars and is attributed to a circumstellar decretion disk, usually considered to be a continuation of the Be phenomenon towards hotter spectral types \citep[see][for further details of the emission characteristics of Be stars]{porter03, rivinius13}. 

Out of the 8 prototypical classical Oe stars proposed by \citet{frost76}, only two were observed in the context of the MiMeS survey: HD\,155806 and HD\,149757, which were both discussed in the previous section and are not found to be magnetic. Additionally, the Be Star Spectra (BeSS) database contains a catalogue of all known $\sim$2000 confirmed or suspected classical and Herbig Oe and Be stars, as well as a collection of more than 100\,000 spectra provided by professional and amateur astronomers \citep{neiner11}. If we include all the stars with an O-type classification present in this database, there is a potential sample of about 70 stars. Unfortunately, only a few of these additional stars were analysed by the MiMeS Survey (HD\,17505, HD\,24912, HD\,37041, HD\,57682). Of the total sample of possible classical Oe stars analysed here, only HD\,155806 exhibits an emission morphology consistent with a classical Oe star. The others either show no obvious emission (HD\,17505), only minor core emission (HD\,24912, HD\,149757), or their emission is attributed to other phenomena (HD\,37041 is nebular; HD\,57682 is magnetospheric - see \citealt{grunhut12c} for further details) during the epoch of observation.  Therefore, with the very limited sample of classical Oe stars available in this work, there is no evidence of a direct connection between magnetism and the classical Oe phenomenon. A more statistically significant study will be required to determine if classical Oe and Be stars are as magnetic or less magnetic than normal O-stars, and this is the subject of a future paper in this series (Neiner et al. in prep); however, a preliminary analysis carried out by the MiMeS collaboration indicates that Be stars have a lower magnetic detection rate relative to non-Be stars \citep{wade15be}.

\subsubsection{ ``Weak-wind'' stars}
``Weak-wind" stars are O-type dwarfs that have anomalously lower observationally derived mass-loss rates (by a factor of $\sim$100), and terminal wind velocities that are also found to be lower than theoretical predictions determined from their luminosity \citep[e.g.,][]{martins04,martins05b,marcolino09,martins12b}. This behaviour occurs for stars with $\log{\rm L}/\rm{L_\odot} < 5.2$, and is found for both solar and lower metallicity objects.  While no current solution exists to explain such behaviour, \citet{martins05b} suggested that magnetic fields may significantly alter the ionisation structure, which would reduce the mass-loss rates and terminal wind velocities. The discovery of a magnetic field in several early-B stars (e.g. $\zeta$\,Cas \citealt{neiner03a}, V2052\,Oph \citealt{neiner03b, neiner12a}) and in the weak-wind O9\,IV star HD\,57682 \citep{grunhut09, grunhut12c} has provided additional support for the possibly important role of magnetic fields to explain this phenomenon.

Included in our survey are six additional targets that have been identified as so-called ``weak wind'' stars by \citet{martins05b} or \citet{marcolino09}: HD\,37468, HD\,38666, HD\,46149, HD\,46202, HD\,66788, HD\,149757. Our observations provide no evidence for the presence of a magnetic field in any of these stars. Results for each star can be found in Appendix~\ref{online_tables_sec}.

The lack of detection in any of the other weak-wind stars in this survey highly suggests that strong, large-scale magnetic fields do not play an important role in this phenomenon. This is further supported by recent MHD simulations of \citet{uddoula08} that have shown that the presence of a strong, large-scale magnetic field can reduce the mass carried away by the stellar wind by around 90\%. This is in poor agreement with the orders of magnitude drop in the observed mass-loss rates. The more likely explanation is related to issues with how mass-loss rates are measured \citep[e.g.][]{huenemoerder12}, or with the theoretical predictions \citep[e.g.][]{muijres12}. 

\subsubsection{Runaway stars}

Recent studies suggest a high incidence of candidate runaways among known magnetic O stars, possibly implying that the detected magnetic fields are generated or acquired during the ejection process \citep[e.g.][]{hubrig11a}. Exploration of the runaway status of all stars analysed here is outside the scope of this current study, but it is worth mentioning that only 3 of the original 7 stars identified as magnetic and runaway candidates by \citet{hubrig11d} should be considered runaway candidates at $>$$1\sigma$ significance (HD\,108, HD\,152408, HD\,191612). HD\,152408, which was previously discussed, shows no evidence of the presence of a magnetic field in this study or by \citet{bagnulo12}. Four additional candidates of this subclass were presented by \citet{hubrig11a}, and of those 4, 3 were analysed in this study and show no evidence of the presence of a magnetic field (HD\,153426, HD\,153919, HD\,154643). Therefore, it is unlikely that magnetic fields generated or acquired during the ejection process are a viable origin for magnetism in O-type stars.

\subsubsection{Of?p stars}
Of the currently identified magnetic O stars, 5 are associated with the peculiar spectral classification ``Of?p". This classification was first introduced by \citet{walborn72, walborn73} according to the presence of C\,{\sc iii} $\lambda 4650$ emission with a strength comparable to the neighbouring N\,{\sc iii} lines. Well-studied Of?p stars are now known to exhibit recurrent, and apparently periodic, spectral variations (in Balmer, He\,{\sc i}, C\,{\sc iii} and Si\,{\sc iii} lines), narrow P Cygni or emission components in the Balmer lines and He\,{\sc i} lines, and UV wind lines weaker than those of typical Of supergiants \citep[see][and references therein]{naze10}. 

Our sample includes all 5 of the known \citep{walborn10} Galactic Of?p stars: HD\,191612, HD\,108, HD\,148937, NGC\,1624-2, and CPD-28\,2561. All but HD\,191612 \citep[detected as magnetic by][and included in the TC]{donati06a} were survey targets. Many of these stars are relatively faint and would not have been considered as suitable survey targets, but were included as a consequence of the known field of HD\,191612.

Our investigation establishes that the Of?p stars represent a distinct class of magnetic O-type stars. It therefore appears that the particular spectral peculiarities that define the Of?p classification are a consequence of their magnetism. The analogous periodic photometric and spectroscopic variability of several Of?p stars discovered in the Magellanic Clouds \citep{naze15,walborn15} provides further support to their magnetic nature (all well-studied magnetic O-type stars are known to exhibit period variability, which is a consequence of rotational modulation of their confined winds).

Detailed investigations of the magnetic field and wind properties of the Of?p stars have been published by \citet{martins10}, \citet{wade11,wade12a, wade12b}, and \citet{hubrig15}.

\section{Summary}

The MiMeS LPs acquired about 500 high-resolution, high-S/N spectropolarimetric Stokes $V$ sequences of O-type stars with ESPaDOnS, Narval, or HARPSpol. An additional $\sim$380 spectropolarimetric Stokes $V$ sequences were collected from the public archives corresponding to each instrument/observatory. In the end we obtained high-resolution spectropolarimetric Stokes $V$ sequences for 97 O-type star systems. 

After co-adding all sequences of a given target obtained on a given night, we ended up with 416 individual polarized spectra of the 97 targets. Mean LSD Stokes $I$, Stokes $V$ and diagnostic null profiles were extracted from each spectrum using a number of different line masks. Ultimately, we concluded that line masks individually tailored for each spectrum were the most suitable for the detection of signal resulting from a Zeeman signature. 

For each of the extracted mean LSD Stokes $I$ profiles, we attempted to constrain their basic spectral properties (e.g. $v_r$, \vsini, \vmac) through the use of an automated tool. These results can be found in Appendix~\ref{online_tables_sec}. In the process of fitting each profile, we encountered several profiles that showed obvious signs of multiplicity or were previously known multi-line spectroscopic systems. In these cases we simultaneously fit the profiles of all components, when possible, and used these fits to extract individual profiles of each star. This procedure also led to the discovery of 2 new possible spectroscopic companions for HD\,46106, and HD\,204827, and the confirmation of a spectroscopic companion for HD\,37041. 

Taking into account the multiple star systems, we therefore performed a magnetic analysis on 108 O-type stars. With a median precision of the $B_\ell$ measurements of 50\,G (25\% of the observations were acquired with $\sigma$$<$20\,G and 70\% with $\sigma$$<$100\,G), this is the most magnetically sensitive study of the largest sample of O-type stars carried out to date. The MiMeS survey discovered or confirmed the presence of a magnetic field in 6 O-type stars: HD\,108, the broad line component of HD\,47129 (A2), HD\,57682, HD\,148937, CPD-28\,2561, and NGC\,1624-2. An additional 3 stars (HD\,36486, HD\,162978, HD\,199579) show evidence suggesting the presence of a magnetic field, but we lack the data to confirm these suspected fields. Based on these results, we derive a magnetic incidence fraction of $7\pm3$\% among the observed O-type stars. This is substantially lower than other previous claims obtained, in general, with low-resolution spectropolarimetry \citep[e.g. $\sim$30\%;][]{hubrig11a}. Furthermore, we were not able to confirm the presence of a magnetic field in 12 stars that were previously claimed to be magnetic by others. This is in good agreement with an independent reanalysis of the low-resolution FORS data by \citet{bagnulo12}, who were not able confirm the majority of magnetic claims among the FORS archival data.

The majority of magnetic O stars are located in the first half of the main sequence ($\zeta$\,Ori\,Aa is the only exception, as it currently resides close to the terminal age of the main sequence). Besides this qualitative observation, we found no correlation between the presence of a magnetic field and the physical properties of the star (e.g. $T_{\rm eff}$, $\log(g)$, $M$, or age). While there is no direct correlation between the rotation period of a star and the presence of a magnetic field, the majority of the confirmed magnetic stars are found to be rotating much slower than non-magnetic stars. The slower rotation likely results from enhanced angular momentum loss due to magnetic coupling with the stellar wind, which is expected to rapidly decrease the angular rotation rate of the star within a few million years \citep[e.g.][]{uddoula09}. In fact, detailed studies of the magnetic O-type stars show convincing evidence for the presence of co-rotating, magnetically structured winds, the presence of which agrees well with theoretical or numerical expectations between the coupling of the magnetic field and wind \citep[e.g.][]{grunhut12c, grunhut13, sundqvist12,wade11, wade12a,wade12b, wade15}.

We also explored the incidence of magnetic fields for certain subclasses of stars that present different phenomenological characteristics (e.g. ``weak-winds", emission). No direct correlation could be found between the phenomena that define most subclasses and the presence of a magnetic field.

The only subclass for which we were able to identify a direct link with the presence of magnetic fields is the peculiar Of?p stars. All 5 of the known Galactic Of?p stars were included in this survey and all were found to be magnetic. We therefore establish that the Of?p stars represent a distinct class of magnetic O-type stars, similar to the intermediate-mass, chemically peculiar Ap/Bp stars.

This paper presents the magnetic results for all O-stars studied as part of the MiMeS Survey. 
In the next paper in this series, we will study the null results in more detail. This will allow us to address issues such as fields below our detection threshold, the possibility of a magnetic desert, and the completeness of our survey. 

\section*{Acknowledgements}
The authors thank the referees, S. Sim{\' o}n-D{\' i}az and G. Mathys, for their valuable suggestions that greatly improved this manuscript. Based on MiMeS LP and archival spectropolarimetric observations obtained at the CFHT which is operated by the National Research Council of Canada, the Institut National des Sciences de l'Univers (INSU) of the Centre National de la Recherche Scientifique (CNRS) of France, and the University of Hawaii; on MiMeS LP and archival observations obtained using the Narval spectropolarimeter at the Observatoire du Pic du Midi (France), which is operated by CNRS/INSU and the University of Toulouse; and on MiMeS LP observations acquired using HARPSpol on the ESO 3.6 m telescope at La Silla Observatory, Program ID 187.D-0917. CFHT, TBL and HARPSpol observations were acquired thanks to generous allocations of observing time within the context of the MiMeS LPs. EA, CN, and the MiMeS collaboration acknowledge financial support from the Programme National de Physique Stellaire (PNPS) of INSU/CNRS. This research has made extensive use of the SIMBAD data base, operated at CDS, Strasbourg, France. The Dunlap Institute is funded through an endowment established by the David Dunlap family and the University of Toronto. We acknowledge the CADC. GAW acknowledges Discovery Grant support from the Natural Sciences and Engineering Research Council (NSERC) of Canada. The authors extend their warm thanks to the staff of the CFHT and TBL for their efforts in support of the MiMeS project.




\bibliographystyle{mnras}
\bibliography{bibtex} 



\appendix

\section{Details of the spectroscopic multiple systems}\label{binary_appendix}
This section provides a summary of the details of the spectroscopic multiple systems. In Fig.~\ref{bin_profs_fig}, we illustrate the quality of our multi-line fitting technique to the LSD profiles of each system. Fig.~\ref{entang_profs_fig} provides a comparison between the observed LSD profile and a synthetic profile corresponding to a single star to illustrate the level of entanglement. Example profiles for HD\,36486 and HD\,199579 are provided in Fig.~\ref{pos_det_profs_fig}.

\begin{itemize}
\item {\bf HD\,1337} (AO\,Cas) is a well-known close binary ($P=3.52$\,d) with two O-star components \citep[e.g.][]{bagnuolo99, palate12}. The LSD profile shows two clearly separated profiles; the RV measurements of each component agree well with previous studies.

\item {\bf HD\,17505} is a triple system composed of three O-type stars, which was previously studied by \citet{hillwig06} and additional references therein. Our LSD profile shows a clearly blended profile with three obvious components. The RV measurements and other aspects of the line profiles of the three components are in good agreement with previous studies.

\item {\bf HD\,35921} (LY\,Aur) is a spectroscopic eclipsing binary with a close visual companion \citep[ADS\,4072B; e.g.][]{stickland94, mason98}. The primary component (A1) is an O star and the secondary (A2) is an early B star. Our LSD profile shows three distinct profiles - two deep and broad profiles (corresponding to the primary and secondary components) and one narrower weak component. The spectroscopic binary is well separated with the primary at positive velocities in our spectrum, in accordance with the orbital solution of \citet{stickland94}. The reported separation of the visual companion \citep[$\sim$0.61 arcsec;][]{mason98} is small enough that it would fall in the instrument's pupil and therefore we included an additional component for this star in our fitting procedure. We suspect that the narrow profile blended with the primary profile at an RV $\sim$-18\,\kms\ comes from the visual companion (component B), although \citet{stickland94} find a value of about $+25$\,\kms. There appears to be a fourth component at very high velocities, but we suspect this is just due to residual telluric features. In fact, similar features can be seen in other profiles for different stars, but their relative strength varies. In this case, the strength of the feature is deep enough that it was attributed to another star.

\item {\bf HD\,36486} ($\delta$\,Ori\,A) is a complex multiple star system consisting of an eclipsing spectroscopic binary with a $\sim$6\,d period (the Aa1 and Aa2), and close visual companion (Ab) \citep[e.g.][ and references therein]{harvin02, richardson15}. Additionally, there are two more angularly separated components B and C. The primary component of the eclipsing binary is a late O-type star and its companion is an early B-type star. There is no evidence in our LSD profiles for the presence of multiple line profiles. Any attempts at fitting our observed profiles with multiple components (using the orbital parameters of \citealt{harvin02}) failed. Given the large magnitude difference in the optical of the two stars in the close binary system \citep[$\Delta V\sim$2.5;][]{harvin02}, we suspect that our LSD profile is dominated by the primary O star. Note that $\delta$\,Ori\,C is a well-known magnetic Bp star and is separated by about 50\,arcsec from $\delta$\,Ori\,A. This separation is sufficiently large that no contamination of the observations by light of $\delta$\,Ori\,c is possible.

\item {\bf HD\,37041} ($\theta^2$\,Ori\,A) is part of a multiple star system with at least four components \citep{mason98}. The brightest component (Aa) belongs to a spectroscopic binary, which is known to have a 21\,d orbital period \citep{aikman74}. The companion (Ab) is not spatially resolved nor detected in previous studies \citep[e.g.][]{sota11}; however, \citet{simon-diaz06} noted that the poor fit of their single-star synthetic line profiles to the wings of the He\,{\sc i} lines in their observed spectrum could be caused by the presence of a companion star. The other components in this system are well separated and should not contribute at all to the observed spectrum of this star. Our LSD profiles show strong evidence for the presence of the companion as a depression in the wings of the primary profile, similar to what was observed by \citeauthor{simon-diaz06}. The parameters for the LSD profile of each component was obtained from the best fit to all observations.

\item {\bf HD\,37043} ($\iota$\,Ori) is part of a 29\,d double-lined, spectroscopic binary system in a highly eccentric orbit \citep{stickland87b, bagnuolo01}. The primary is a late O-type star and the secondary is an early-type B star. Our observed LSD profile shows the secondary clearly blended with the blue wing of the primary star. We fixed the RV values to the orbital velocities from \citet{stickland87b} to constrain our fits to the entangled profile. The resulting EW ratios of the two components are in good agreement with results of \citet{bagnuolo01}.

\item {\bf HD\,37366} is a double-lined spectroscopic system in a 32\,d orbit with a faint visual companion \citep[see][and references therein]{boyajian07}.  The primary star is classified as a late O-type star, while the secondary is an early B-type star. Our LSD profile shows two well-separated profiles - a narrow profile at negative velocities and a very broad profile at positive velocities. The RV measurements and widths of the profiles are consistent with the study of \citet{boyajian07}.

\item {\bf HD\,37468} ($\sigma$\,Ori\,AB) is a massive triple system composed of the recently confirmed double-lined spectroscopic binary (the Aa and Ab components) and the well-known visual companion $\sigma$\,Ori\,B that is separated by less than 0.3 arcsec from the binary system \citep[][and references therein]{simon-diaz11}. The binary system is comprised of an O-type star and an early B-type companion \citep{simon-diaz15}. Our LSD profile shows the clear presence of a narrow line profile entangled with a much broader profile. We fit our LSD profile without any constraints and found that our best-fitting values are consistent with \citet{simon-diaz15}. $\sigma$\,Ori\,B was not included in our original fits. Our attempts to include this component resulted in a failure for our fits to converge. Further constraining the models using the orbital and rotational velocities determined by \citet{simon-diaz15} did not improve the situation. We therefore neglected component B. This should have minimal impact on the results, as this component represents about 20\% of the overall EW, or about 10\% of the EW of the Aa and Ab components.

\item {\bf HD\,46106} is not a previously known spectroscopic binary. Our LSD profile shows strong evidence for two components - one narrow line component and one broad line component. Our RV measurement differs substantially from a previous measurement of \citet{duflot95} and the line asymmetry also differs from the profiles of \citet{sota11}, both key points that further support our binarity claim. 

\item {\bf HD\,46149} is a recently discovered spectroscopic binary composed of a late O-type star and an early B-type star \citep{mahy09}. Our LSD profiles show evidence for both a broad and narrow line component. We conducted our fits without any constraints and the resulting RV measurements are in good agreement with \citet{mahy09}.

\item {\bf HD\,47129} (Plaskett's star) is a well-known multiple star system consisting of a spectroscopic binary in addition to 3 visual companions at distances of 36 mas, 0.78 arcsec and 1.12 arcsec \citep{turner08,sana14}. The spectroscopic binary system was previously found to be composed of two O star components, one with narrow lines (A1) and the other showing relatively broad lines (A2), with similar RV variations \citep[e.g.][]{plaskett22, linder08}; however, the recent work of Grunhut et al. (in prep) argues that the broad line component shows only weak RV variations consistent with a mean RV of about 15\,\kms. We therefore constrained the broad line component to have a fixed RV of 15\,\kms, but allowed all other values to vary. The resulting RV measurements for the narrow line component are in good agreement with previous studies \citep[e.g.][]{linder08}. The spectral types used in this work are from \citeauthor{linder08}.

\item {\bf HD\,47839} (15\,Mon) is a known speckle (with a separation of 0.035 arcsec) and single-lined spectroscopic binary system \citep{gies93}. The primary star has a spectral type of O7\,V, while the secondary is in the range of B0-O9 (\citeauthor{gies93} suspect a spectral type of O9.5\,Vn, considering the breadth of the potential secondary's lines). Our LSD profiles show evidence for stronger absorption in the red wing compared to the blue wing, which we attribute to the presence of the companion with broad lines - a result that supports the findings of \citeauthor{gies93}. We constrained our fits by adopting the RV values corresponding to the orbital solution of \citeauthor{gies93}. Given that these stars are of similar temperature, we use the EW ratio of the best-fitting profiles to establish a flux ratio of $\Delta m = 1$, which agrees well with the lower limit of the flux ratio from a similar analysis of He\,{\sc i} lines and the speckle observations as measured by \citeauthor{gies93}.

\item {\bf HD\,48099} is a double-lined spectroscopic binary with a short $\sim$3\,d period. The system is composed of two O stars \citep{mahy10}. Both of our LSD profiles show evidence for the blending of two profiles. We used the predicted RV values and the rotational velocities of \citet{mahy10} to constrain our fits to the observed LSD profiles. The resulting best-fitting profiles are in good agreement with the results presented by \citet{mahy10} and provide a reasonable fit to the observed profiles.

\item {\bf HD\,54662} is a long period ($\sim$560\,d) double-lined O-O star spectroscopic binary \citep{boyajian07}. Our LSD profile shows clear evidence for the presence of both a narrow and broad line component. We carried out an unconstrained fit to the LSD profile and the resulting RV measurements and line broadening for the narrow line component are consistent with previous studies \citep{conti77, boyajian07}. We find a higher value for the line broadening of the broad line component than \citeauthor{boyajian07}. The EW ratios of the two fits are in good agreement with the flux ratio found by \citeauthor{boyajian07}, given that the temperatures of the two stars are similar.

\item {\bf HD\,93250} is a recently discovered interferometric binary system with a separation of less than 2$\times10^{-3}$ arcsec. \citet{sana11} carried out a detailed investigation and found no evidence for the presence of a spectroscopic companion. It was also shown that two similar profiles separated by 50\,\kms\ could reproduce the apparently single line profile. Our LSD profile shows no convincing evidence for the presence of two blended profiles. We note that the intensity of the red wing of our LSD profile is slightly elevated relative to the blue wing, which may be an indication of a spectroscopic companion. Considering the degeneracy of potential fits, we did not attempt to fit this profile.

\item {\bf HD\,153426} is a known radial velocity variable star that is suspected to be an SB2 \citep{crampton72}. Our first LSD profile obtained on 2011-07-03 shows clear evidence for extended absorption in the red wing that we attribute to the presence of a very broad profile, in addition to the narrow line component. The presence of this broad profile is not apparent in the second LSD profile obtained on 2012-06-22. We fit LSD profiles assuming maximum contribution from the broad line component. This provides a good fit to our first profile, but does not fit the second profile well. 

\item {\bf HD\,155889} is a known speckle binary with a 0.19 arcsec separation \citep[e.g.][]{mason98}. Our LSD profile exhibits enhanced absorption in the red wing of the line profile, which we attribute to the presence of a spectroscopic companion, which may or may not be the speckle companion. Our fits to the LSD profile assume maximum contribution of the broad line companion.

\item {\bf HD\,164794} (9\,Sgr) is a recently uncovered spectroscopic O-O star binary in an eccentric orbit, with a long (8.6\,yr) orbital period. The study of \citet{rauw12} determined both the orbital and spectroscopic parameters of the secondary star. They found both components to have similar brightness and similar line widths. Our spectra were obtained at epochs where the stars have similar RVs and their spectra are highly blended. We attempted to fit our LSD profiles using the RV solution and estimated rotational velocities of the previous study, but the final results are too degenerate to find a reasonably unique solution. We therefore do not attempt to disentangle the profiles for this system and analysed the single blended profile instead. The spectral types noted in this work are from \citet{rauw12}.

\item {\bf HD\,165052} is a well-known SB2 system composed of two O-type stars in a close 3\,d orbit \citep{conti74a, morrison78a,stickland97b}. This system was recently investigated as part of the study of \citet{linder07}, who found the two components to have similar spectral types and a luminosity ratio of about 1.55, in favour of the primary. This study also determined the rotational broadening parameters of each component. Our observation does not show clear evidence for two separate profiles; however, this is easily explained as we observed this star at conjunction and thus both profiles are well blended. We first attempted to fit the LSD profile using RV values for each component based on the orbital solution of \citeauthor{linder07}, as well as their rotational broadening measurements. While the resulting solution provided a reasonable fit to the observed LSD profile, we found a much better fit when only constraining the RV of each component. In the latter case the total line broadening for each component is considerably higher than what is reported by \citeauthor{linder07} and we find the primary star to exhibit broader lines. The EW ratio of the two profiles are in good agreement with results of \citeauthor{linder07} with the primary to secondary EW ratio of 1.59.

\item {\bf HD\,167771} is a well-known spectroscopic binary, although the secondary component was only revealed many years after the first RV variation measurements by \citet{stickland97c}. This study found that the EW ratio (primary/secondary) of the cross-correlation function profiles from {\it IUE} data was about 1.7, with velocity widths of 115 and 85\,\kms\ for the primary and secondary, respectively. Unfortunately, our observation was taken at a phase when the two profiles are heavily blended and the observed LSD profiles appear to be that of a single star. We attempted to fit the observation using the orbital solution and rotational velocities of each component from \citeauthor{stickland97c}. The results were in poor agreement with the data. We next tried to fit the LSD profile using just the RV constraints from \citeauthor{stickland97c}. We obtained a much better fit, but the total line broadening was about 30\,\kms\ higher for the primary and about 40\,\kms\ higher for the secondary. We furthermore found that the EW ratio of the two profiles was closer to 1.5 than the expected 1.7. Considering the degeneracy involved when fitting these highly blended profiles we chose not to utilise our fits in any measurements.

\item {\bf HD\,190918} is a well-known WR+O binary system with several faint, nearby visual companions \citep[e.g.][]{hoffleit83, mason01}. The WR+O binary is in a long 112\,d orbit \citep{underhill93b}. Our observed spectrum shows strong emission lines consistent with the presence of the WR star in addition to clear absorption lines resulting from the O star companion. Our line mask was tailored to only include absorption lines contributed from the O star. As the WR star does not contribute to the absorption line spectrum profile, the LSD profiles extracted for this star should be considered as originating from a single star.

\item {\bf HD\,191201} is a double-lined spectroscopic O-O star binary with an $\sim$8\,d orbital period in a circular orbit \citep{burkholder97}. Our data shows two well-separated profiles and the RV measurements of each profile are consistent with the orbital solution presented by \citeauthor{burkholder97} - the primary component is at positive RV and the secondary at negative RV. We did not use any constraints when fitting the LSD profile. 

\item {\bf HD\,193322} is a hierarchical multiple star system consisting of at least 6 components \citep[see][and references therein]{brummelaar11}. The B, C and D components are found at angular separations greater than 2.68 arcsec and are not observed as part of our spectra \citep{turner08}. The Aa component is a late O-type star in a long period orbit (35\,yr) with another binary system consisting of a late O-type star (Ab1) and likely an early B-type star (Ab2). Our LSD line profile exhibits a narrow component with extended absorption in the wings of the profile. According to the study of \citeauthor{brummelaar11}, based on the relative brightness of each component and the previously determined rotational velocities, we suspect that we are only sensitive to components Aa (the broad line component, evident in the wings) and Ab1 (the narrow component). We constrained the fits to our LSD profile using the rotational broadening values of \citeauthor{brummelaar11}.

\item {\bf HD\,193443} is a spectroscopic binary recently studied by \citet{mahy13}, composed of two late O-type stars.  Our LSD profile does not show obvious signs of binarity, although it is slightly asymmetric with an extended red wing. Using the RV solution of \citeauthor{mahy13} we conducted our fit to the LSD profile. The best-fitting profile provides a good fit to the observation and, given that the temperatures of the two stars are similar, the resulting EW ratio of the individual profiles are in good agreement with the relative brightness ratio of the primary to secondary of about 4, as found by \citeauthor{mahy13}.

\item {\bf HD\,199579} is a spectroscopic binary with a faint \citep[$\Delta V\sim2.5$;][]{williams01} secondary. The primary component is an O-type star and the companion is likely a B-type star. This star was observed only once and our LSD profile appears dominated by the primary. We find  no obvious signs of the secondary component in our spectrum.

\item {\bf HD\,204827} is a known speckle and {\it Hipparcos} binary with a separation of about 0.9 arcsec \citep[e.g.][]{mason98}. This system was not a previously known SB2, although it is a well-known single-line spectroscopic binary \citep{petrie61}. Our LSD profile shows clear evidence for the presence of a strong broad profile (Aa) with a weak contribution from an additional profile. This secondary profile may belong to the visual companion (Ab). Assuming that the stars have similar spectral types, the EW ratio of the two components suggest the dominant profile contributes roughly 95\% of the total optical light. This is in good agreement with the {\it Hipparcos} magnitude difference of the visual companion \citep{perryman97}. 

\item {\bf HD\,206267} is a spectroscopic triple system \citep{stickland95}. The brightest component and faintest component are O-type stars locked in a 3.7\,d orbit, while the intermediate bright star is found at a constant velocity and has a spectral type of OB \citep{burkholder97}. Our LSD profile shows no evidence of multiple components. We attempted to constrain the fit using parameters from \citeauthor{stickland95} and \citeauthor{burkholder97}, but the individual components are too entangled in our LSD profile to find a reasonably non-degenerate solution.

\item {\bf HD\,209481} (14\,Cep, LZ\,Cep) is a well-known double-lined spectroscopic binary consisting of two late O-type stars in a close 3\,d orbit \citep[see][and references therein]{mahy11b}. We have several observations of this star, and the LSD profiles of some reveal two well-separated profiles. We constrained the profile parameters of the two components using these easily separable profiles, and then used these constraints for the more entangled observations. To further constrain the fits, we also utilised the RV solution of \citeauthor{mahy11b}.

\end{itemize}

\begin{figure*}
\centering
\includegraphics[width=2.3in]{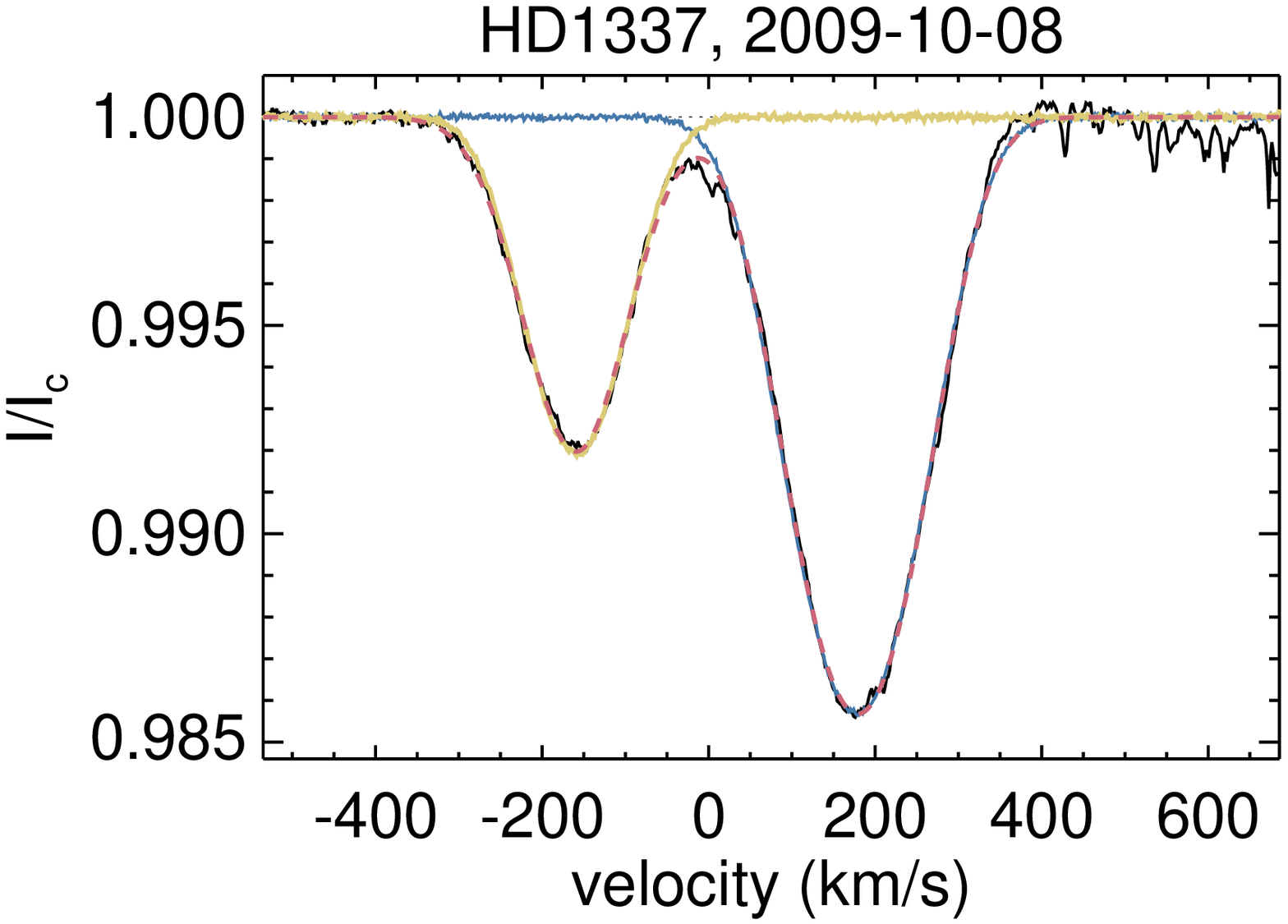}
\includegraphics[width=2.3in]{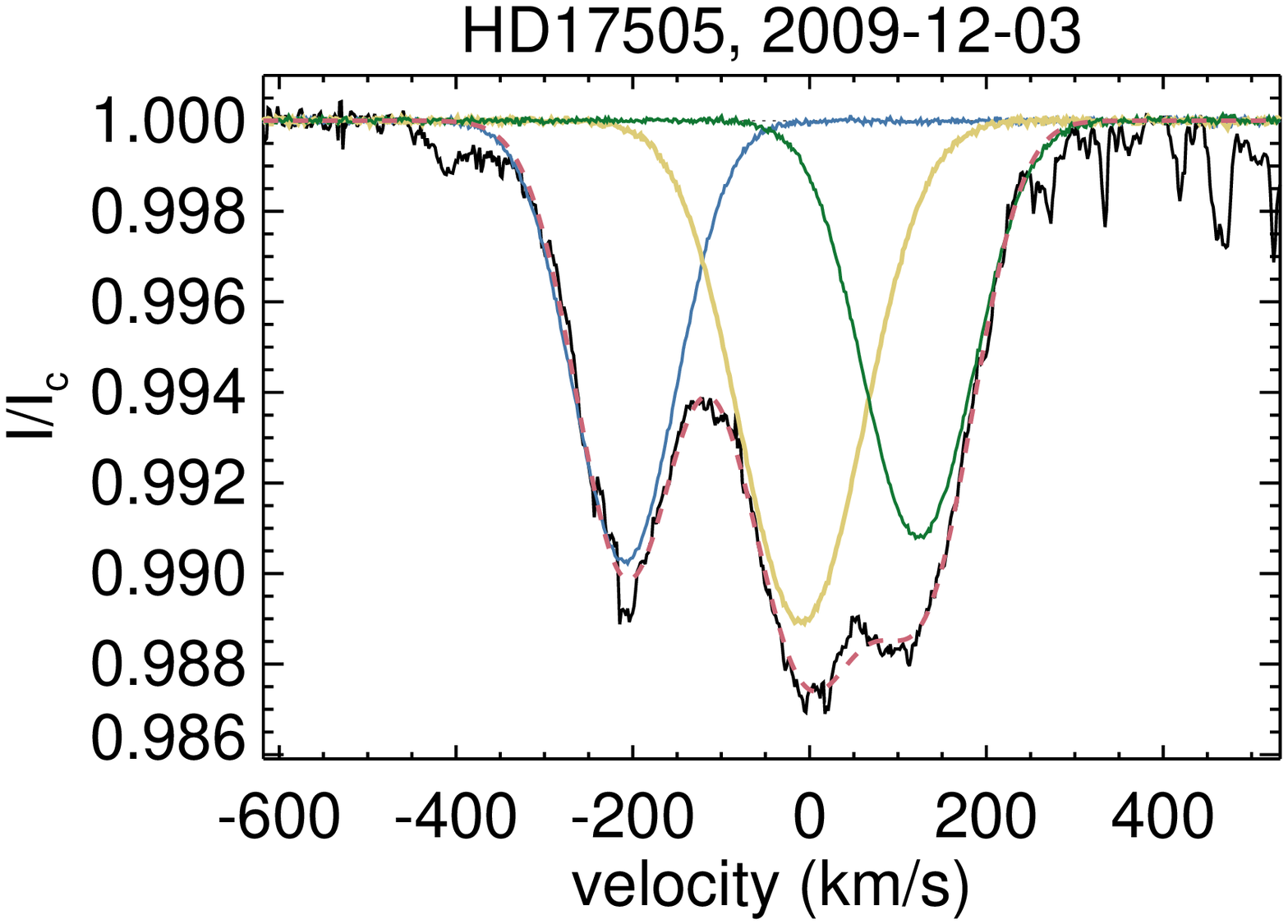}
\includegraphics[width=2.3in]{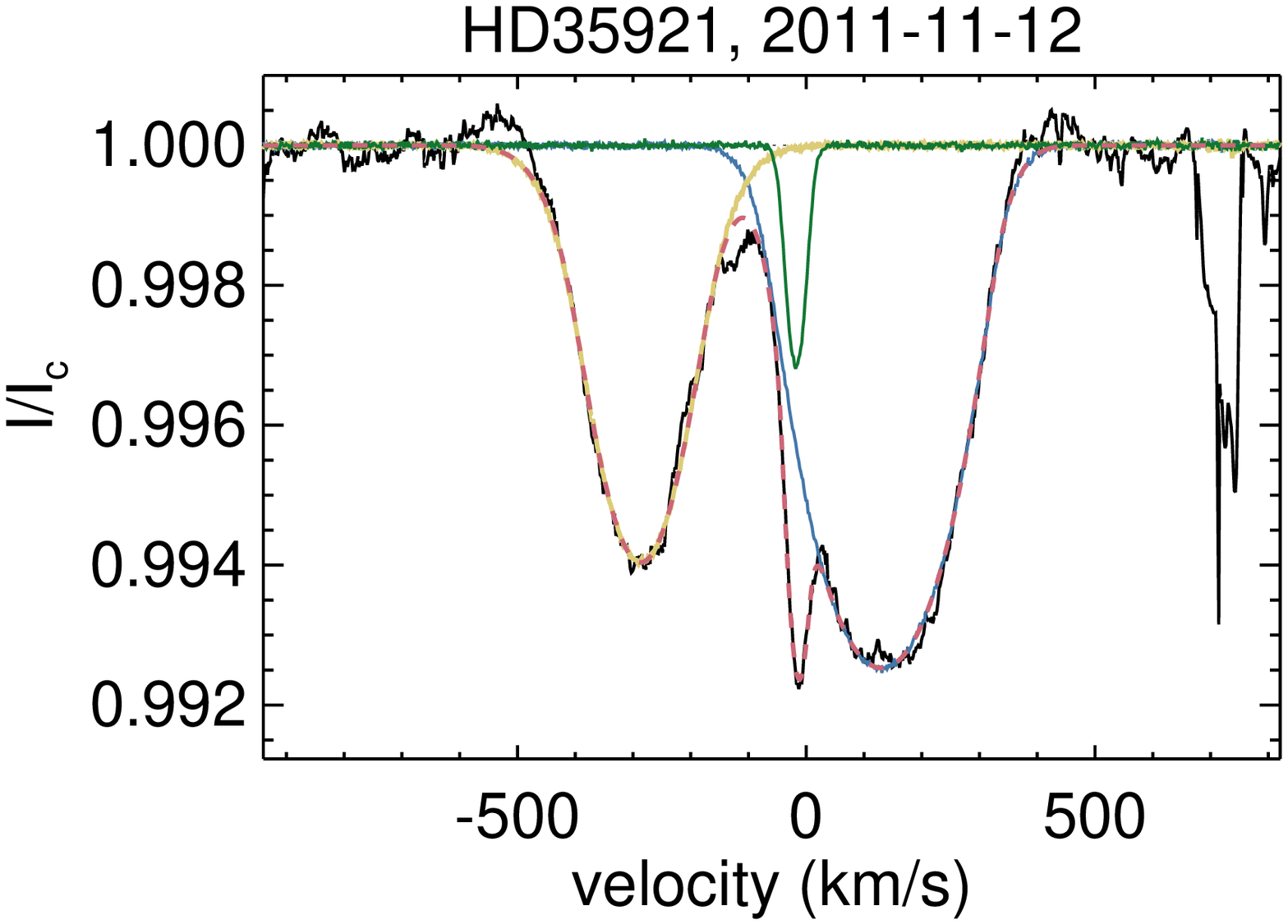}\\
\includegraphics[width=2.3in]{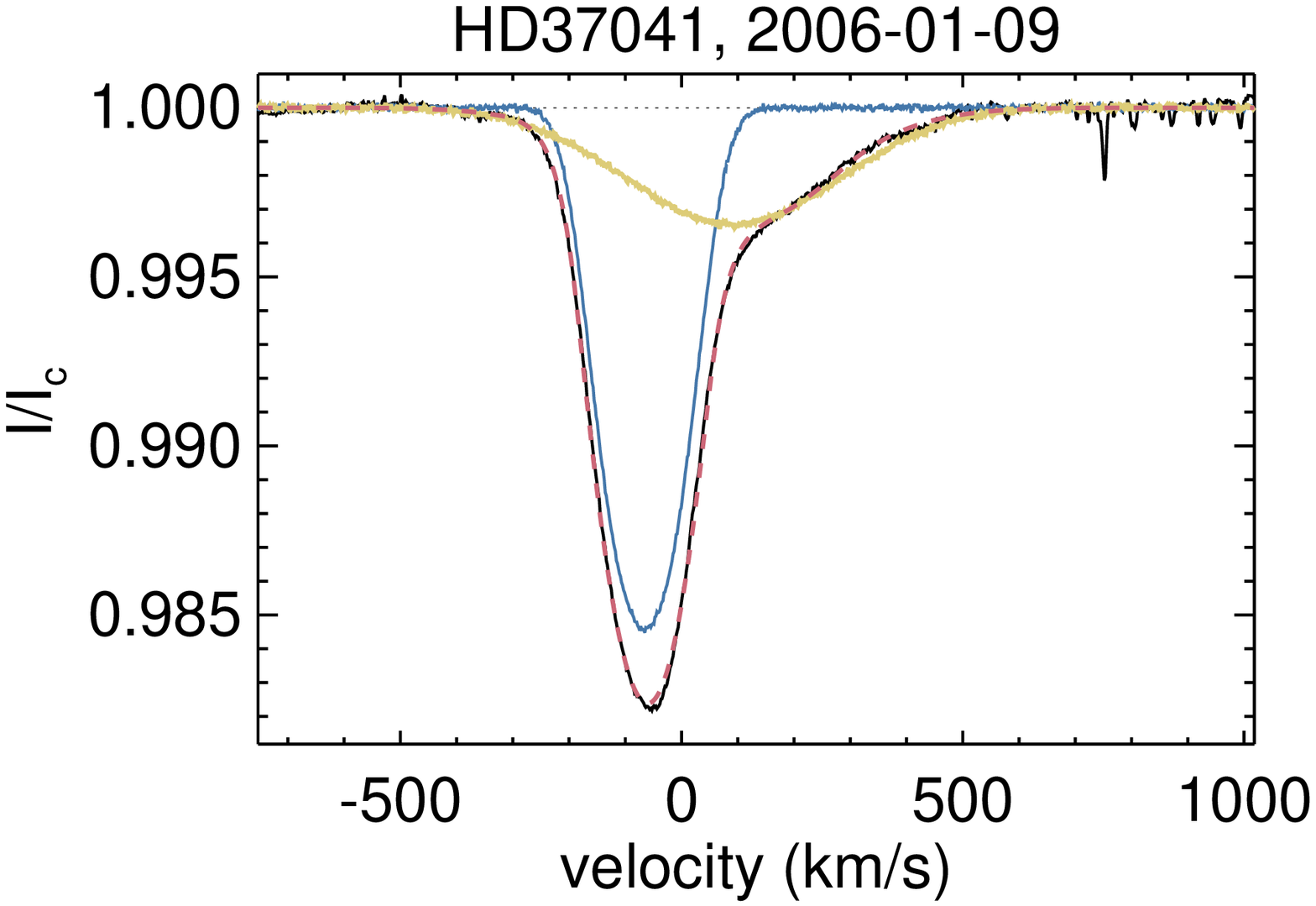}
\includegraphics[width=2.3in]{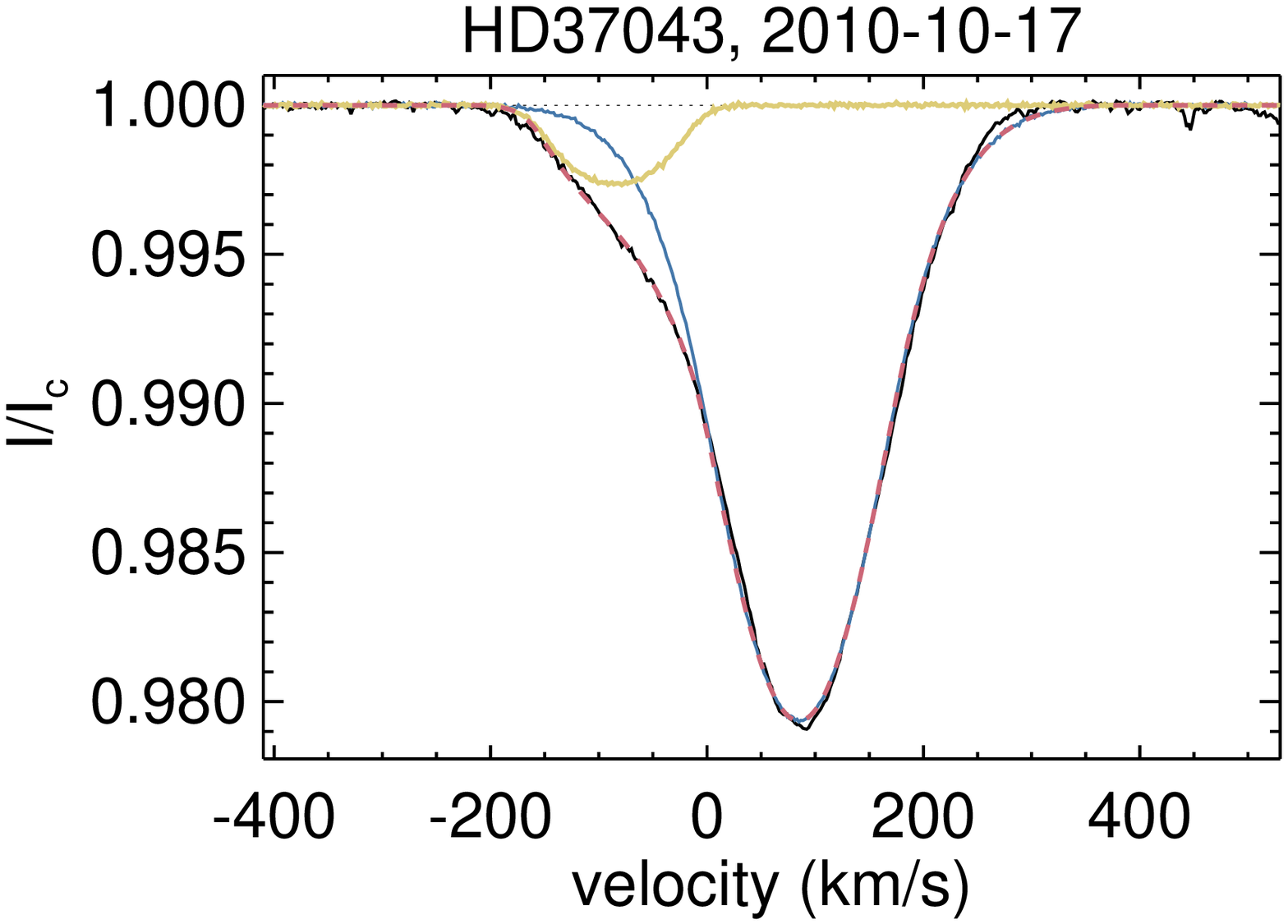}
\includegraphics[width=2.3in]{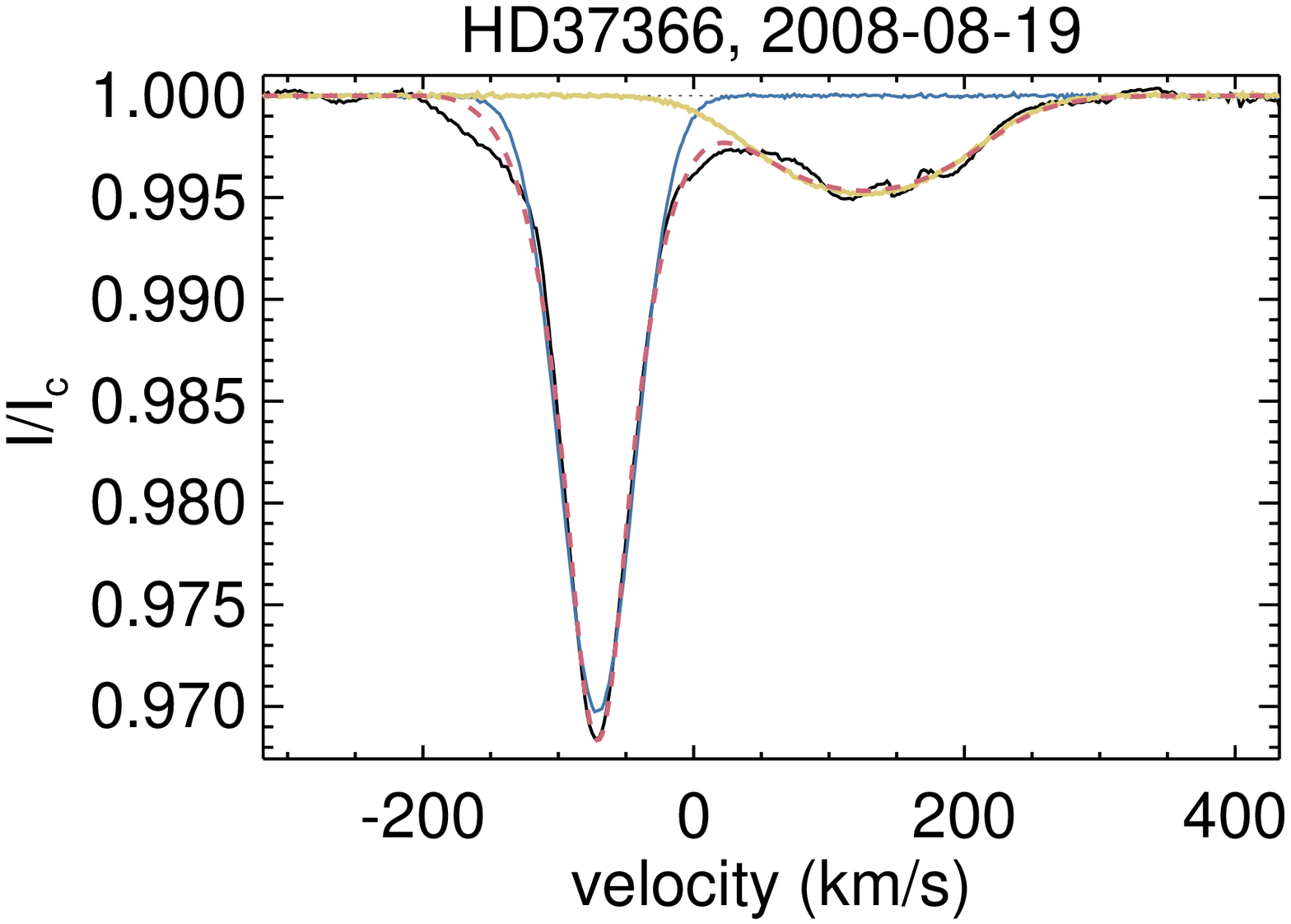}\\
\includegraphics[width=2.3in]{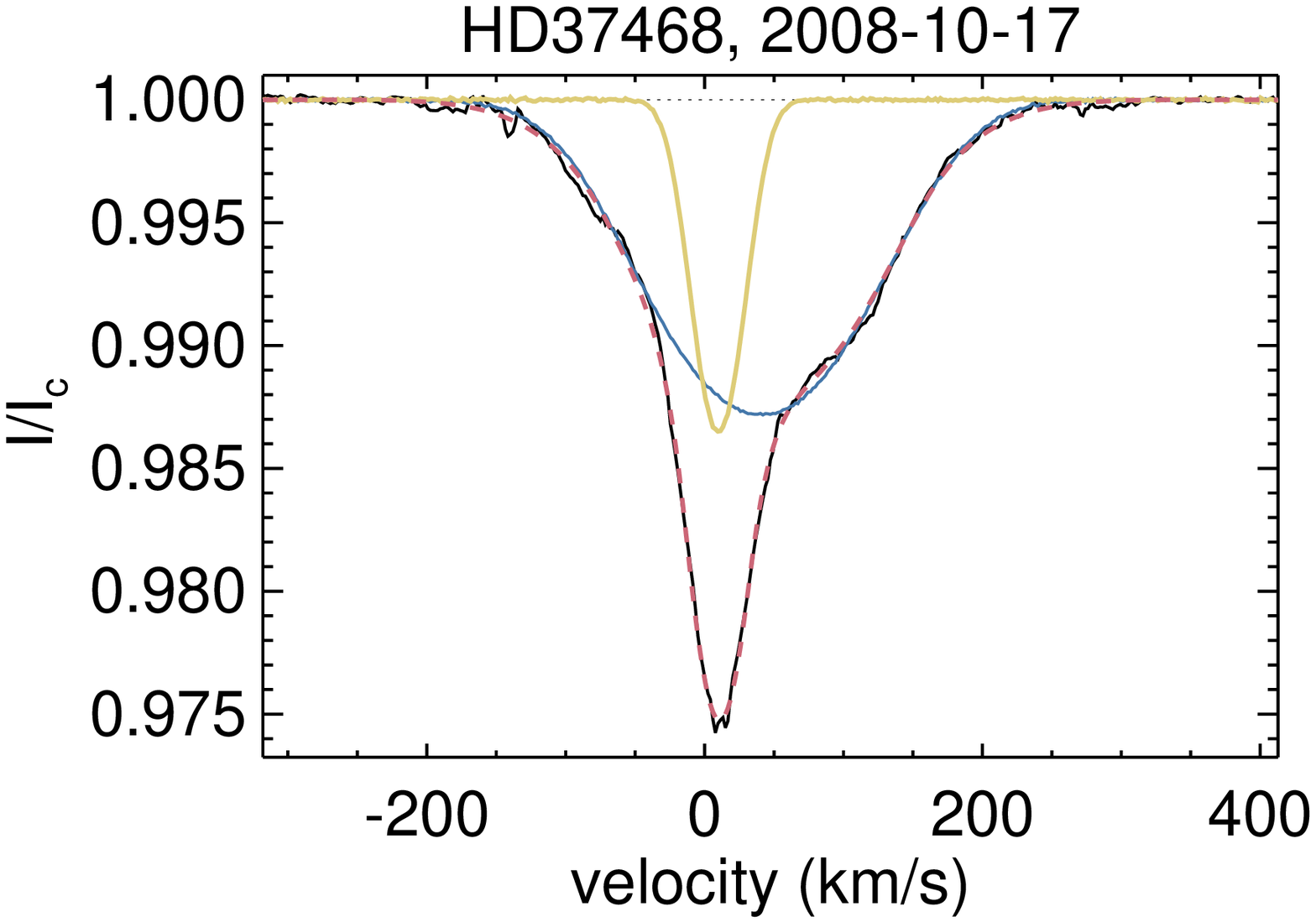}
\includegraphics[width=2.3in]{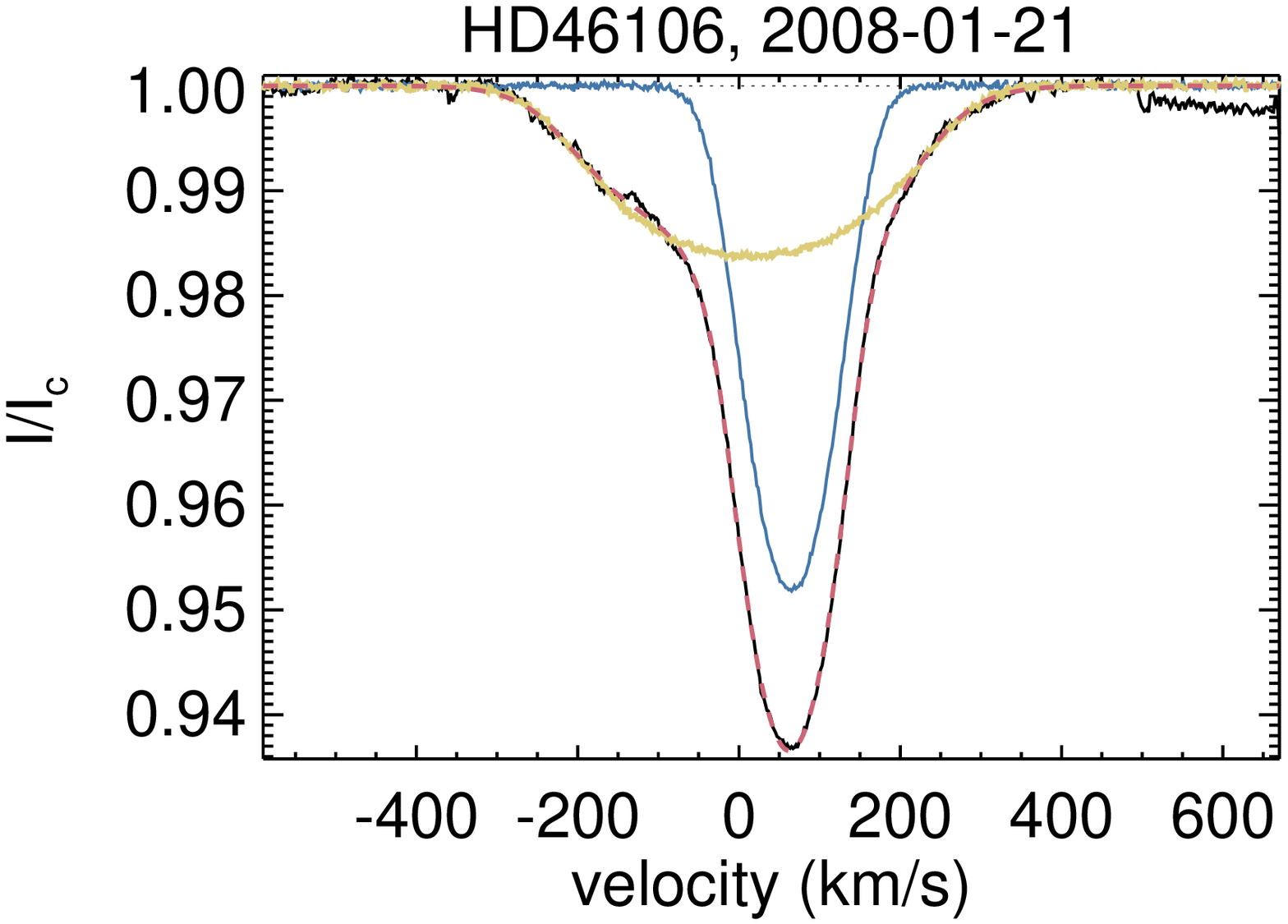}
\includegraphics[width=2.3in]{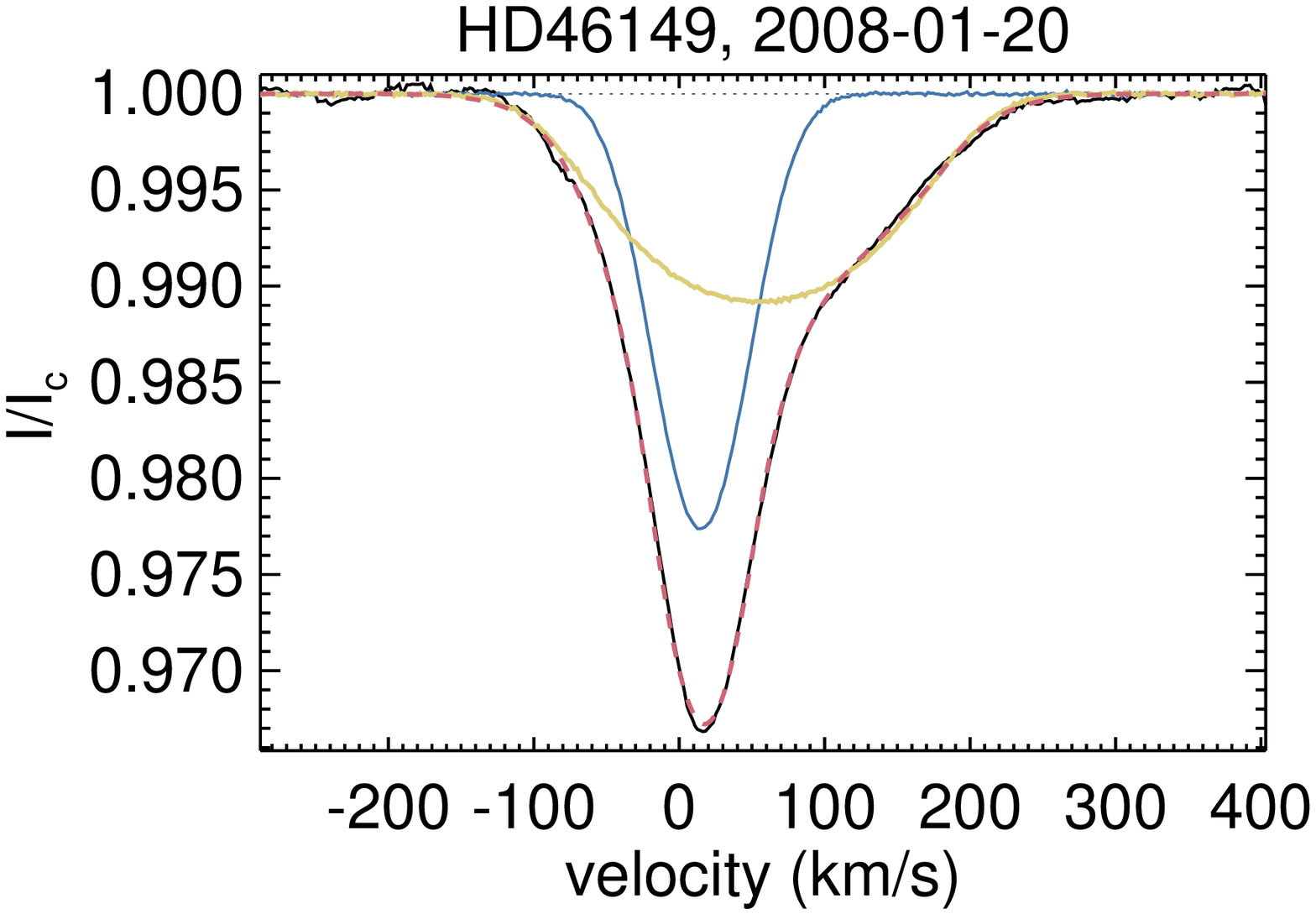}\\
\includegraphics[width=2.3in]{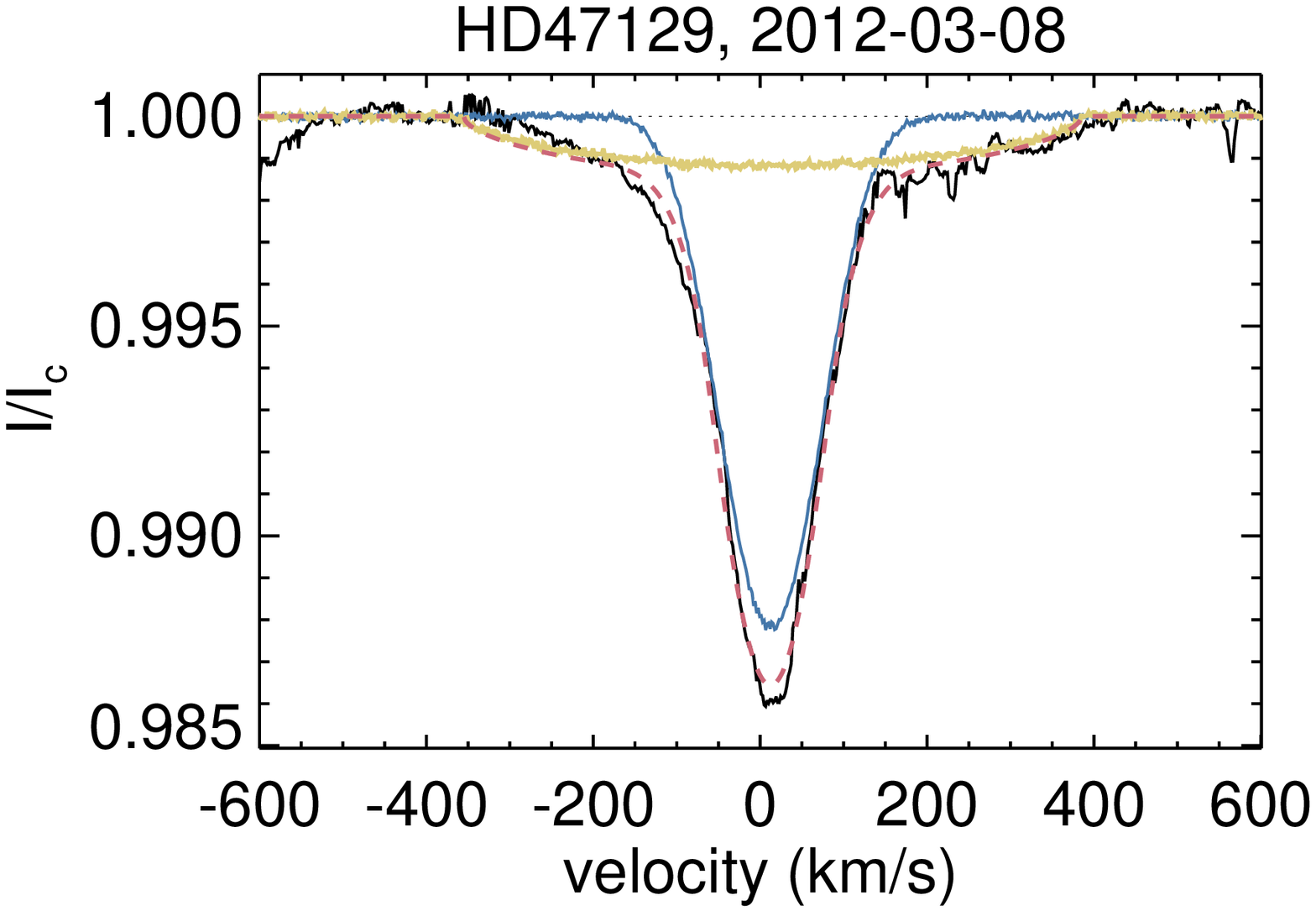}
\includegraphics[width=2.3in]{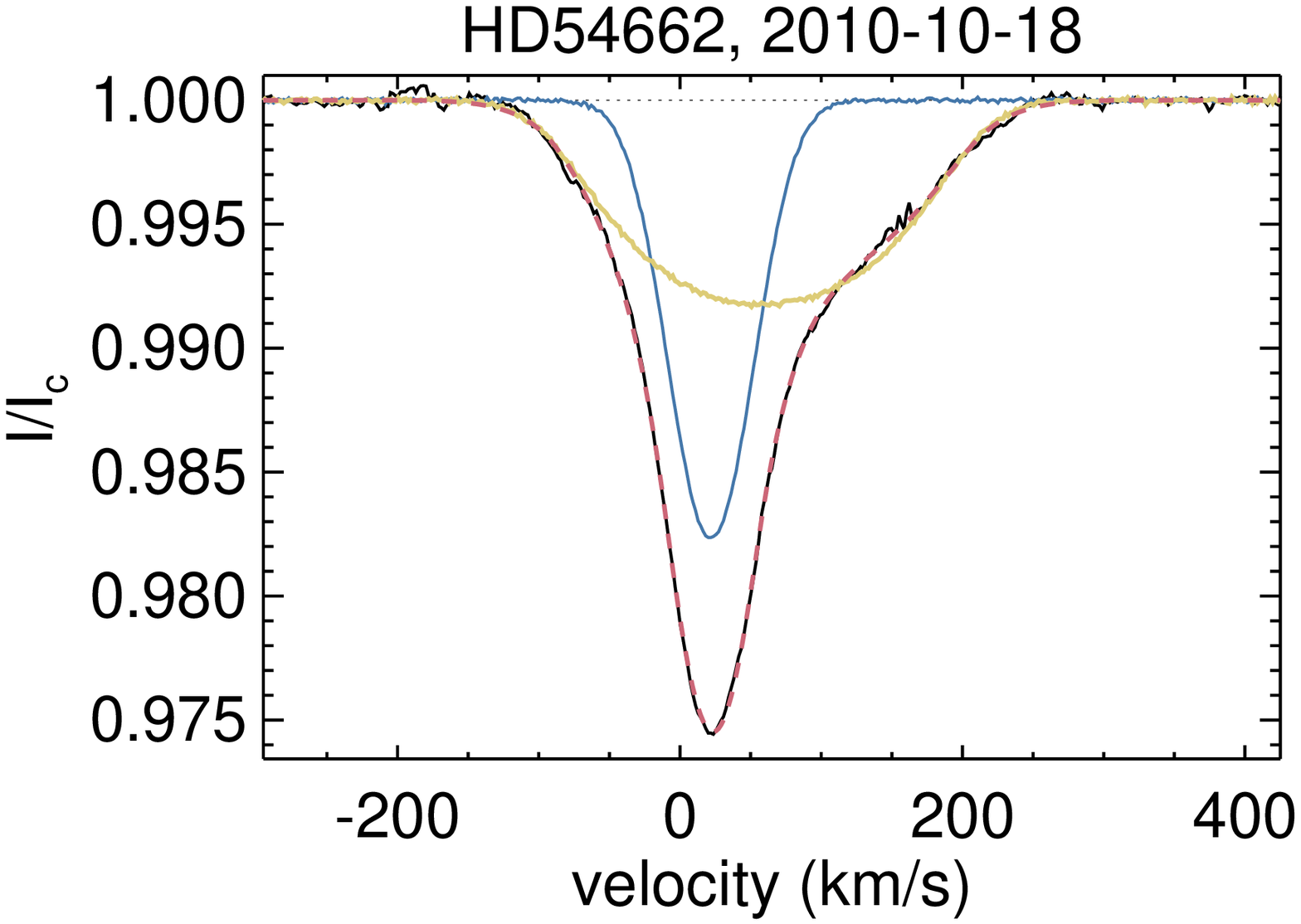}
\includegraphics[width=2.3in]{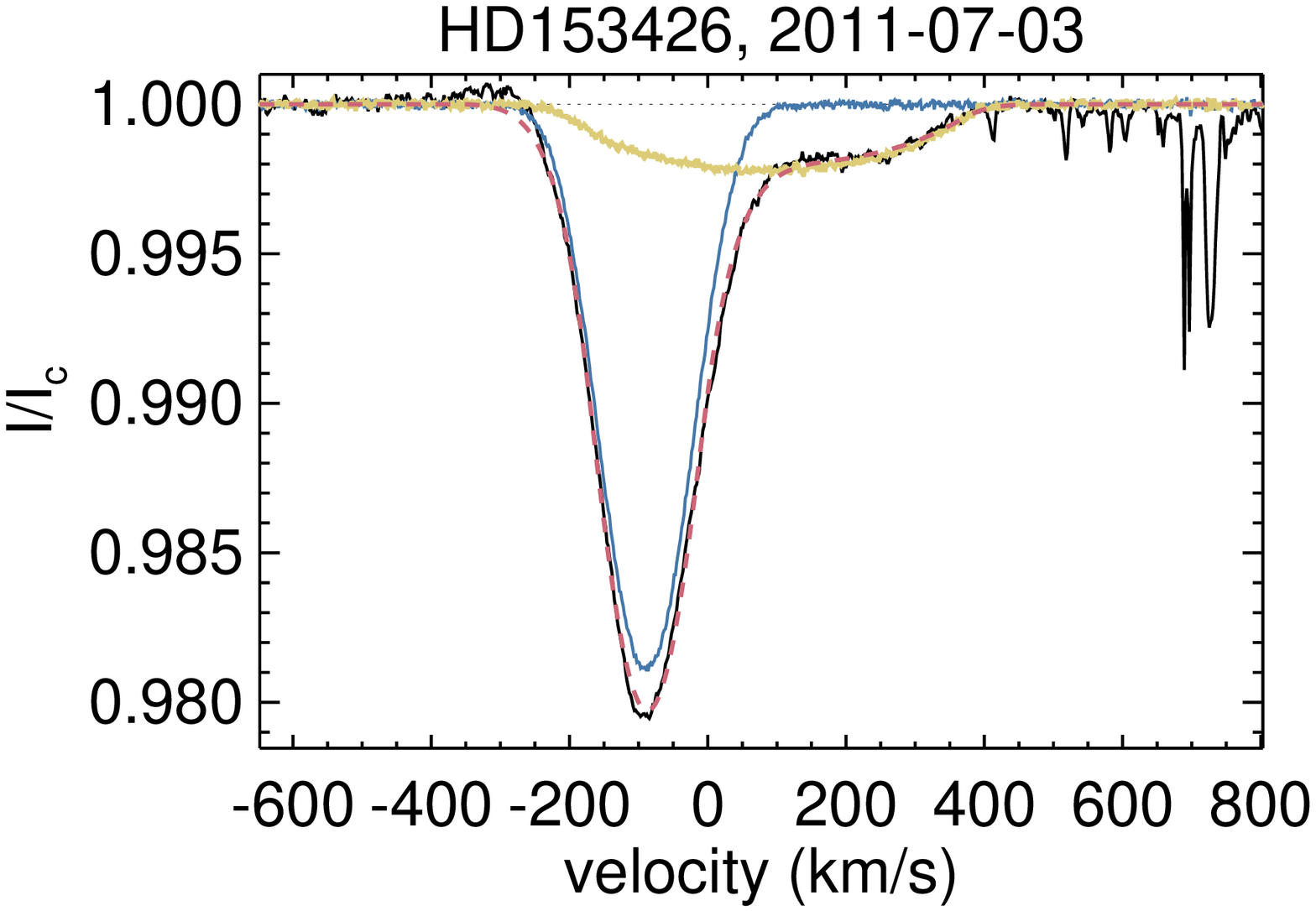}\\
\includegraphics[width=2.3in]{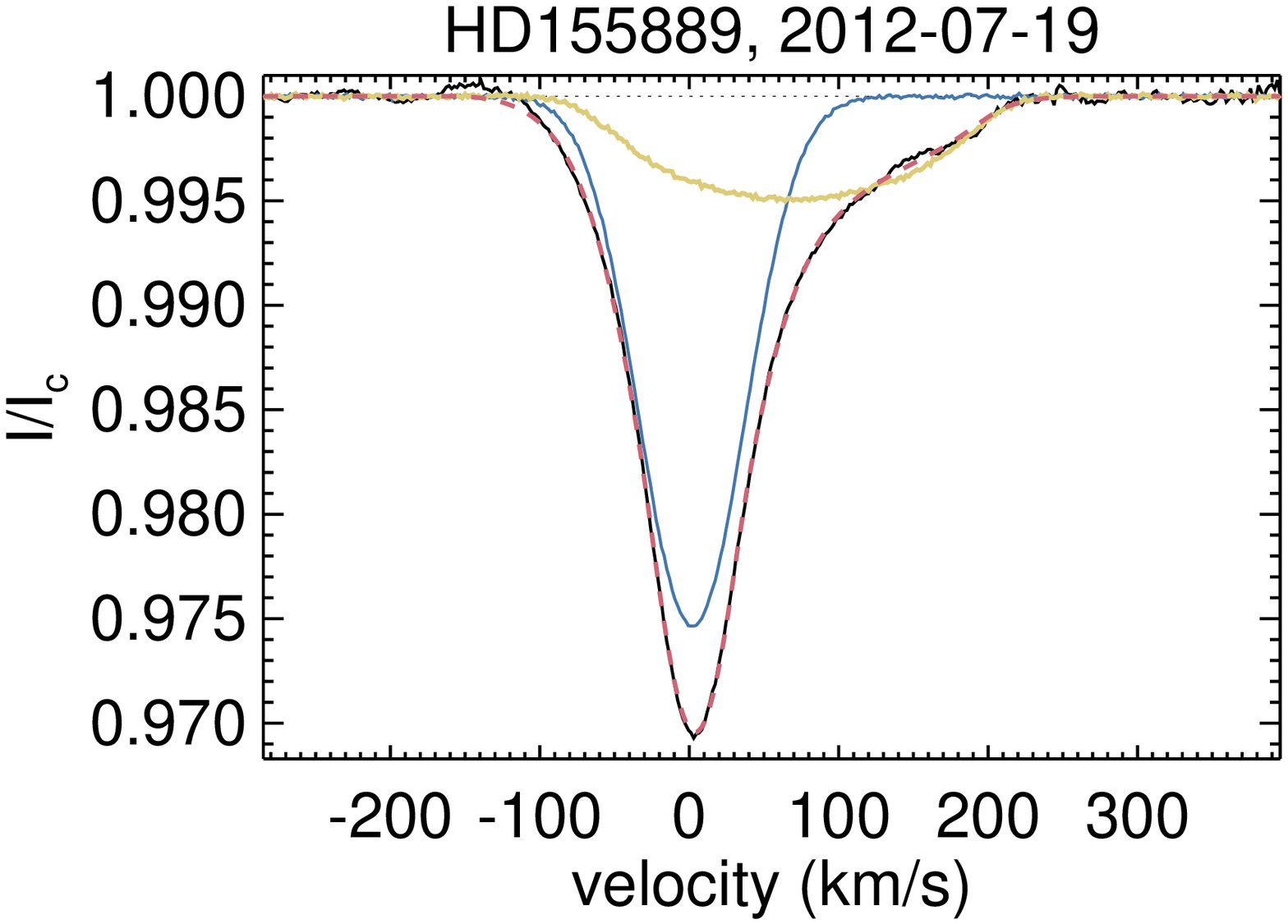}
\includegraphics[width=2.3in]{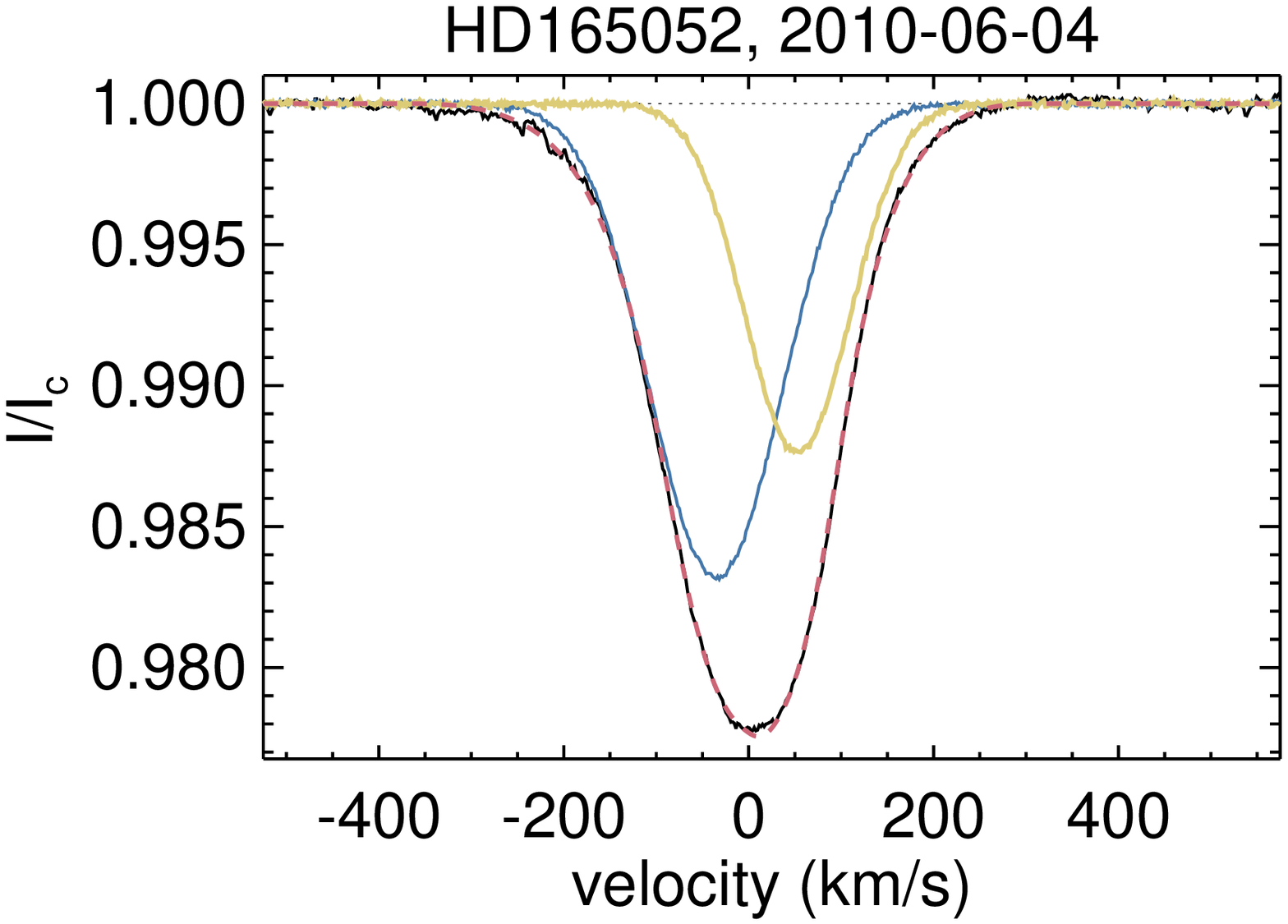}
\includegraphics[width=2.3in]{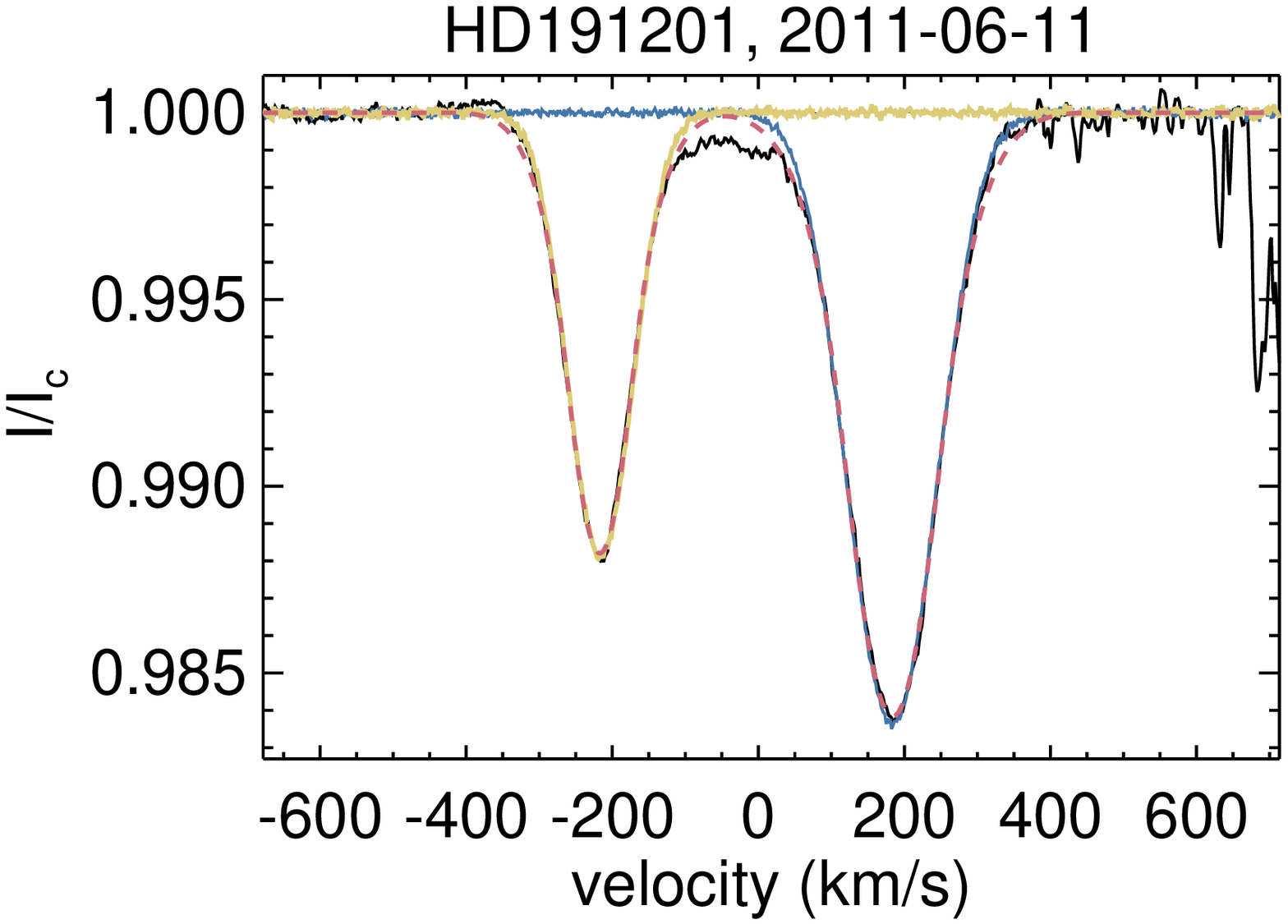}\\
\caption{Example unpolarized Stokes $I$ LSD profiles for multi-line spectroscopic binaries for which we were able to fit individual profiles to each component. The solid black line is the observed LSD profile, the solid blue line is the fit to the primary profile, solid yellow line is the fit to the secondary profile, the solid green line is the fit to the tertiary profile (where applicable), and the dashed red line is the combined fit to the observed line profile. The name and observation date have been provided for each profile.}
\label{bin_profs_fig}
\end{figure*}

\begin{figure*}
\centering
\includegraphics[width=2.3in]{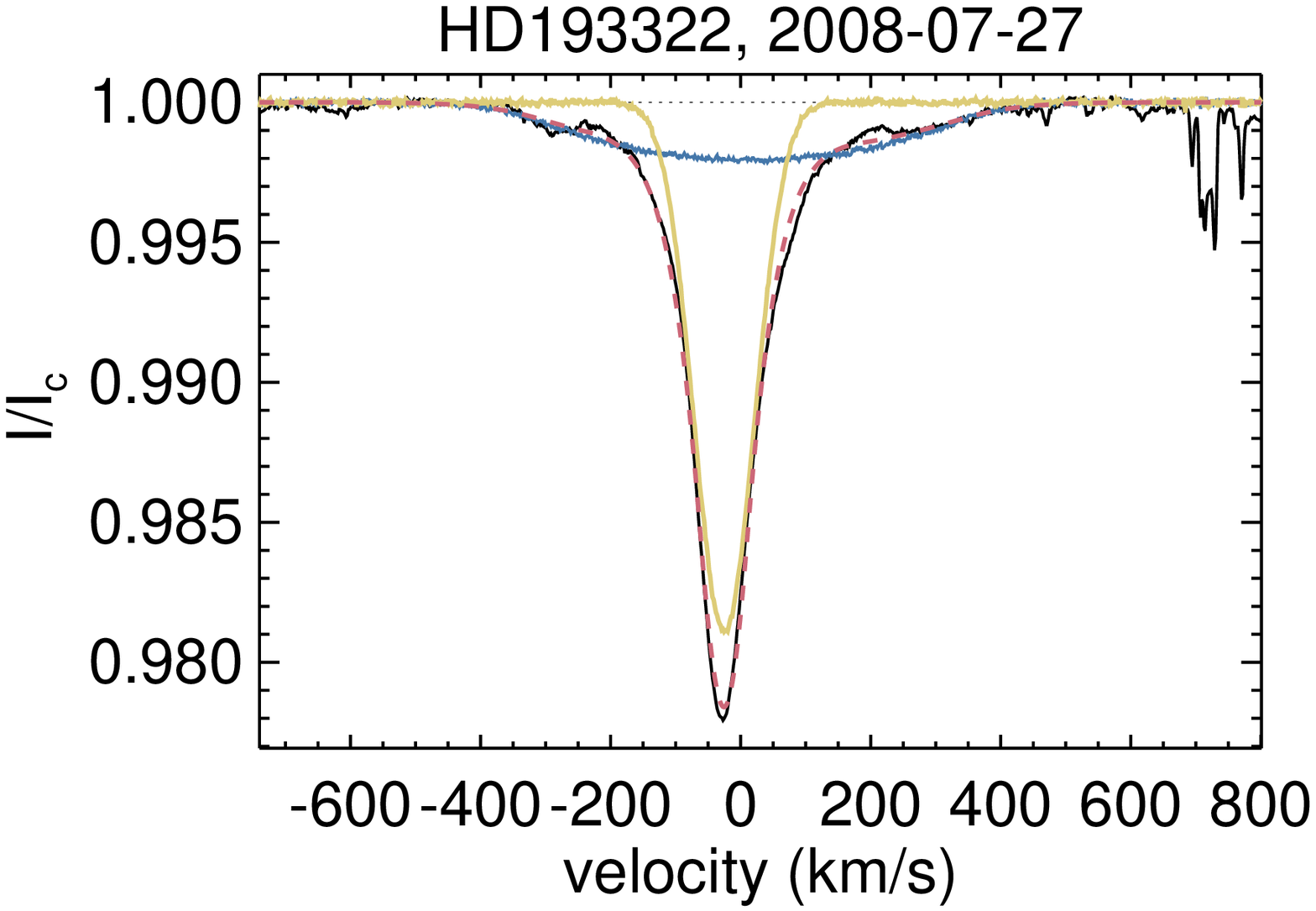}
\includegraphics[width=2.3in]{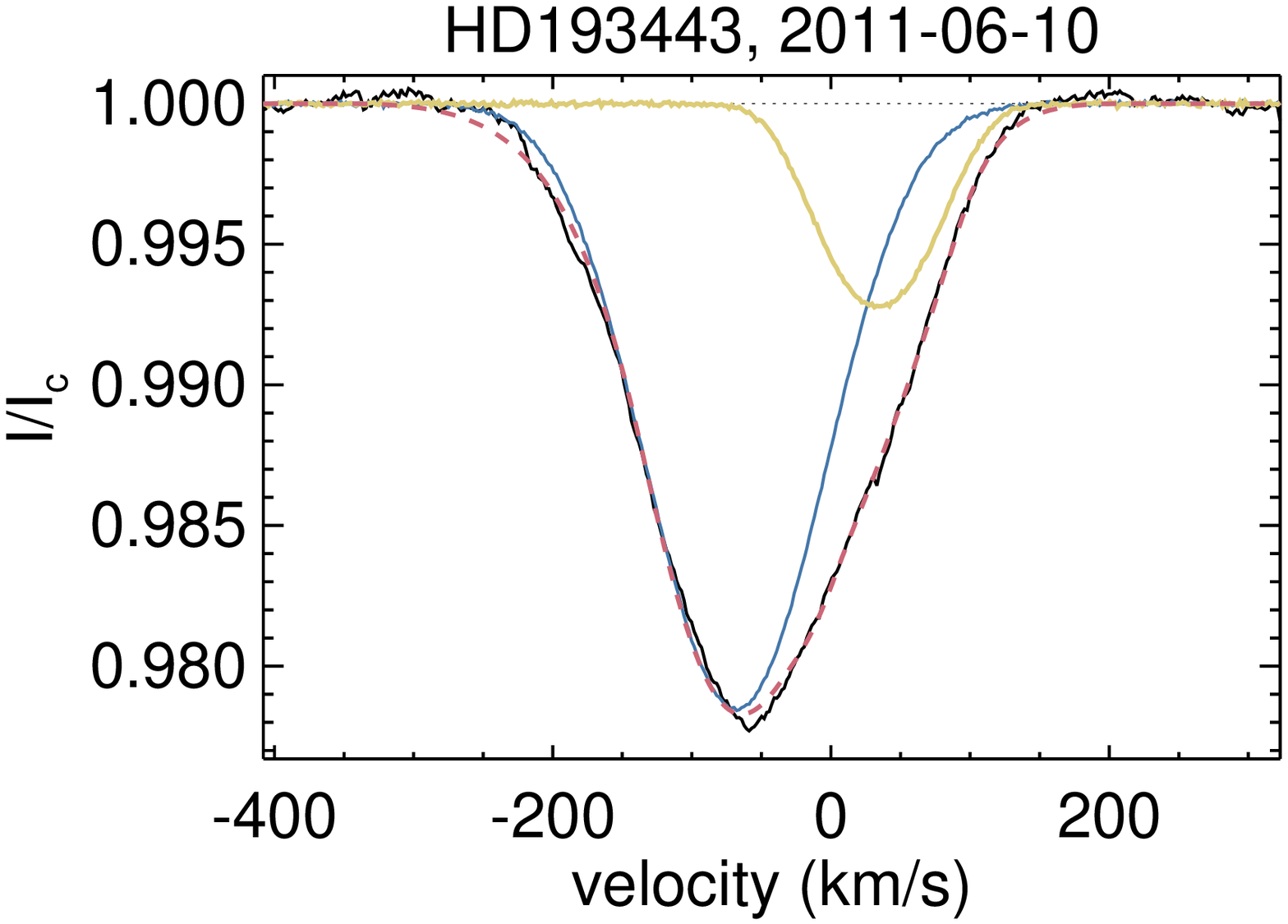}
\includegraphics[width=2.3in]{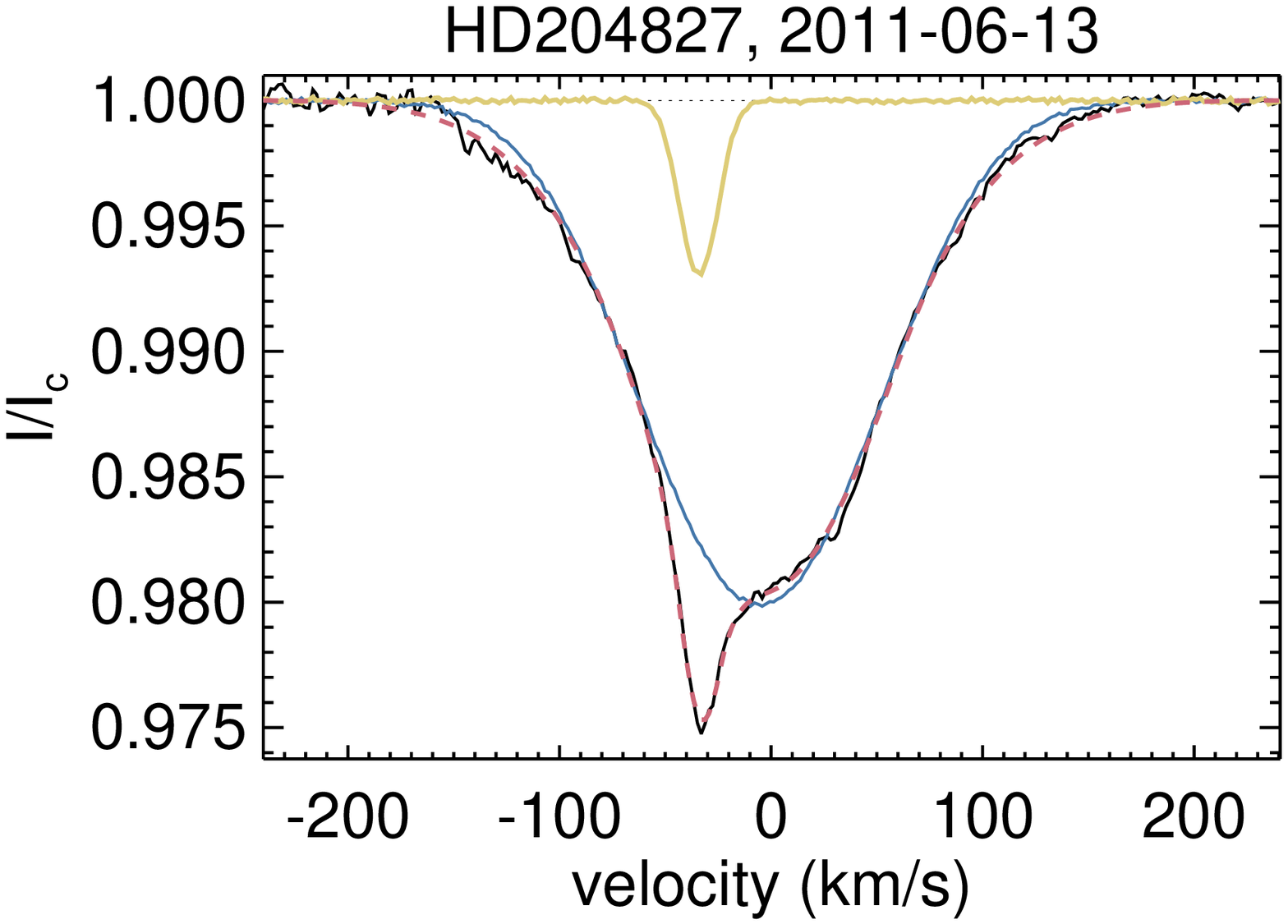}\\
\includegraphics[width=2.3in]{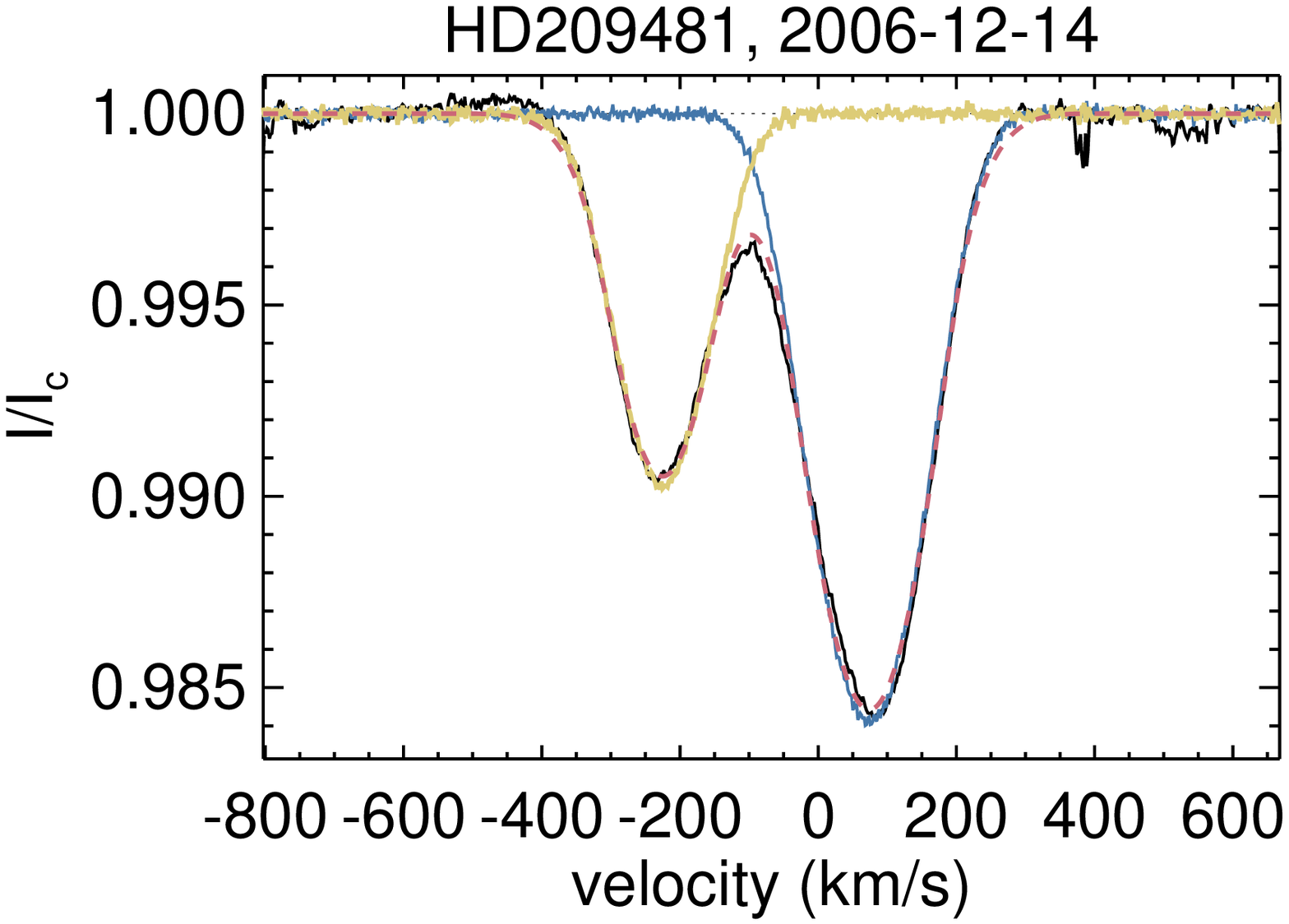}
\contcaption{}
\end{figure*}

\begin{figure*}
\centering
\includegraphics[width=2.3in]{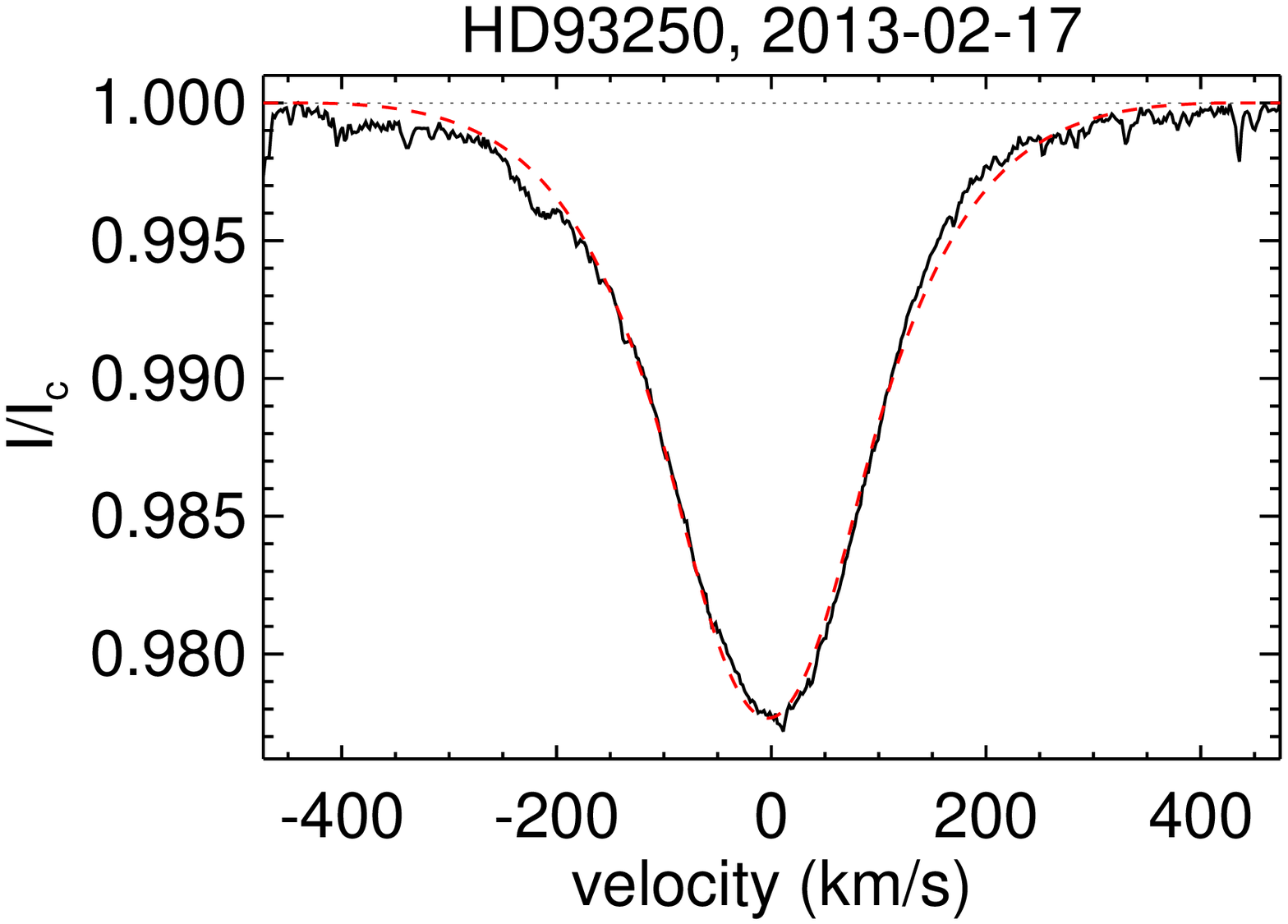}
\includegraphics[width=2.3in]{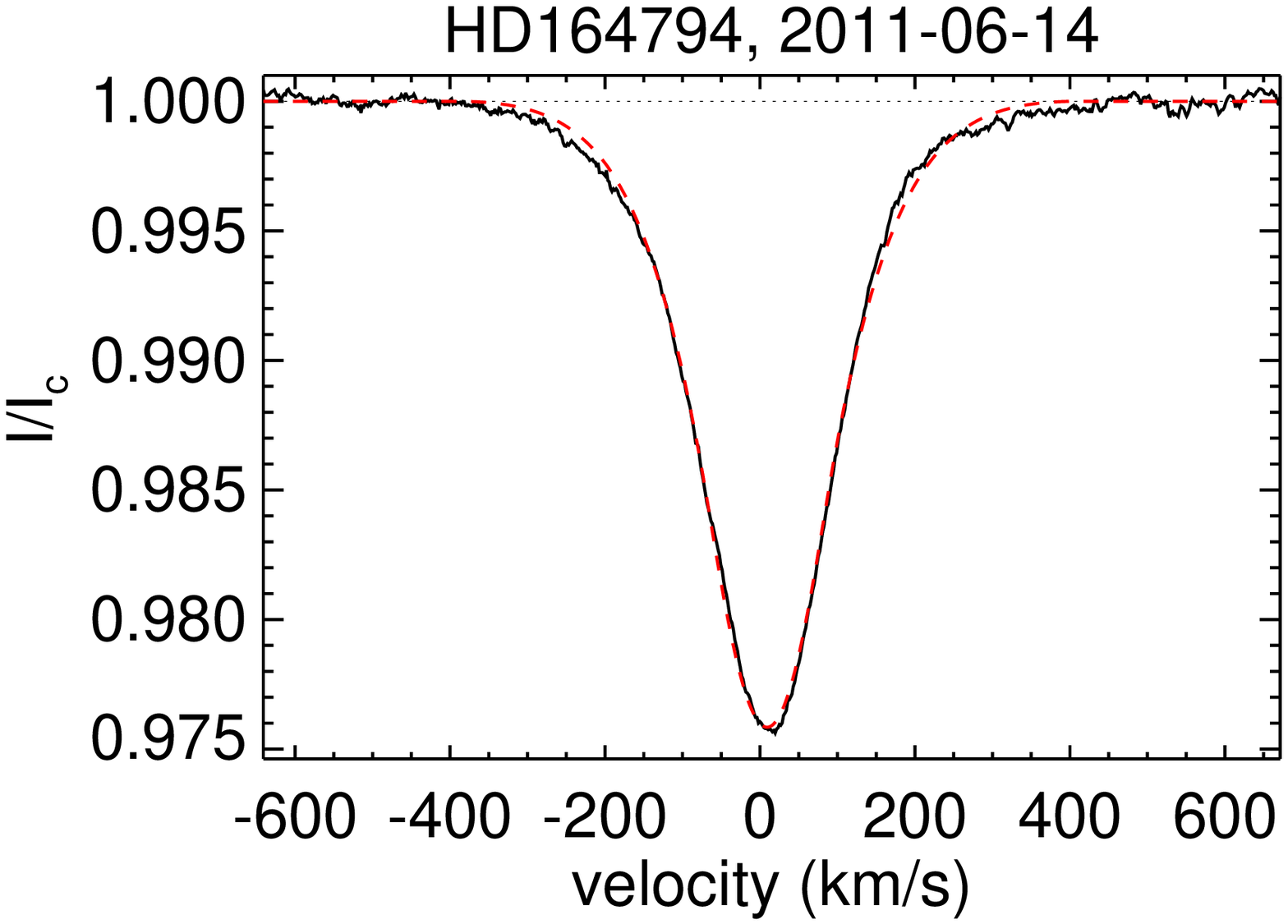}
\includegraphics[width=2.3in]{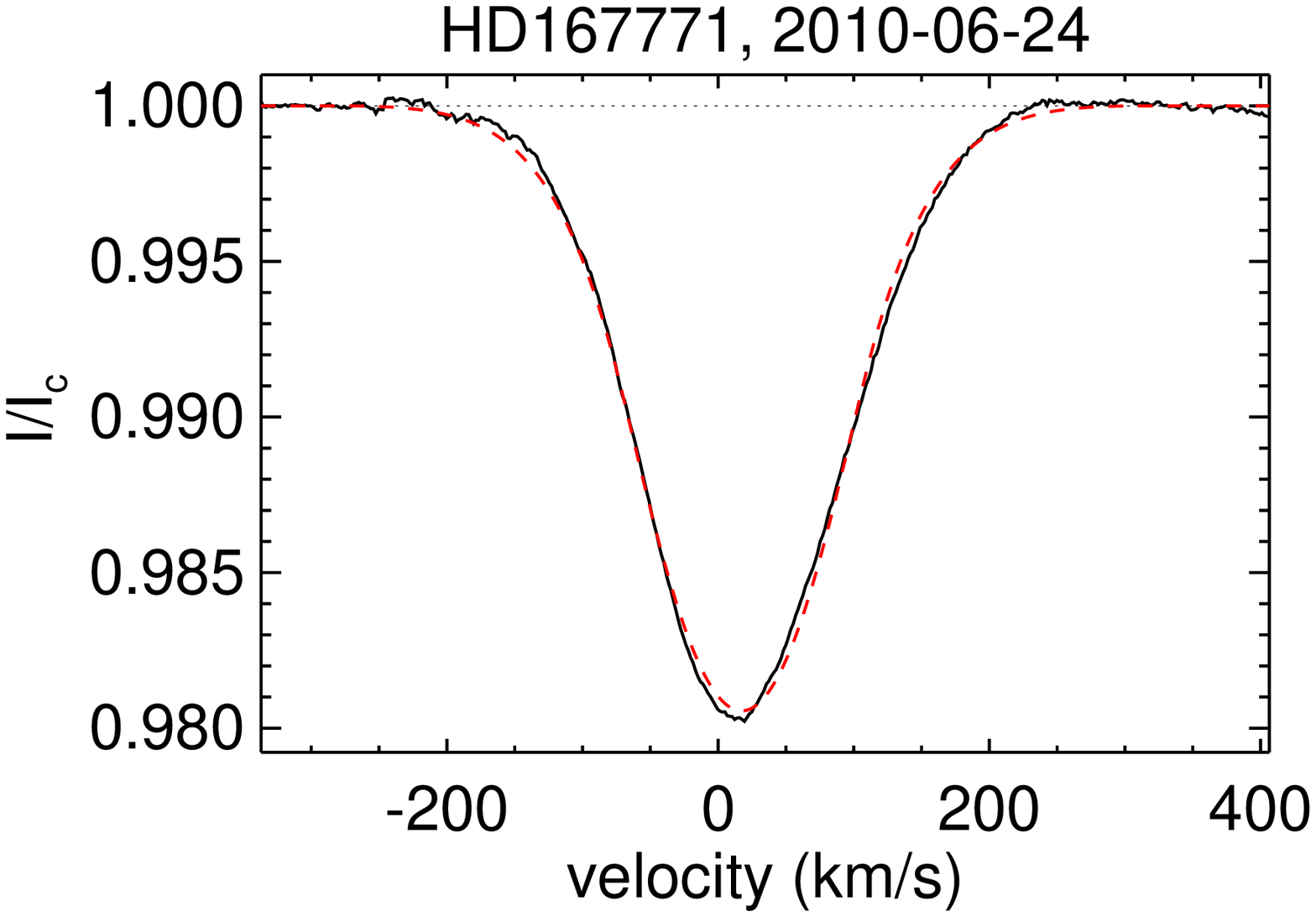}\\
\includegraphics[width=2.3in]{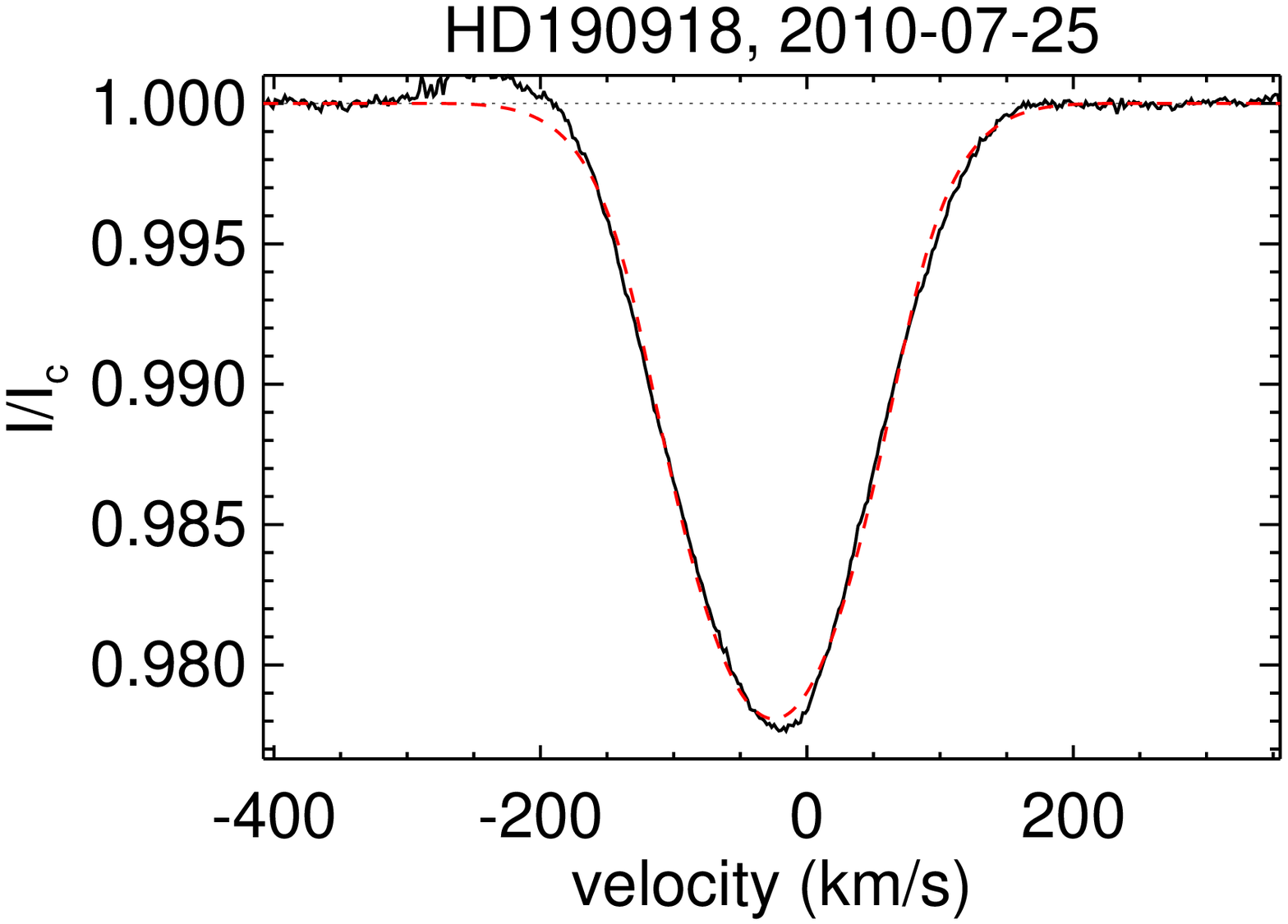}
\includegraphics[width=2.3in]{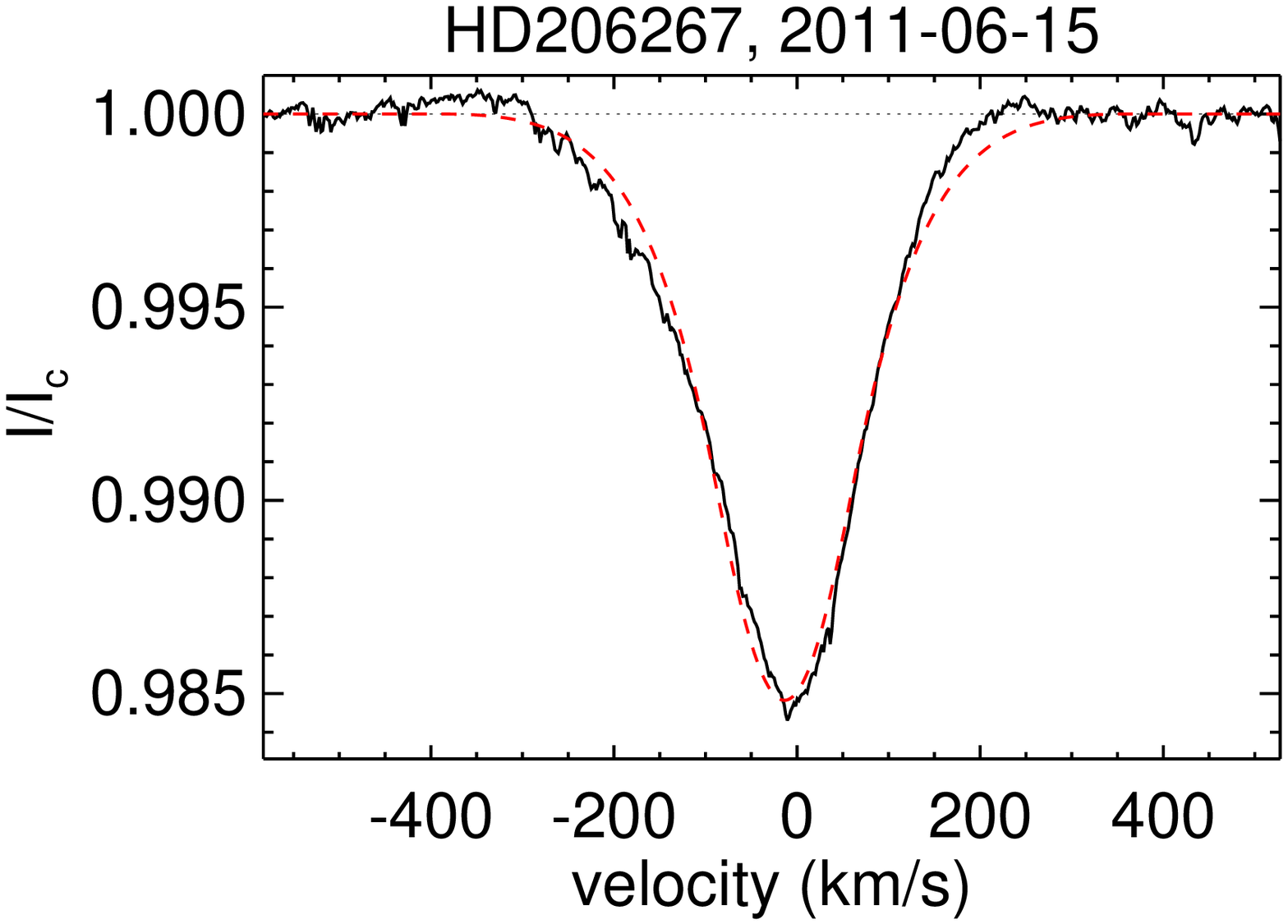}
\caption{Example unpolarized Stokes $I$ LSD profiles for multi-line spectroscopic binaries for which we were unable to fit individual profiles to each component. The solid black line is the observed LSD profile, the dashed red line is a single star synthetic profile fit to the entangled observation. The name and observation date have been provided for each star.}
\label{entang_profs_fig}
\end{figure*}

\section{Assessment of LSD noise characteristics}\label{lsd_noise_sect}
We describe here the details of an analysis used to characterise the reliability of the LSD uncertainties. On the one hand, if the uncertainties are over-estimated this affects our ability to detect signal and therefore identify magnetic stars. On the other hand, if the uncertainties are under-estimated, this would result in a larger number of spurious detections. In order to address this problem we compared the FAPs measured from the observed null profiles to a theoretical FAP distribution, from the single star population only. The theoretical distribution was obtained by generating a series of random profiles with the same noise characteristics as the observed profiles and then measuring the FAP of each theoretical null profile. The individual pixel uncertainties used for the theoretical profiles were created assuming a Gaussian distribution with a standard deviation equal to the uncertainty found in the observed LSD profile. 100 theoretical profiles generated with random noise were obtained for each observation to create the theoretical distribution.

\begin{figure}
\centering
\includegraphics[width=3.2in]{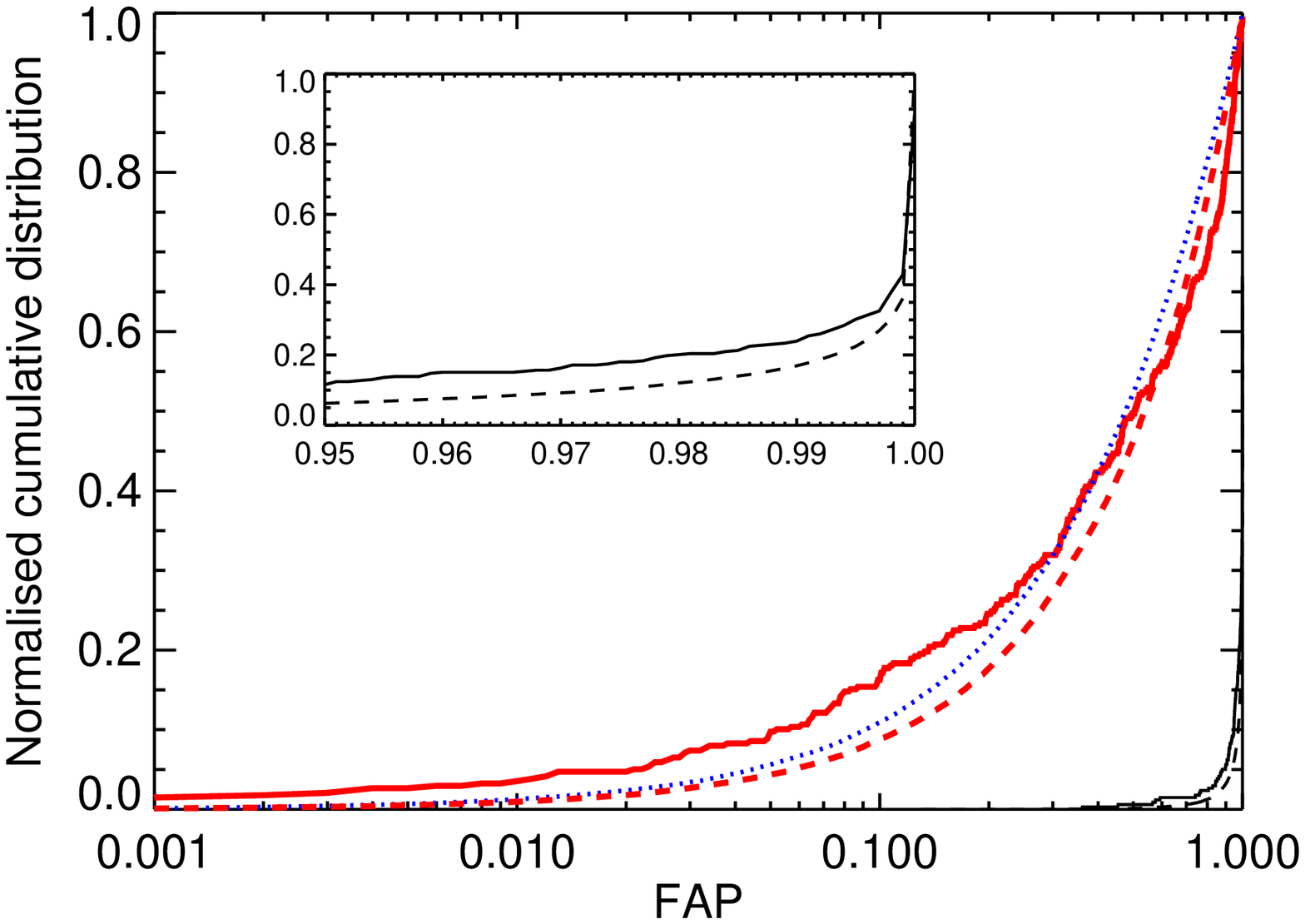}\\
\caption{Comparison of cumulative distributions of the FAP. The FAP distribution obtained from the observed unbinned null profiles (thin black) and observed binned null profiles (thick red) are compared to a theoretical null distribution (dotted blue). The theoretical null distribution is the same when using the uncertainties obtained from the observed binned or unbinned profiles. Also shown are theoretical null distributions where the individual pixel uncertainties have been decreased by 20\% for the unbinned profiles (thin black dashed) and decreased by 3\% for the binned profiles (thick red dashed). The inset provides an expanded view of a small region of the unbinned distributions to highlight the differences. The corrected distributions significantly improve the agreement between the observed and theoretical distributions.}
\label{fap_dist_fig}
\end{figure}

In Fig.~\ref{fap_dist_fig}, we compare the observed and theoretical cumulative distributions for both the binned and unbinned profiles. The obtained FAP distribution from the observed unbinned profiles is significantly different from the predicted distribution - the observed sample contains a significantly larger fraction of high-FAP values compared to the theoretical distribution. One possible interpretation of this mismatch is that the uncertainties established from the LSD procedure are incorrect. We therefore proceeded to recompute the FAP distribution by adjusting all LSD uncertainties by a fixed value. A two-sided KS test was used to find a fixed value that provided the best agreement between the noise-adjusted distribution and the predicted null distribution. The results from this exercise suggest that the LSD uncertainties are {\it over-estimated} by about 20\%. We carried out this same procedure for the binned profiles and the results were very different - in this case the uncertainties were over-estimated by about 3\%. We repeated this same analysis with the LSD profiles generated by the \citet{donati97} LSD code. The results are similar, but the LSD profiles generated with the \citeauthor{donati97} code result in an over-estimation by about 10\% for the unbinned profiles and 6\% for the binned profiles. 

\begin{figure}
\centering
\includegraphics[width=3.2in]{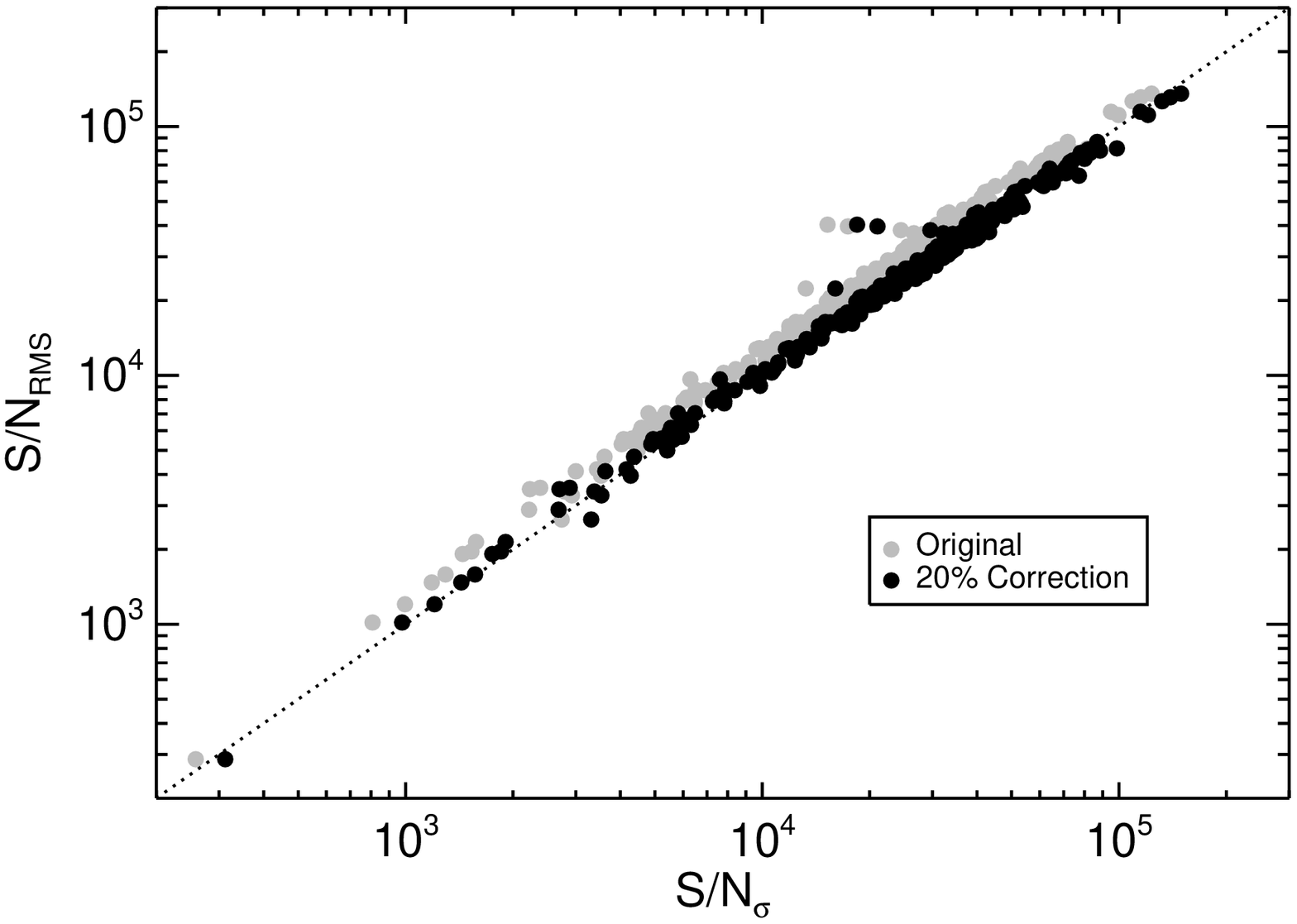}
\caption{Comparison of the S/N level estimated from the LSD profiles ($<1/\sigma>$), versus the S/N estimated from the RMS of the diagnostic null profile. The grey points correspond to the original measurements, while the black points correspond to the same measurements, but increasing the S/N by 20\%, as found from the noise analysis discussed in the text. The dotted line corresponds to a one-to-one relation.}
\label{lsd_snr_comp_fig}
\end{figure}

Another check to test the reliability of the LSD uncertainties was to compare the mean S/N estimated from the pixel uncertainties ($1/\sigma$) with the S/N estimated from the root-mean-square (RMS) deviation from no signal in the null profiles. In the case of pure Gaussian noise, the RMS S/N should equal the $\sigma$ S/N. In Fig.~\ref{lsd_snr_comp_fig}, we compare the S/N obtained from each method. The results show that there is a systematic offset between the two S/N measurements - the S/N obtained from the RMS measurements is consistently higher than the S/N obtained from the mean uncertainty. These results are consistent with the previous analysis. Increasing the S/N measurements from the uncertainties by about 20\% brings the two S/N measurements into much better agreement. 

The level of disagreement of the uncertainties on the order of 10-20\%, as found in this work, is consistent with other magnetometry studies \citep[e.g.][]{wade00}. The unbinned profiles are used to compute $B_\ell$ and $N_\ell$ values, and so these measurements, and their uncertainties, should be considered accurate to within about 10-20\%. This has little impact on the results presented in this work as the $B_\ell$ values are not used to establish whether a Zeeman signature is detected or not. The binned profiles are used to establish the statistical significance of a detected Zeeman signature, and, in this case, the noise characteristics of these profiles agree well with theoretical predictions.

\section{Summary of results}\label{online_tables_sec}
This section provides a summary of all measurements for each observation of each star. Table~\ref{single_star_tab} presents the results for all presumably single O stars, while Table~\ref{binary_star_tab} presents the results for all stars that are either part of known multi-line spectroscopic systems, or where there is sufficient evidence to suggest that the stars are part of multi-line spectroscopic systems.

\begin{landscape}
\begin{table}
\centering
\caption{Table of observations and magnetic results for spectroscopically single stars. Included are the names adopted in this analysis, the common name, the spectral type and the luminosity class, the adopted line mask, the observation date, the heliocentric Julian date at mid exposure (2450000+), the exposure time of the polarimetric sequence, the instrument used (ESPaDOnS (E), Narval (N), or HARPSpol (H)), the peak S/N in the co-added spectrum within the range of 500 to 600\,nm, the optimal velocity width determined for the binned LSD profile, the detection flag (definite detection (DD), marginal detection (MD) or non-detection (ND)), the integration range corresponding to the line profile, the longitudinal magnetic field strength ($B_\ell$), the longitudinal field strength measured from the diagnostic null profile ($N_\ell$), and the uncertainty of the longitudinal field ($\sigma$). Also given are the results from the line fitting routine: the radial velocity ($v_r$), the projected rotational velocity (\vsini), additional non-rotational velocity considered macroturbulence (\vmac), and the median total line broadening $v_{\rm tot}$, computed by summing \vsini\ and \vmac\ in quadrature. Some observations consisted of multiple sequences with different exposure times. We list all sequences, but only provide details for the combined spectra.}\label{single_star_tab}
\begin{tabular}{c@{\hskip 0.05in}c@{\hskip 0.05in}c@{\hskip 0.025in}cc@{\hskip 0.05in}r@{\hskip 0.05in}rc@{\hskip 0.05in}r@{\hskip 0.05in}r@{\hskip 0.05in}r@{\hskip 0.05in}r@{\hskip 0.05in}r@{\hskip 0.05in}r@{\hskip 0.05in}r@{\hskip 0.05in}r@{\hskip 0.05in}r@{\hskip 0.05in}r@{\hskip 0.05in}r@{\hskip 0.05in}r}
\hline
Name & Common & Spec &  Mask & Date & \multicolumn{1}{c}{HJD} &\multicolumn{1}{c}{Exp} & Ins & \multicolumn{1}{c}{S/N} & Width & Det & Int lims & $B_\ell$  & $N_\ell$ & $\sigma$ & $v_r$ & \vsini & \vmac & \multicolumn{1}{c}{$v_{\rm tot}$} \\
\ & name & type & \ & \ & \ & \multicolumn{1}{c}{(s)} & \ & \ & (\kms) & flag & (\kms) & (G) & (G) & (G) & (\kms) & (\kms) & (\kms) & (\kms) \\
\hline
HD\,108 & \ & O8f?p\,var & t35000g40 & 2007-10-15 & 4389.4322 & $3\times4\times900$ & N & 930 & 16.2 &ND & -300,50 & -52 & 15 & 22 & -95 & 115 & 98 & 151 \\ 
\ & \ & \ & \ & 2007-10-16 & 4390.5286 & $3\times4\times900$ & N & 831 & 18.0 &ND & -300,50 & -5 & -34 & 27 & -91 & 122 & 86 & 149 \\ 
\ & \ & \ & \ & 2007-10-18 & 4392.3743 & $3\times4\times900$ & N & 609 & 12.6 &ND & -300,50 & -53 & 13 & 33 & -94 & 121 & 87 & 149 \\ 
\ & \ & \ & \ & 2007-10-19 & 4393.4062 & $3\times4\times900$ & N & 978 & 7.2 &ND & -300,50 & -61 & -5 & 20 & -93 & 122 & 96 & 155 \\ 
\ & \ & \ & \ & 2007-10-20 & 4394.3890 & $3\times4\times900$ & N & 958 & 18.0 &MD & -300,50 & -65 & 31 & 20 & -97 & 124 & 89 & 153 \\ 
\ & \ & \ & \ & 2007-10-21 & 4395.4171 & $3\times4\times900$ & N & 962 & 14.4 &MD & -300,50 & -69 & -9 & 20 & -97 & 123 & 96 & 156 \\
\hline
\end{tabular}
\end{table}
\end{landscape}
\begin{landscape}
\begin{table}
\centering
\contcaption{}
\begin{tabular}{c@{\hskip 0.05in}c@{\hskip 0.05in}c@{\hskip 0.025in}cc@{\hskip 0.05in}r@{\hskip 0.05in}rc@{\hskip 0.05in}r@{\hskip 0.05in}r@{\hskip 0.05in}r@{\hskip 0.05in}r@{\hskip 0.05in}r@{\hskip 0.05in}r@{\hskip 0.05in}r@{\hskip 0.05in}r@{\hskip 0.05in}r@{\hskip 0.05in}r@{\hskip 0.05in}r@{\hskip 0.05in}r}
\hline
Name & Common & Spec &  Mask & Date & \multicolumn{1}{c}{HJD} &\multicolumn{1}{c}{Exp} & Ins & \multicolumn{1}{c}{S/N} & Width & Det & Int lims & $B_\ell$  & $N_\ell$ & $\sigma$ & $v_r$ & \vsini & \vmac & \multicolumn{1}{c}{$v_{\rm tot}$} \\
\ & name & type & \ & \ & \ & \multicolumn{1}{c}{(s)} & \ & \ & (\kms) & flag & (\kms) & (G) & (G) & (G) & (\kms) & (\kms) & (\kms) & (\kms) \\
\hline
HD\,13745 & \ & O9.7\,II(n) & t30000g35 & 2012-08-21 & 6161.6453 & $1\times4\times1200$ & N & 488 & 10.8 &ND & -230,225 & -21 & 60 & 82 & -23 & 184 & 64 & 195 \\ 
\ & \ & \ & \ & 2012-08-22 & 6162.6066 & $1\times4\times1200$ & N & 405 & 16.2 &ND & -230,225 & -132 & 116 & 105 & -20 & 159 & 112 & 195 \\ 
\ & \ & \ & \ & 2012-08-23 & 6163.6452 & $1\times4\times1200$ & N & 478 & 21.6 &ND & -230,225 & 108 & 103 & 72 & -18 & 179 & 92 & 201 \\ 
HD\,14633 & \ & ON8.5\,V & t34000g40 & 2009-10-08 & 5113.8973 & $1\times4\times1300$ & E & 869 & 19.8 &ND & -230,155 & -7 & 32 & 37 & -35 & 123 & 86 & 150 \\ 
HD\,24431 & \ & O9\,III & t32000g35 & 2011-11-06 & 5873.0370 & $2\times4\times720$ & E & 1144 & 14.4 &ND & -180,150 & 13 & 5 & 24 & -14 & 58 & 114 & 128 \\ 
HD\,24534 & X\,Per & O9.5npe & t32000g40 & 2010-12-13 & 5544.7754 & $2\times4\times790$ & E & 1741 & 12.6 &ND & -265,240 & 308 & -86 & 111 & -1 & 205 & 93 & 225 \\
\hline
\end{tabular}
\end{table}
\end{landscape}
\begin{landscape}
\begin{table}
\centering
\contcaption{}
\begin{tabular}{c@{\hskip 0.05in}c@{\hskip 0.05in}c@{\hskip 0.025in}cc@{\hskip 0.05in}r@{\hskip 0.05in}rc@{\hskip 0.05in}r@{\hskip 0.05in}r@{\hskip 0.05in}r@{\hskip 0.05in}r@{\hskip 0.05in}r@{\hskip 0.05in}r@{\hskip 0.05in}r@{\hskip 0.05in}r@{\hskip 0.05in}r@{\hskip 0.05in}r@{\hskip 0.05in}r@{\hskip 0.05in}r}
\hline
Name & Common & Spec &  Mask & Date & \multicolumn{1}{c}{HJD} &\multicolumn{1}{c}{Exp} & Ins & \multicolumn{1}{c}{S/N} & Width & Det & Int lims & $B_\ell$  & $N_\ell$ & $\sigma$ & $v_r$ & \vsini & \vmac & \multicolumn{1}{c}{$v_{\rm tot}$} \\
\ & name & type & \ & \ & \ & \multicolumn{1}{c}{(s)} & \ & \ & (\kms) & flag & (\kms) & (G) & (G) & (G) & (\kms) & (\kms) & (\kms) & (\kms) \\
\hline
\ & \ & \ & \ & 2008-10-14 & 4754.5248 & $2\times4\times90$ & N & 885 & 16.2 &ND & -110,205 & 46 & 15 & 26 & 34 & 62 & 91 & 110 \\ 
\ & \ & \ & \ & 2008-10-26 & 4766.4996 & $7\times4\times90$ & N & 1925 & 9.0 &ND & -110,205 & -13 & -26 & 12 & 35 & 62 & 89 & 108 \\ 
\ & \ & \ & \ & 2009-03-15 & 4906.3005 & $1\times4\times150$ & N & 975 & 9.0 &ND & -110,205 & 1 & 48 & 23 & 32 & 62 & 88 & 108 \\ 
\ & \ & \ & \ & 2009-03-17 & 4908.3350 & $1\times4\times150$ & N & 1147 & 7.2 &ND & -110,205 & -20 & 8 & 20 & 31 & 62 & 91 & 110 \\ 
\ & \ & \ & \ & 2010-10-16 & 5486.9385 & $4\times4\times70$ & E & 2513 & 16.2 &ND & -110,205 & 8 & 6 & 10 & 34 & 59 & 90 & 107 \\ 
HD\,36879 & \ & O7\,V(n)((f)) & t37000g40 & 2009-09-05 & 5081.0801 & $1\times4\times900$ & E & 747 & 18.0 &ND & -220,275 & 36 & -78 & 78 & 32 & 187 & 94 & 209 \\
\hline
\end{tabular}
\end{table}
\end{landscape}
\begin{landscape}
\begin{table}
\centering
\contcaption{}
\begin{tabular}{c@{\hskip 0.05in}c@{\hskip 0.05in}c@{\hskip 0.025in}cc@{\hskip 0.05in}r@{\hskip 0.05in}rc@{\hskip 0.05in}r@{\hskip 0.05in}r@{\hskip 0.05in}r@{\hskip 0.05in}r@{\hskip 0.05in}r@{\hskip 0.05in}r@{\hskip 0.05in}r@{\hskip 0.05in}r@{\hskip 0.05in}r@{\hskip 0.05in}r@{\hskip 0.05in}r@{\hskip 0.05in}r}
\hline
Name & Common & Spec &  Mask & Date & \multicolumn{1}{c}{HJD} &\multicolumn{1}{c}{Exp} & Ins & \multicolumn{1}{c}{S/N} & Width & Det & Int lims & $B_\ell$  & $N_\ell$ & $\sigma$ & $v_r$ & \vsini & \vmac & \multicolumn{1}{c}{$v_{\rm tot}$} \\
\ & name & type & \ & \ & \ & \multicolumn{1}{c}{(s)} & \ & \ & (\kms) & flag & (\kms) & (G) & (G) & (G) & (\kms) & (\kms) & (\kms) & (\kms) \\
\hline
\ & \ & \ & \ & 2010-12-30 & 5561.0732 & $2\times4\times415$ & E & 957 & 1.8 &{\bf DD} & -60,110 & 92 & -6 & 14 & 22 & 5 & 62 & 62 \\ 
HD\,66788 & \ & O8\,V & t35000g40 & 2010-01-23 & 5220.9878 & $1\times4\times600$ & E & 189 & 14.4 &ND & 15,280 & 53 & -122 & 90 & 149 & 27 & 87 & 91 \\ 
\ & \ & \ & \ & 2010-01-31 & 5228.9922 & $1\times4\times600$ & E & 219 & 9.0 &ND & 15,280 & -179 & -74 & 114 & 148 & 27 & 83 & 88 \\ 
HD\,66811 & $\zeta$\,Pup & O4\,If & t40000g40 & 2011-12-13 & 5909.7247 & $5\times4\times120$ & H & 1940 & 18.0 &MD & -345,300 & 27 & -51 & 22 & -16 & 175 & 198 & 264 \\ 
\ & \ & \ & \ & 2012-02-13 & 5971.8333 & $30\times4\times20$ & E & 6195 & 32.4 &ND & -345,300 & -5 & -2 & 15 & -14 & 200 & 158 & 255 \\ 
HD\,69106 & \ & O9.7\,In & t30000g35 & 2012-01-09 & 5937.0428 & $1\times4\times1710$ & E & 1183 & 27.0 &ND & -340,375 & 113 & 255 & 92 & 10 & 311 & 81 & 321 \\
\hline
\end{tabular}
\end{table}
\end{landscape}
\begin{landscape}
\begin{table}
\centering
\contcaption{}
\begin{tabular}{c@{\hskip 0.05in}c@{\hskip 0.05in}c@{\hskip 0.025in}cc@{\hskip 0.05in}r@{\hskip 0.05in}rc@{\hskip 0.05in}r@{\hskip 0.05in}r@{\hskip 0.05in}r@{\hskip 0.05in}r@{\hskip 0.05in}r@{\hskip 0.05in}r@{\hskip 0.05in}r@{\hskip 0.05in}r@{\hskip 0.05in}r@{\hskip 0.05in}r@{\hskip 0.05in}r@{\hskip 0.05in}r}
\hline
Name & Common & Spec &  Mask & Date & \multicolumn{1}{c}{HJD} &\multicolumn{1}{c}{Exp} & Ins & \multicolumn{1}{c}{S/N} & Width & Det & Int lims & $B_\ell$  & $N_\ell$ & $\sigma$ & $v_r$ & \vsini & \vmac & \multicolumn{1}{c}{$v_{\rm tot}$} \\
\ & name & type & \ & \ & \ & \multicolumn{1}{c}{(s)} & \ & \ & (\kms) & flag & (\kms) & (G) & (G) & (G) & (\kms) & (\kms) & (\kms) & (\kms) \\
\hline
\ & \ & \ & \ & 2011-08-20 & 5794.4007 & $1\times4\times45$ & N & 751 & 34.2 &ND & -420,420 & 181 & -197 & 155 & 0 & 375 & 97 & 388 \\ 
\ & \ & \ & \ & 2011-08-21 & 5795.3316 & $1\times4\times45$ & N & 1021 & 30.6 &ND & -420,420 & 143 & 110 & 122 & 3 & 351 & 157 & 385 \\ 
\ & \ & \ & \ & 2011-08-22 & 5796.3507 & $1\times4\times45$ & N & 634 & 36.0 &ND & -420,420 & 146 & -235 & 198 & -1 & 373 & 78 & 381 \\ 
\ & \ & \ & \ & 2011-08-23 & 5797.3309 & $1\times4\times45$ & N & 1051 & 5.4 &ND & -420,420 & -63 & 40 & 120 & -1 & 367 & 91 & 378 \\ 
\ & \ & \ & \ & 2011-08-26 & 5800.3249 & $1\times4\times45$ & N & 517 & 28.8 &ND & -420,420 & 1 & -137 & 266 & 2 & 363 & 129 & 386 \\ 
\ & \ & \ & \ & 2011-08-27 & 5801.3478 & $1\times4\times45$ & N & 593 & 30.6 &ND & -420,420 & 193 & 131 & 205 & -1 & 363 & 130 & 386 \\
\hline
\end{tabular}
\end{table}
\end{landscape}
\begin{landscape}
\begin{table}
\centering
\contcaption{}
\begin{tabular}{c@{\hskip 0.05in}c@{\hskip 0.05in}c@{\hskip 0.025in}cc@{\hskip 0.05in}r@{\hskip 0.05in}rc@{\hskip 0.05in}r@{\hskip 0.05in}r@{\hskip 0.05in}r@{\hskip 0.05in}r@{\hskip 0.05in}r@{\hskip 0.05in}r@{\hskip 0.05in}r@{\hskip 0.05in}r@{\hskip 0.05in}r@{\hskip 0.05in}r@{\hskip 0.05in}r@{\hskip 0.05in}r}
\hline
Name & Common & Spec &  Mask & Date & \multicolumn{1}{c}{HJD} &\multicolumn{1}{c}{Exp} & Ins & \multicolumn{1}{c}{S/N} & Width & Det & Int lims & $B_\ell$  & $N_\ell$ & $\sigma$ & $v_r$ & \vsini & \vmac & \multicolumn{1}{c}{$v_{\rm tot}$} \\
\ & name & type & \ & \ & \ & \multicolumn{1}{c}{(s)} & \ & \ & (\kms) & flag & (\kms) & (G) & (G) & (G) & (\kms) & (\kms) & (\kms) & (\kms) \\
\hline
HD\,167263 & 16\,Sgr & O9.5\,II-IIIn & t31000g35 & 2011-06-14 & 5728.0772 & $1\times4\times500$ & E & 1200 & 25.2 &ND & -230,258 & -74 & -39 & 44 & 5 & 51 & 201 & 207 \\ 
HD\,167264 & 15\,Sgr & O9.7\,Iab & t30000g35 & 2005-06-20 & 3543.0215 & $3\times4\times300$ & E & 1263 & 5.4 &ND & -155,170 & -14 & -3 & 19 & 19 & 78 & 92 & 120 \\ 
\ & \ & \ & \ & 2009-07-25 & 5038.3841 & $1\times4\times675$ & N & 917 & 12.6 &ND & -155,170 & 1 & -8 & 26 & -1 & 60 & 104 & 120 \\ 
\ & \ & \ & \ & 2009-07-26 & 5039.3971 & $1\times4\times600$ & N & 539 & 16.2 &ND & -155,170 & 4 & -19 & 43 & 0 & 1 & 141 & 141 \\ 
\ & \ & \ & \ & 2009-07-27 & 5040.3804 & $1\times4\times600$ & N & 788 & 10.8 &ND & -155,170 & -4 & -8 & 30 & -1 & 62 & 102 & 119 \\ 
\ & \ & \ & \ & 2009-07-28 & 5041.3726 & $1\times4\times600$ & N & 943 & 12.6 &ND & -155,170 & -17 & 25 & 24 & 0 & 61 & 101 & 118 \\
\hline
\end{tabular}
\end{table}
\end{landscape}
\begin{landscape}
\begin{table}
\centering
\contcaption{}
\begin{tabular}{c@{\hskip 0.05in}c@{\hskip 0.05in}c@{\hskip 0.025in}cc@{\hskip 0.05in}r@{\hskip 0.05in}rc@{\hskip 0.05in}r@{\hskip 0.05in}r@{\hskip 0.05in}r@{\hskip 0.05in}r@{\hskip 0.05in}r@{\hskip 0.05in}r@{\hskip 0.05in}r@{\hskip 0.05in}r@{\hskip 0.05in}r@{\hskip 0.05in}r@{\hskip 0.05in}r@{\hskip 0.05in}r}
\hline
Name & Common & Spec &  Mask & Date & \multicolumn{1}{c}{HJD} &\multicolumn{1}{c}{Exp} & Ins & \multicolumn{1}{c}{S/N} & Width & Det & Int lims & $B_\ell$  & $N_\ell$ & $\sigma$ & $v_r$ & \vsini & \vmac & \multicolumn{1}{c}{$v_{\rm tot}$} \\
\ & name & type & \ & \ & \ & \multicolumn{1}{c}{(s)} & \ & \ & (\kms) & flag & (\kms) & (G) & (G) & (G) & (\kms) & (\kms) & (\kms) & (\kms) \\
\hline
\ & \ & \ & \ & 2007-11-09 & 4414.4893 & $1\times4\times550$ & N & 787 & 16.2 &ND & -175,170 & 27 & -26 & 31 & -17 & 69 & 106 & 126 \\ 
\ & \ & \ & \ & 2007-11-13 & 4418.2738 & $1\times4\times200$ & N & 459 & 16.2 &ND & -175,170 & -94 & 25 & 51 & -17 & 73 & 112 & 134 \\ 
\ & \ & \ & \ & 2007-12-21 & 4639.5936 & $3\times4\times275$ & E & 2157 & 18.0 &ND & -175,170 & -25 & 5 & 11 & -17 & 74 & 104 & 128 \\ 
\ & \ & \ & \ & 2008-06-21 & 4639.5936 & $5\times4\times450$ & N & 1572 & 7.2 &ND & -175,170 & 19 & -25 & 14 & -9 & 84 & 105 & 134 \\ 
\ & \ & \ & \ & 2008-06-22 & 4640.5958 & $5\times4\times450$ & N & 2341 & 14.4 &ND & -175,170 & 16 & -13 & 10 & -15 & 65 & 108 & 126 \\ 
\ & \ & \ & \ & 2008-06-25 & 4643.6168 & $3\times4\times450$ & N & 2029 & 14.4 &ND & -175,170 & 9 & -28 & 12 & -15 & 72 & 100 & 123 \\
\hline
\end{tabular}
\end{table}
\end{landscape}
\begin{landscape}
\begin{table}
\centering
\contcaption{}
\begin{tabular}{c@{\hskip 0.05in}c@{\hskip 0.05in}c@{\hskip 0.025in}cc@{\hskip 0.05in}r@{\hskip 0.05in}rc@{\hskip 0.05in}r@{\hskip 0.05in}r@{\hskip 0.05in}r@{\hskip 0.05in}r@{\hskip 0.05in}r@{\hskip 0.05in}r@{\hskip 0.05in}r@{\hskip 0.05in}r@{\hskip 0.05in}r@{\hskip 0.05in}r@{\hskip 0.05in}r@{\hskip 0.05in}r}
\hline
Name & Common & Spec &  Mask & Date & \multicolumn{1}{c}{HJD} &\multicolumn{1}{c}{Exp} & Ins & \multicolumn{1}{c}{S/N} & Width & Det & Int lims & $B_\ell$  & $N_\ell$ & $\sigma$ & $v_r$ & \vsini & \vmac & \multicolumn{1}{c}{$v_{\rm tot}$} \\
\ & name & type & \ & \ & \ & \multicolumn{1}{c}{(s)} & \ & \ & (\kms) & flag & (\kms) & (G) & (G) & (G) & (\kms) & (\kms) & (\kms) & (\kms) \\
\hline
\ & \ & \ & \ & 2007-10-15 & 4389.3354 & $3\times4\times300$ & N & 1596 & 10.8 &ND & -105,85 & 17 & 3 & 9 & -11 & 27 & 43 & 51 \\ 
\ & \ & \ & \ & 2007-10-16 & 4390.4334 & $3\times4\times300$ & N & 1377 & 10.8 &ND & -105,85 & -11 & 9 & 10 & -9 & 26 & 46 & 53 \\ 
\ & \ & \ & \ & 2007-10-17 & 4391.4930 & $3\times4\times500$ & N & 2188 & 9.0 &ND & -105,85 & -3 & -5 & 7 & -11 & 25 & 45 & 52 \\ 
\ & \ & \ & \ & 2007-10-18 & 4392.2742 & $3\times4\times300$ & N & 1139 & 9.0 &ND & -105,85 & 11 & 1 & 12 & -10 & 27 & 47 & 54 \\ 
\ & \ & \ & \ & 2007-10-19 & 4393.2939 & $3\times4\times500$ & N & 1993 & 5.4 &ND & -105,85 & 9 & -1 & 7 & -9 & 25 & 45 & 52 \\ 
\ & \ & \ & \ & 2007-10-20 & 4394.2781 & $3\times4\times500$ & N & 2022 & 5.4 &ND & -105,85 & 9 & 6 & 7 & -11 & 28 & 42 & 50 \\
\hline
\end{tabular}
\end{table}
\end{landscape}
\begin{landscape}
\begin{table}
\centering
\contcaption{}
\begin{tabular}{c@{\hskip 0.05in}c@{\hskip 0.05in}c@{\hskip 0.025in}cc@{\hskip 0.05in}r@{\hskip 0.05in}rc@{\hskip 0.05in}r@{\hskip 0.05in}r@{\hskip 0.05in}r@{\hskip 0.05in}r@{\hskip 0.05in}r@{\hskip 0.05in}r@{\hskip 0.05in}r@{\hskip 0.05in}r@{\hskip 0.05in}r@{\hskip 0.05in}r@{\hskip 0.05in}r@{\hskip 0.05in}r}
\hline
Name & Common & Spec &  Mask & Date & \multicolumn{1}{c}{HJD} &\multicolumn{1}{c}{Exp} & Ins & \multicolumn{1}{c}{S/N} & Width & Det & Int lims & $B_\ell$  & $N_\ell$ & $\sigma$ & $v_r$ & \vsini & \vmac & \multicolumn{1}{c}{$v_{\rm tot}$} \\
\ & name & type & \ & \ & \ & \multicolumn{1}{c}{(s)} & \ & \ & (\kms) & flag & (\kms) & (G) & (G) & (G) & (\kms) & (\kms) & (\kms) & (\kms) \\
\hline
\ & \ & \ & \ & 2012-02-08 & 5966.7229 & $1\times4\times600$ & E & 88 & 9.0 &{\bf DD} & -105,40 & 3493 & 141 & 629 & -35 & 10 & 84 & 85 \\ 
\ & \ & \ & \ & 2012-03-24 & 6011.3300 & $1\times4\times1200$ & N & 78 & 5.4 &MD & -105,40 & 3296 & -23 & 690 & -32 & 14 & 72 & 73 \\ 
\ & \ & \ & \ & 2012-09-26 & 6197.9751 & $2\times4\times1350$ & E & 222 & 7.2 &ND & -105,40 & -23 & -206 & 215 & -31 & 14 & 69 & 70 \\ 
\ & \ & \ & \ & 2012-09-29 & 6200.9997 & $2\times4\times1350$ & E & 204 & 9.0 &ND & -105,40 & 31 & -46 & 262 & -31 & 16 & 67 & 69 \\ 
\ & \ & \ & \ & 2012-12-08 & 6270.7804 & $2\times4\times1350$ & E & 207 & 1.8 &{\bf DD} & -105,40 & 3316 & 143 & 329 & -33 & 14 & 85 & 86 \\ 
\ & \ & \ & \ & 2012-12-23 & 6285.8219 & $3\times4\times1350$ & E & 260 & 1.8 &{\bf DD} & -105,40 & 4269 & -69 & 243 & -35 & 14 & 82 & 83 \\ 
\ & \ & \ & \ & 2012-12-31 & 6293.7615 & $2\times4\times1350$ & E & 253 & 1.8 &{\bf DD} & -105,40 & 3536 & -50 & 249 & -33 & 13 & 81 & 82 \\ 
\ & \ & \ & \ & 2013-01-02 & 6295.8571 & $1\times4\times1350$ & E & 70 & 3.6 &ND & -105,40 & 4348 & 406 & 2721 & -30 & 17 & 90 & 91 \\ 
\hline
\end{tabular}
\end{table}
\end{landscape}
\begin{landscape}
\begin{table}
\centering

\caption{Table of observations and magnetic results for spectroscopic binary systems. Included are the name adopted in this analysis, the common name, the spectral type, the luminosity class, the adopted line mask, the observation date, the heliocentric Julian date at mid exposure (2450000+), the exposure time of the polarimetric sequence, the instrument used (ESPaDOnS (E), Narval (N), or HARPSpol (H)), the peak S/N in the co-added spectrum within the range of 500 to 600\,nm, the designation of the spectroscopic component, the optimal velocity width determined for the binned LSD profile, the detection flag (definite detection (DD), marginal detection (MD) or non-detection (ND)), the integration range corresponding to the line profile, the longitudinal magnetic field strength ($B_\ell$), the longitudinal field strength measured from the diagnostic null profile ($N_\ell$), and the uncertainty of the longitudinal field ($\sigma$). Also given are the results from the line fitting routine: the radial velocity ($v_r$), the projected rotational velocity (\vsini), additional non-rotational velocity considered macroturbulence (\vmac, and the total line broadening $v_{\rm tot}$., computed by adding \vsini\ and \vmac\ in quadrature. We were unable to disentangle individual profiles for the last 6 stars listed. Some observations consisted of multiple sequences with different exposure times. We list all sequences, but only provide details for the combined spectra.}\label{binary_star_tab}
\begin{tabular}{c@{\hskip 0.05in}c@{\hskip 0.05in}c@{\hskip 0.025in}cc@{\hskip 0.05in}r@{\hskip 0.05in}rc@{\hskip 0.05in}r@{\hskip 0.05in}c@{\hskip 0.05in}r@{\hskip 0.05in}r@{\hskip 0.05in}r@{\hskip 0.05in}r@{\hskip 0.05in}r@{\hskip 0.05in}r@{\hskip 0.05in}r@{\hskip 0.05in}r@{\hskip 0.05in}r@{\hskip 0.05in}r@{\hskip 0.05in}r}
\hline
Name & Common & Spec &  Mask & Date & \multicolumn{1}{c}{HJD}& \multicolumn{1}{c}{Exp} & Ins & \multicolumn{1}{c}{S/N} & Spec & Width & Det & Int lims & $B_\ell$  & $N_\ell$ & $\sigma$ & $v_r$ & \vsini & \vmac & \multicolumn{1}{c}{$v_{\rm tot}$} \\
\ & name & type & \ & \ & \ & \multicolumn{1}{c}{(s)} & \ & \ & Comp & (\kms) & flag & (\kms) & (G) & (G) & (G) & (\kms) & (\kms) & (\kms) & (\kms) \\
\hline
HD\,1337 & AO\,Cas & O9.5\,II(n) & t31000g35 & 2009-10-08 & 5113.8454 & $1\times4\times720$ & E & 1280 & A &10.8 & ND & -6,363 & 46 & -37 & 39 & 178 & 118 & 113 & 164 \\ 
\ & \ & \ & \ & \ & \ & \ & \ & \ & B &14.4 & ND & -319,0 & -46 & -134 & 74 & -160 & 82 & 99 & 129 \\ 
HD\,17505 & \ & O6.5\,IIIn((f)) & t37000g35 & 2009-12-03 & 5169.7075 & $1\times4\times600$ & E & 660 & Aa1 &10.8 & ND & -348,-67 & 86 & -9 & 126 & -209 & 42 & 129 & 136 \\ 
\ & \ & \ & \ & \ & \ & \ & \ & \ & Ab &14.4 & ND & -179,176 & -314 & 61 & 128 & -10 & 53 & 154 & 163 \\ 
\ & \ & \ & \ & \ & \ & \ & \ & \ & Aa2 &14.4 & ND & 1,271 & -74 & -44 & 121 & 123 & 62 & 126 & 141 \\ 
HD\,35921 & \ & O9.5\,II & t31000g35 & 2011-11-12 & 5878.8720 & $1\times4\times940$ & E & 1215 & A &23.4 & ND & -114,372 & -102 & -44 & 75 & 130 & 202 & 79 & 217 \\ 
\ & \ & \ & \ & \ & \ & \ & \ & \ & B &21.6 & ND & -492,-81 & 76 & -48 & 104 & -287 & 122 & 117 & 169 \\ 
\ & \ & \ & \ & \ & \ & \ & \ & \ & C &3.6 & ND & -53,19 & 16 & -45 & 74 & -18 & 27 & 16 & 31 \\ 
HD\,37041 & $\theta^2$\,Ori\,A & O9.5\,Ivp & t32000g40 & 2006-01-09 & 3746.0046 & $4\times4\times1.7$ & E & 1467 & Aa &14.4 & ND & -240,104 & 2 & 19 & 39 & -68 & 119 & 86 & 147 \\ 
\ & \ & \ & \ & \ & \ & \ & \ & \ & Ab &36.0 & ND & -344,514 & -45 & 357 & 285 & 92 & 202 & 374 & 425 \\ 
\ & \ & \ & \ & 2007-03-06 & 4166.8592 & $3\times4\times400$ & E & 2558 & Aa &18.0 & ND & -256,96 & 9 & -46 & 22 & -80 & 119 & 90 & 149 \\ 
\ & \ & \ & \ & \ & \ & \ & \ & \ & Ab &45.0 & ND & -339,586 & 180 & 196 & 181 & 123 & 202 & 320 & 379 \\ 
\ & \ & \ & \ & 2012-02-02 & 5960.8833 & $4\times4\times250$ & E & 2051 & Aa &14.4 & ND & -65,286 & 42 & 28 & 28 & 108 & 119 & 90 & 149 \\ 
\ & \ & \ & \ & \ & \ & \ & \ & \ & Ab &28.8 & ND & -353,343 & 574 & -9 & 185 & -6 & 202 & 207 & 289 \\ 
HD\,37043 & $\iota$\,Ori & O9\,III\,var & t32000g35 & 2010-10-17 & 5487.9486 & $12\times4\times30$ & E & 2793 & A &18.0 & ND & -77,256 & 24 & 9 & 11 & 90 & 79 & 111 & 136 \\ 
\ & \ & \ & \ & \ & \ & \ & \ & \ & B &23.4 & ND & -189,267 & 681 & 144 & 273 & 39 & 201 & 57 & 209 \\ 
HD\,37366 & \ & O9.5\,IV & t30000g35 & 2008-08-19 & 4699.1091 & $1\times4\times1600$ & E & 1057 & A &7.2 & ND & -149,6 & -11 & -15 & 13 & -72 & 20 & 61 & 65 \\ 
\ & \ & \ & \ & \ & \ & \ & \ & \ & B &10.8 & ND & -34,285 & 45 & -12 & 96 & 125 & 109 & 82 & 136 \\ 
HD\,37468 & $\sigma$\,Ori & O9.7\,III & t30000g35 & 2008-10-17 & 4757.9598 & $5\times4\times75$ & E & 2790 & Aa &19.8 & ND & -149,229 & -7 & 1 & 19 & 40 & 117 & 106 & 158 \\ 
\ & \ & \ & \ & \ & \ & \ & \ & \ & Ab &5.4 & ND & -41,60 & 5 & -11 & 10 & 9 & 26 & 32 & 41 \\ 
HD\,46106 & \ & O9.7\,II-III & t30000g35 & 2008-01-21 & 4487.8443 & $1\times4\times900$ & E & 605 & A &9.0 & ND & -66,195 & -23 & 65 & 73 & 64 & 85 & 69 & 109 \\ 
\ & \ & \ & \ & \ & \ & \ & \ & \ & B &10.8 & ND & -300,330 & -425 & 567 & 291 & 14 & 253 & 117 & 278 \\ 
HD\,46149 & \ & O8.5\,V & t34000g40 & 2007-03-09 & 4169.9140 & $2\times4\times450$ & E & 547 & A &7.2 & ND & -37,137 & -94 & -1 & 45 & 49 & 40 & 61 & 73 \\ 
\ & \ & \ & \ & \ & \ & \ & \ & \ & B &19.8 & ND & -149,213 & 18 & 2 & 108 & 32 & 140 & 76 & 159 \\ 
\ & \ & \ & \ & 2008-01-20 & 4486.7552 & $3\times4\times860$ & E & 1126 & A &5.4 & ND & -74,104 & -1 & 14 & 21 & 16 & 40 & 60 & 72 \\ 
\ & \ & \ & \ & \ & \ & \ & \ & \ & B &19.8 & ND & -126,239 & -67 & 40 & 50 & 56 & 140 & 78 & 160 \\ 
\ & \ & \ & \ & 2012-02-03 & 5961.8011 & $1\times4\times1140$ & E & 986 & A &9.0 & ND & -40,135 & -32 & -7 & 27 & 47 & 40 & 61 & 73 \\ 
\ & \ & \ & \ & \ & \ & \ & \ & \ & B &12.6 & ND & -142,210 & 86 & -51 & 60 & 34 & 140 & 66 & 155 \\ 
HD\,47129 & Plaskett's~star & O8fp\,var & t32000g40 & 2012-02-03 & 5961.8587 & $2\times4\times600$ & E & 2010 & A1 &10.8 & ND & 82,352 & 29 & -34 & 37 & 217 & 77 & 66 & 102 \\ 
\ & \ & \ & \ & \ & \ & \ & \ & \ & A2 &45.0 & MD & -447,473 & 640 & 221 & 404 & 15 & 370 & 156 & 402 \\ 
\ & \ & \ & \ & 2012-02-08 & 5966.8775 & $4\times4\times600$ & E & 2350 & A1 &10.8 & ND & -119,145 & 27 & 31 & 24 & 12 & 77 & 105 & 130 \\ 
\ & \ & \ & \ & \ & \ & \ & \ & \ & A2 &36.0 & ND & -364,386 & -149 & -510 & 297 & 15 & 370 & 0 & 370 \\ 
\ & \ & \ & \ & 2012-02-09 & 5967.7681 & $2\times4\times600$ & E & 1560 & A1 &14.4 & ND & -209,55 & 13 & -6 & 39 & -77 & 77 & 94 & 122 \\ 
\ & \ & \ & \ & \ & \ & \ & \ & \ & A2 &36.0 & ND & -398,431 & -543 & 165 & 217 & 15 & 370 & 0 & 370 \\ 
\ & \ & \ & \ & 2012-02-11 & 5969.7754 & $4\times4\times600$ & E & 2911 & A1 &7.2 & ND & -289,-67 & 21 & 3 & 18 & -177 & 77 & 75 & 107 \\ 
\ & \ & \ & \ & \ & \ & \ & \ & \ & A2 &23.4 & ND & -355,388 & 98 & -90 & 202 & 15 & 370 & 0 & 370 \\ 
\ & \ & \ & \ & 2012-03-13 & 6000.4172 & $1\times4\times1200$ & N & 977 & A1 &5.4 & ND & -249,-10 & -13 & 68 & 66 & -130 & 77 & 69 & 103 \\ 
\ & \ & \ & \ & \ & \ & \ & \ & \ & A2 &10.8 & ND & -353,379 & -1242 & 964 & 447 & 15 & 370 & 0 & 370 \\

\hline
\end{tabular}
\end{table}
\end{landscape}
\begin{landscape}
\begin{table}
\centering

\contcaption{}
\begin{tabular}{c@{\hskip 0.05in}c@{\hskip 0.05in}c@{\hskip 0.025in}cc@{\hskip 0.05in}r@{\hskip 0.05in}rc@{\hskip 0.05in}r@{\hskip 0.05in}c@{\hskip 0.05in}r@{\hskip 0.05in}r@{\hskip 0.05in}r@{\hskip 0.05in}r@{\hskip 0.05in}r@{\hskip 0.05in}r@{\hskip 0.05in}r@{\hskip 0.05in}r@{\hskip 0.05in}r@{\hskip 0.05in}r@{\hskip 0.05in}r}
\hline
Name & Common & Spec &  Mask & Date & \multicolumn{1}{c}{HJD}& \multicolumn{1}{c}{Exp} & Ins & \multicolumn{1}{c}{S/N} & Spec & Width & Det & Int lims & $B_\ell$  & $N_\ell$ & $\sigma$ & $v_r$ & \vsini & \vmac & \multicolumn{1}{c}{$v_{\rm tot}$} \\
\ & name & type & \ & \ & \ & \multicolumn{1}{c}{(s)} & \ & \ & Comp & (\kms) & flag & (\kms) & (G) & (G) & (G) & (\kms) & (\kms) & (\kms) & (\kms) \\
\hline
\ & \ & \ & \ & 2012-03-14 & 6001.3654 & $1\times4\times1200$ & N & 1049 & A1 &10.8 & ND & -166,43 & -8 & 121 & 52 & -62 & 77 & 47 & 90 \\ 
\ & \ & \ & \ & \ & \ & \ & \ & \ & A2 &43.2 & ND & -510,539 & -269 & 190 & 564 & 15 & 370 & 240 & 441 \\ 
\ & \ & \ & \ & 2012-03-23 & 6010.3588 & $1\times4\times1200$ & N & 587 & A1 &14.4 & ND & -166,167 & 93 & 28 & 131 & -1 & 77 & 110 & 134 \\ 
\ & \ & \ & \ & \ & \ & \ & \ & \ & A2 &25.2 & ND & -353,385 & 59 & -310 & 844 & 15 & 370 & 0 & 370 \\ 
\ & \ & \ & \ & 2012-03-25 & 6012.3442 & $1\times4\times1200$ & N & 1192 & A1 &12.6 & ND & -271,-22 & 26 & -141 & 52 & -148 & 77 & 70 & 104 \\ 
\ & \ & \ & \ & \ & \ & \ & \ & \ & A2 &34.2 & ND & -497,534 & 776 & -484 & 551 & 15 & 370 & 220 & 430 \\ 
\ & \ & \ & \ & 2012-09-24 & 6196.0910 & $2\times4\times900$ & E & 2492 & A1 &10.8 & ND & -31,275 & 18 & 21 & 35 & 122 & 77 & 107 & 132 \\ 
\ & \ & \ & \ & \ & \ & \ & \ & \ & A2 &34.2 & ND & -357,386 & -515 & -181 & 279 & 15 & 370 & 0 & 370 \\ 
\ & \ & \ & \ & 2012-09-26 & 6198.0940 & $2\times4\times900$ & E & 2365 & A1 &12.6 & ND & -200,41 & -20 & 1 & 27 & -80 & 77 & 86 & 116 \\ 
\ & \ & \ & \ & \ & \ & \ & \ & \ & A2 &32.4 & ND & -377,406 & 109 & -101 & 184 & 15 & 370 & 0 & 370 \\ 
\ & \ & \ & \ & 2012-09-27 & 6199.0822 & $2\times4\times900$ & E & 2335 & A1 &10.8 & MD & -242,-17 & 50 & 6 & 29 & -129 & 77 & 67 & 102 \\ 
\ & \ & \ & \ & \ & \ & \ & \ & \ & A2 &14.4 & MD & -373,403 & 230 & -166 & 193 & 15 & 370 & 0 & 370 \\ 
\ & \ & \ & \ & 2012-09-30 & 6202.1011 & $2\times4\times900$ & E & 2283 & A1 &7.2 & ND & -247,3 & 38 & 31 & 39 & -123 & 77 & 77 & 109 \\ 
\ & \ & \ & \ & \ & \ & \ & \ & \ & A2 &34.2 & MD & -355,383 & -304 & 184 & 165 & 15 & 370 & 0 & 370 \\ 
\ & \ & \ & \ & 2012-11-28 & 6261.1258 & $2\times4\times900$ & E & 1858 & A1 &7.2 & ND & -136,91 & -34 & 16 & 42 & -22 & 77 & 63 & 99 \\ 
\ & \ & \ & \ & \ & \ & \ & \ & \ & A2 &43.2 & ND & -443,480 & 103 & -25 & 323 & 15 & 370 & 53 & 374 \\ 
\ & \ & \ & \ & 2012-11-29 & 6262.0632 & $3\times4\times900$ & E & 2040 & A1 &9.0 & ND & -58,183 & -88 & 51 & 45 & 62 & 77 & 87 & 116 \\ 
\ & \ & \ & \ & \ & \ & \ & \ & \ & A2 &43.2 & ND & -366,500 & -204 & -622 & 391 & 15 & 370 & 183 & 413 \\ 
\ & \ & \ & \ & 2012-12-01 & 6264.1275 & $2\times4\times900$ & E & 1045 & A1 &12.6 & ND & 86,354 & 501 & 246 & 171 & 219 & 77 & 92 & 120 \\ 
\ & \ & \ & \ & \ & \ & \ & \ & \ & A2 &45.0 & ND & -425,462 & 565 & -86 & 1626 & 15 & 370 & 0 & 370 \\ 
\ & \ & \ & \ & 2012-12-09 & 6272.1203 & $2\times4\times900$ & E & 2195 & A1 &10.8 & ND & -287,-55 & -5 & 2 & 30 & -171 & 77 & 63 & 99 \\ 
\ & \ & \ & \ & \ & \ & \ & \ & \ & A2 &30.6 & ND & -346,388 & 4 & 78 & 207 & 15 & 370 & 0 & 370 \\ 
\ & \ & \ & \ & 2012-12-20 & 6282.9362 & $2\times4\times900$ & E & 2194 & A1 &9.0 & ND & -53,181 & -48 & 21 & 28 & 63 & 77 & 72 & 105 \\ 
\ & \ & \ & \ & \ & \ & \ & \ & \ & A2 &27.0 & ND & -345,400 & 709 & -1185 & 879 & 15 & 370 & 0 & 370 \\ 
\ & \ & \ & \ & 2012-12-21 & 6284.1161 & $2\times4\times900$ & E & 2222 & A1 &10.8 & ND & -148,55 & 31 & 0 & 27 & -48 & 77 & 64 & 100 \\ 
\ & \ & \ & \ & \ & \ & \ & \ & \ & A2 &34.2 & ND & -386,424 & 248 & 135 & 263 & 15 & 370 & 0 & 370 \\ 
\ & \ & \ & \ & 2012-12-25 & 6288.0983 & $1\times4\times900$ & E & 1195 & A1 &7.2 & ND & -272,-55 & 17 & -12 & 75 & -163 & 77 & 61 & 98 \\ 
\ & \ & \ & \ & \ & \ & \ & \ & \ & A2 &34.2 & ND & -389,422 & -351 & 738 & 607 & 15 & 370 & 0 & 370 \\ 
\ & \ & \ & \ & 2012-12-26 & 6289.1047 & $2\times4\times900$ & E & 2090 & A1 &10.8 & ND & -184,28 & 6 & 21 & 34 & -79 & 77 & 79 & 110 \\ 
\ & \ & \ & \ & \ & \ & \ & \ & \ & A2 &27.0 & ND & -449,484 & 39 & 228 & 319 & 15 & 370 & 0 & 370 \\ 
\ & \ & \ & \ & 2012-12-27 & 6289.9822 & $2\times4\times900$ & E & 2539 & A1 &9.0 & ND & -134,109 & -41 & 13 & 24 & -13 & 77 & 106 & 131 \\ 
\ & \ & \ & \ & \ & \ & \ & \ & \ & A2 &36.0 & ND & -424,458 & -105 & -7 & 235 & 15 & 370 & 0 & 370 \\ 
HD\,47839 & 15\,Mon & O7\,V((f))\,var & t37000g40 & 2006-12-10 & 4080.5001 & $1\times4\times270$ & N & 796 & A &10.8 & ND & -89,156 & -23 & -2 & 30 & 32 & 53 & 86 & 101 \\ 
\ & \ & \ & \ & \ & \ & \ & \ & \ & B &19.8 & ND & -102,273 & -578 & 88 & 540 & 85 & 183 & 34 & 186 \\ 
\ & \ & \ & \ & 2006-12-15 & 4085.4876 & $1\times4\times700$ & N & 1503 & A &12.6 & ND & -91,154 & -1 & 27 & 18 & 30 & 53 & 86 & 101 \\ 
\ & \ & \ & \ & \ & \ & \ & \ & \ & B &18.0 & ND & -120,278 & 41 & 206 & 178 & 78 & 183 & 34 & 186 \\ 
\ & \ & \ & \ & 2007-09-09 & 4353.6886 & $1\times4\times740$ & N & 1120 & A &12.6 & ND & -89,156 & 2 & -22 & 27 & 32 & 53 & 86 & 101 \\ 
\ & \ & \ & \ & \ & \ & \ & \ & \ & B &14.4 & ND & -121,267 & -312 & -172 & 195 & 73 & 183 & 34 & 186 \\ 
\ & \ & \ & \ & 2007-09-10 & 4354.6775 & $1\times4\times1000$ & N & 1383 & A &10.8 & ND & -89,156 & 2 & 3 & 20 & 32 & 53 & 86 & 101 \\ 
\ & \ & \ & \ & \ & \ & \ & \ & \ & B &21.6 & ND & -120,278 & -125 & 43 & 146 & 79 & 183 & 34 & 186 \\ 
\ & \ & \ & \ & 2007-09-11 & 4355.6849 & $1\times4\times800$ & N & 990 & A &12.6 & ND & -91,156 & -20 & -2 & 30 & 32 & 53 & 86 & 101 \\ 
\ & \ & \ & \ & \ & \ & \ & \ & \ & B &12.6 & ND & -127,275 & 209 & 124 & 200 & 71 & 183 & 34 & 186 \\ 
\ & \ & \ & \ & 2007-10-20 & 4394.6986 & $4\times4\times200$ & N & 1589 & A &12.6 & ND & -89,154 & -3 & 6 & 16 & 32 & 53 & 86 & 101 \\ 
\ & \ & \ & \ & \ & \ & \ & \ & \ & B &18.0 & ND & -107,284 & -61 & 233 & 151 & 87 & 183 & 34 & 186 \\ 
\hline
\end{tabular}
\end{table}
\end{landscape}
\begin{landscape}
\begin{table}
\centering

\contcaption{}
\begin{tabular}{c@{\hskip 0.05in}c@{\hskip 0.05in}c@{\hskip 0.025in}cc@{\hskip 0.05in}r@{\hskip 0.05in}rc@{\hskip 0.05in}r@{\hskip 0.05in}c@{\hskip 0.05in}r@{\hskip 0.05in}r@{\hskip 0.05in}r@{\hskip 0.05in}r@{\hskip 0.05in}r@{\hskip 0.05in}r@{\hskip 0.05in}r@{\hskip 0.05in}r@{\hskip 0.05in}r@{\hskip 0.05in}r@{\hskip 0.05in}r}
\hline
Name & Common & Spec &  Mask & Date & \multicolumn{1}{c}{HJD}& \multicolumn{1}{c}{Exp} & Ins & \multicolumn{1}{c}{S/N} & Spec & Width & Det & Int lims & $B_\ell$  & $N_\ell$ & $\sigma$ & $v_r$ & \vsini & \vmac & \multicolumn{1}{c}{$v_{\rm tot}$} \\
\ & name & type & \ & \ & \ & \multicolumn{1}{c}{(s)} & \ & \ & Comp & (\kms) & flag & (\kms) & (G) & (G) & (G) & (\kms) & (\kms) & (\kms) & (\kms) \\
\hline
\ & \ & \ & \ & 2007-10-23 & 4397.7020 & $4\times4\times200$ & N & 1696 & A &12.6 & ND & -91,156 & 1 & 7 & 15 & 32 & 53 & 86 & 101 \\ 
\ & \ & \ & \ & \ & \ & \ & \ & \ & B &14.4 & ND & -120,296 & 329 & 50 & 172 & 89 & 183 & 34 & 186 \\ 
\ & \ & \ & \ & 2012-02-02 & 5960.9956 & $3\times4\times160$ & E & 2098 & A &7.2 & ND & -89,147 & -2 & 1 & 12 & 30 & 53 & 86 & 101 \\ 
\ & \ & \ & \ & \ & \ & \ & \ & \ & B &7.2 & ND & 21,318 & 108 & 61 & 264 & 167 & 183 & 34 & 186 \\ 
HD\,48099 & \ & O6.5\,V(n)((f)) & t38000g40 & 2006-12-14 & 4084.4811 & $1\times4\times960$ & N & 630 & A &18.0 & ND & -121,240 & 59 & -35 & 66 & 58 & 91 & 118 & 149 \\ 
\ & \ & \ & \ & \ & \ & \ & \ & \ & B &9.0 & ND & -168,24 & 64 & 41 & 100 & -73 & 51 & 61 & 80 \\ 
\ & \ & \ & \ & 2012-02-03 & 5961.9608 & $2\times4\times840$ & E & 2024 & A &9.0 & ND & -140,220 & 4 & -36 & 22 & 39 & 91 & 118 & 149 \\ 
\ & \ & \ & \ & \ & \ & \ & \ & \ & B &9.0 & ND & -151,42 & -12 & -25 & 37 & -54 & 51 & 61 & 80 \\ 
HD\,54662 & \ & O7\,V((f))z\,var? & t37000g40 & 2010-10-18 & 5489.0726 & $2\times4\times720$ & E & 1331 & A &5.4 & ND & -61,105 & -23 & -29 & 25 & 22 & 36 & 57 & 67 \\ 
\ & \ & \ & \ & \ & \ & \ & \ & \ & B &16.2 & ND & -126,245 & 11 & -30 & 59 & 58 & 148 & 70 & 163 \\ 
HD\,153426 & \ & O9\,II-III & t32000g35 & 2011-07-03 & 5746.8157 & $1\times4\times800$ & E & 695 & A &14.4 & ND & -257,79 & 57 & 58 & 55 & -90 & 88 & 106 & 138 \\ 
\ & \ & \ & \ & \ & \ & \ & \ & \ & B &27.0 & ND & -236,403 & 422 & 110 & 442 & 85 & 295 & 75 & 304 \\ 
\ & \ & \ & \ & 2012-06-22 & 6101.8866 & $1\times4\times740$ & E & 825 & A &16.2 & ND & -208,123 & -2 & -86 & 52 & -44 & 88 & 106 & 138 \\ 
\ & \ & \ & \ & \ & \ & \ & \ & \ & B &32.4 & ND & -354,342 & -316 & 608 & 736 & -9 & 295 & 75 & 304 \\ 
HD\,155889 & \ & O9.5\,IV & t32000g40 & 2012-07-19 & 6128.7479 & $1\times4\times420$ & H & 277 & A &7.2 & ND & -94,98 & -7 & 42 & 37 & 2 & 29 & 76 & 82 \\ 
\ & \ & \ & \ & \ & \ & \ & \ & \ & B &12.6 & ND & -85,223 & -378 & 224 & 164 & 68 & 135 & 47 & 143 \\ 
HD\,165052 & \ & O5.5Vz+O8V & t38000g40 & 2010-06-04 & 5351.9373 & $1\times4\times1100$ & E & 1190 & A &14.4 & ND & -239,166 & -3 & 27 & 55 & -37 & 63 & 152 & 165 \\ 
\ & \ & \ & \ & \ & \ & \ & \ & \ & B &14.4 & ND & -106,213 & 10 & 61 & 65 & 53 & 52 & 120 & 131 \\ 
HD\,191201 & \ & O9.5\,III+B0\,IV & t30000g35 & 2011-06-11 & 5725.0925 & $1\times4\times1400$ & E & 1047 & A &10.8 & ND & 20,346 & 12 & 3 & 51 & 184 & 72 & 113 & 134 \\ 
\ & \ & \ & \ & \ & \ & \ & \ & \ & B &12.6 & ND & -338,-97 & -51 & 19 & 54 & -218 & 49 & 87 & 100 \\ 
HD\,193322 & \ & O9\,IV(n) & t31000g35 & 2008-07-27 & 4675.9022 & $2\times4\times500$ & E & 1237 & Aa &21.6 & ND & -445,477 & 278 & 38 & 299 & 15 & 350 & 183 & 395 \\ 
\ & \ & \ & \ & \ & \ & \ & \ & \ & Ab1 &9.0 & ND & -155,102 & 27 & -17 & 22 & -27 & 40 & 98 & 106 \\ 
HD\,193443 & \ & O9\,III & t32000g35 & 2011-06-10 & 5724.0826 & $1\times4\times700$ & E & 907 & A &16.2 & ND & -241,103 & 8 & 5 & 39 & -69 & 73 & 121 & 141 \\ 
\ & \ & \ & \ & \ & \ & \ & \ & \ & B &10.8 & ND & -64,134 & -43 & -7 & 77 & 34 & 62 & 57 & 84 \\ 
HD\,204827 & \ & O9.7\,III & t30000g35 & 2011-06-13 & 5727.0551 & $1\times4\times900$ & E & 662 & Aa &16.2 & ND & -148,140 & 98 & 17 & 49 & -5 & 71 & 98 & 120 \\ 
\ & \ & \ & \ & \ & \ & \ & \ & \ & Ab &3.6 & ND & -56,-11 & 45 & 58 & 54 & -35 & 10 & 18 & 21 \\ 
HD\,209481 & 14\,Cep & O9\,IV(n)\,var & t31000g35 & 2006-12-14 & 4084.2744 & $1\times4\times600$ & N & 867 & A &16.2 & ND & -121,268 & 14 & 60 & 74 & 74 & 122 & 109 & 164 \\ 
\ & \ & \ & \ & \ & \ & \ & \ & \ & B &16.2 & ND & -394,-58 & -36 & 51 & 120 & -226 & 92 & 105 & 140 \\ 
\ & \ & \ & \ & 2009-07-20 & 5033.4828 & $1\times4\times675$ & N & 667 & A &21.6 & ND & -179,231 & -73 & 124 & 97 & 27 & 127 & 115 & 171 \\ 
\ & \ & \ & \ & \ & \ & \ & \ & \ & B &5.4 & ND & -275,75 & 160 & -168 & 134 & -100 & 93 & 110 & 144 \\ 
\ & \ & \ & \ & 2009-07-24 & 5037.4851 & $1\times4\times825$ & N & 1191 & A &21.6 & ND & -306,105 & 85 & -32 & 48 & -98 & 127 & 115 & 171 \\ 
\ & \ & \ & \ & \ & \ & \ & \ & \ & B &9.0 & ND & 38,385 & 213 & 24 & 83 & 212 & 93 & 110 & 144 \\ 
\ & \ & \ & \ & 2009-07-25 & 5038.4921 & $1\times4\times675$ & N & 916 & A &19.8 & ND & -180,231 & 59 & 2 & 66 & 26 & 127 & 115 & 171 \\ 
\ & \ & \ & \ & \ & \ & \ & \ & \ & B &10.8 & ND & -271,80 & -51 & 60 & 107 & -95 & 93 & 110 & 144 \\ 
\ & \ & \ & \ & 2009-07-26 & 5039.4922 & $1\times4\times675$ & N & 804 & A &19.8 & ND & -161,253 & 192 & 67 & 62 & 46 & 127 & 115 & 171 \\ 
\ & \ & \ & \ & \ & \ & \ & \ & \ & B &18.0 & ND & -329,20 & -225 & 9 & 90 & -154 & 93 & 110 & 144 \\ 
\ & \ & \ & \ & 2009-07-27 & 5040.4895 & $1\times4\times675$ & N & 315 & A &19.8 & ND & -300,107 & 142 & -112 & 180 & -97 & 127 & 115 & 171 \\ 
\ & \ & \ & \ & \ & \ & \ & \ & \ & B &14.4 & ND & 40,384 & -459 & -307 & 322 & 212 & 93 & 110 & 144 \\ 
\ & \ & \ & \ & 2009-07-28 & 5041.4801 & $1\times4\times675$ & N & 1058 & A &18.0 & ND & -195,216 & 26 & -9 & 54 & 10 & 127 & 115 & 171 \\ 
\ & \ & \ & \ & \ & \ & \ & \ & \ & B &10.8 & ND & -241,110 & 5 & 1 & 76 & -66 & 93 & 110 & 144 \\ 
\ & \ & \ & \ & 2009-07-29 & 5042.4962 & $1\times4\times675$ & N & 1199 & A &12.6 & ND & -152,261 & 23 & -62 & 48 & 54 & 127 & 115 & 171 \\ 
\ & \ & \ & \ & \ & \ & \ & \ & \ & B &12.6 & ND & -348,0 & -21 & 128 & 70 & -174 & 93 & 110 & 144 \\ 
\ & \ & \ & \ & 2009-07-30 & 5043.4763 & $1\times4\times675$ & N & 1037 & A &9.0 & ND & -302,110 & 85 & -17 & 60 & -96 & 127 & 115 & 171 \\ 
\ & \ & \ & \ & \ & \ & \ & \ & \ & B &14.4 & ND & 27,376 & -62 & -49 & 100 & 201 & 93 & 110 & 144 \\ 
\hline
\end{tabular}
\end{table}
\end{landscape}
\begin{landscape}
\begin{table}
\centering

\contcaption{}
\begin{tabular}{c@{\hskip 0.05in}c@{\hskip 0.05in}c@{\hskip 0.025in}cc@{\hskip 0.05in}r@{\hskip 0.05in}rc@{\hskip 0.05in}r@{\hskip 0.05in}c@{\hskip 0.05in}r@{\hskip 0.05in}r@{\hskip 0.05in}r@{\hskip 0.05in}r@{\hskip 0.05in}r@{\hskip 0.05in}r@{\hskip 0.05in}r@{\hskip 0.05in}r@{\hskip 0.05in}r@{\hskip 0.05in}r@{\hskip 0.05in}r}
\hline
Name & Common & Spec &  Mask & Date & \multicolumn{1}{c}{HJD}& \multicolumn{1}{c}{Exp} & Ins & \multicolumn{1}{c}{S/N} & Spec & Width & Det & Int lims & $B_\ell$  & $N_\ell$ & $\sigma$ & $v_r$ & \vsini & \vmac & \multicolumn{1}{c}{$v_{\rm tot}$} \\
\ & name & type & \ & \ & \ & \multicolumn{1}{c}{(s)} & \ & \ & Comp & (\kms) & flag & (\kms) & (G) & (G) & (G) & (\kms) & (\kms) & (\kms) & (\kms) \\
\hline
\ & \ & \ & \ & 2009-07-31 & 5044.4961 & $1\times4\times675$ & N & 1105 & A &19.8 & ND & -208,201 & 2 & -39 & 57 & -4 & 127 & 115 & 171 \\ 
\ & \ & \ & \ & \ & \ & \ & \ & \ & B &18.0 & ND & -213,134 & 40 & -17 & 86 & -40 & 93 & 110 & 144 \\ 
\ & \ & \ & \ & 2009-08-03 & 5047.4605 & $1\times4\times675$ & N & 968 & A &21.6 & ND & -228,180 & 39 & -69 & 69 & -24 & 127 & 115 & 171 \\ 
\ & \ & \ & \ & \ & \ & \ & \ & \ & B &18.0 & ND & -171,182 & 71 & 5 & 89 & 6 & 93 & 110 & 144 \\ 
\ & \ & \ & \ & 2011-06-21 & 5735.0269 & $3\times4\times400$ & E & 2417 & A &19.8 & ND & -265,145 & 3 & 2 & 22 & -60 & 127 & 115 & 171 \\ 
\ & \ & \ & \ & \ & \ & \ & \ & \ & B &18.0 & ND & -66,282 & -3 & -6 & 39 & 108 & 93 & 110 & 144 \\ 
HD\,36486 & $\delta$\,Ori\,A & O9.5\,IINwk & t30000g35 & 2008-10-23 & 4763.4967 & $1\times4\times50$ & N & 866 & A &19.8 & ND & -100,280 & 58 & 16 & 41 & 95 & 126 & 96 & 158 \\ 
\ & \ & \ & \ & 2008-10-24 & 4764.5328 & $10\times4\times50$ & N & 6093 & A &10.8 & {\bf DD} & -220,220 & 17 & 13 & 9 & 0 & 117 & 114 & 163 \\ 
HD\,93250 & \ & O4\,IIIfc: & t42000g35 & 2013-02-17 & 6341.6051 & $2\times4\times900$ & H & 313 & - &14.4 & ND & -271,267 & 138 & 9 & 122 & -3 & 86 & 203 & 220 \\ 
HD\,164794 & 9\,Sgr & O3.5\,V((f*))+O5-5.5\,V((f)) & t43000g45 & 2005-06-19 & 3541.0074 & $5\times4\times300$ & E & 979 & - &19.8 & ND & -234,257 & -30 & -12 & 66 & 12 & 73 & 188 & 202 \\ 
\ & \ & \ & \ & 2005-06-20 & 3542.0184 & $4\times4\times300$ & E & 999 & - &12.6 & ND & -243,264 & -62 & 65 & 66 & 9 & 74 & 195 & 209 \\ 
\ & \ & \ & \ & 2005-06-23 & 3544.9829 & $3\times4\times300$ & E & 1016 & - &18.0 & ND & -243,264 & 34 & -45 & 63 & 10 & 74 & 195 & 209 \\ 
\ & \ & \ & \ & 2011-05-25 & 5707.8434 & $1\times4\times600$ & H & 588 & - &9.0 & ND & -236,252 & -1 & 83 & 48 & 0 & 81 & 185 & 202 \\ 
\ & \ & \ & \ & - & - & $2\times4\times900$ & H &  - &  - & - & -,- & - & - & - & - & - & - & - & - \\ 
\ & \ & \ & \ & 2011-06-14 & 5728.0338 & $2\times4\times600$ & E & 1920 & - &16.2 & ND & -240,257 & 65 & 19 & 34 & 9 & 77 & 188 & 203 \\ 
HD\,167771 & HR\,6841 & O7\,III(f) & t36000g40 & 2010-06-24 & 5373.0014 & $2\times4\times800$ & E & 1576 & - &16.2 & ND & -162,198 & 22 & 3 & 20 & 17 & 91 & 118 & 149 \\ 
HD\,190918 & \ & WN5o+O9\,I & t30000g35 & 2008-07-30 & 4677.8197 & $2\times4\times490$ & E & 1070 & B &18.0 & ND & -217,165 & -147 & 2 & 60 & -10 & 110 & 97 & 147 \\ 
\ & \ & \ & \ & 2010-07-25 & 5404.0084 & $1\times4\times1300$ & E & 866 & B &14.4 & ND & -193,186 & 56 & 127 & 79 & -25 & 104 & 87 & 136 \\ 
HD\,199579 & HR\,8023 & O6.5\,V((f))z & t38000g40 & 2008-08-14 & 4754.8639 & $4\times4\times600$ & E & 2029 & A(?) &12.6 & {\bf DD} & -159,183 & -32 & -9 & 19 & 10 & 60 & 126 & 140 \\ 
HD\,206267 & \ & O6.5\,V((f)) & t38000g40 & 2011-06-15 & 5729.0643 & $6\times4\times450$ & E & 1824 & - &16.2 & ND & -239,211 & 54 & 8 & 46 & -14 & 84 & 165 & 185 \\ 
\hline
\end{tabular}
\end{table}
\end{landscape}


\bsp	
\label{lastpage}
\end{document}